\documentclass[twocolumn,amstext,amssymb,superscriptaddress,floatfix,showpacs,nofootinbib]{revtex4-1}

\usepackage{CJK}
\usepackage[utf8]{inputenc}
\usepackage[table,dvipsnames]{xcolor}
\usepackage{amsmath}
\usepackage{slashed}
\usepackage{physics}
\usepackage{tabularx}
\usepackage{graphicx}
\usepackage{dcolumn}
\usepackage{bm}
\usepackage{hyperref}
\usepackage{natbib}
\usepackage{braket}
\usepackage{booktabs}
\usepackage{feynmp} 
\usepackage{multirow}
\usepackage{pifont}
\usepackage{float}
\usepackage{makecell}
\usepackage{placeins}
\usepackage{chngcntr}

\usepackage{hyperref}

\newcommand{\xmark}{\ding{55}}
\newcommand{\cmark}{{\color{blue}\ding{51}}}
\newcommand{\atMF}{\ensuremath{\left|_{M_F}\right.}}

\newcolumntype{?}{!{\vrule width 1pt}}
\newcolumntype{`}{!{\vrule width 1.5pt}}
\newcommand{\1}{\mathbf{1}}

\makeatletter
\def\l@subsubsection#1#2{}
\makeatother

\definecolor{Gray}{gray}{0.85}
\definecolor{LightGreen}{rgb}{0.88, 1, 0.88}
\definecolor{Blue}{rgb}{0,1,1}
\definecolor{Lime}{rgb}{0,1,0}
\definecolor{LightCyan}{rgb}{0.88,1,1}
\definecolor{LightRed}{rgb}{1, 0.85, 0.85}
\definecolor{Red}{rgb}{1, 0, 0}
\definecolor{LightYellow}{rgb}{1, 1, 0.85}
\definecolor{Yellow}{rgb}{1,1,0.05}
\definecolor{LightBlue}{rgb}{0.87, 0.94, 1}
\definecolor{white}{gray}{1}
\definecolor{black}{gray}{0}

\definecolor{lightgray}{gray}{0.91}
\renewcommand{\arraystretch}{1.4}

\def\di{\displaystyle}
\def\beq{\begin{equation}}
\def\eeq{\end{equation}}\def\eq#1{(\ref{#1})}
\def\eq#1{(\ref{#1})}

\renewcommand{\figurename}{{\bf Figure}}
\renewcommand{\thefigure}{{\bf\arabic{figure}}}
\renewcommand{\tablename}{{\bf Table}}
\renewcommand{\thetable}{{\bf\arabic{table}}}

\renewcommand{\thesection}{{\bf \Roman{section}}}
\renewcommand{\thesubsection}{{\Alph{subsection}}}

\newcolumntype{C}{>{$}c<{$}}
\AtBeginDocument{
\heavyrulewidth=.16em
\lightrulewidth=.1em
\cmidrulewidth=.03em
\belowrulesep=.4ex
\belowbottomsep=0pt
\aboverulesep=.4ex
\abovetopsep=0pt
\cmidrulesep=\doublerulesep
\cmidrulekern=.5em
\defaultaddspace=.5em
}

\makeatletter
    \def\CT@@do@color{%
      \global\let\CT@do@color\relax
            \@tempdima\wd\z@
            \advance\@tempdima\@tempdimb
            \advance\@tempdima\@tempdimc
    \advance\@tempdimb\tabcolsep
    \advance\@tempdimc\tabcolsep
    \advance\@tempdima2\tabcolsep
            \kern-\@tempdimb
            \leaders\vrule
                    \hskip\@tempdima\@plus  1fill
            \kern-\@tempdimc
            \hskip-\wd\z@ \@plus -1fill }
    \makeatother

\begin{document} 
{DO-TH 20/08}\\

\title{Model Building from  Asymptotic Safety with Higgs and Flavor Portals}
\author{Gudrun~Hiller}
\author{Clara~Hormigos-Feliu}
\affiliation{Fakult\"at Physik, TU Dortmund, Otto-Hahn-Str.4, D-44221 Dortmund, Germany}
\author{Daniel~F.~Litim}
\author{Tom~Steudtner}
\affiliation{Department of Physics and Astronomy, University of Sussex, Brighton, BN1 9QH, U.K.}
\begin{abstract}
 We perform a comprehensive search for Standard Model extensions inspired by asymptotic safety.
 Our  models feature a   singlet matrix scalar field, three generations of vector-like  leptons, 
and direct links to the Higgs and flavor  sectors via 
 new Yukawa and portal couplings.  A novel feature is that the enlarged scalar sector  may spontaneously break
 lepton flavor universality.
We provide a complete
two-loop renormalization group analysis 
of the running gauge, Yukawa, and quartic couplings to find ultraviolet fixed points 
and the BSM critical surface of parameters, $i.e.$ the set  of  boundary conditions at the TeV scale for which 
models remain well-behaved and predictive up to the Planck scale without encountering Landau poles or  instabilities.
This includes  templates for asymptotically safe Standard Model extensions 
 which match  
 the measured values of gauge couplings and the Higgs, top, and bottom masses.
We further detail the  phenomenology of our models covering production, decay,  fermion mixing,  
anomalous magnetic moments, effects from scalar mixing and chiral enhancement, and constraints on model parameters from data.   
Signatures at proton-proton and lepton colliders such as lepton flavor violation and displaced vertices, and the prospect for 
electric dipole moments or charged lepton-flavor-violating type 
processes,  are also indicated.
\end{abstract}

\maketitle
\tableofcontents
\flushbottom

\section{\bf Introduction and Basic Setup} 
\label{sec:setup}

\subsection{Motivation and Background}
Ultraviolet (UV) fixed points play a central role for fundamental quantum field theories.
They ensure that running couplings remain finite and well-defined  even at highest energies
such  that cross sections or scattering amplitudes stay well-behaved.
Important examples are given by {asymptotic freedom} of non-abelian gauge interactions and the strong nuclear force, where the 
fixed point is non-interacting   \cite{Gross:1973id, Politzer:1973fx}.
UV fixed points may also be interacting, a scenario known as 
asymptotic safety, and conjectured a while ago  both in particle physics  \cite{Bailin:1974bq} and quantum gravity  \cite{Weinberg:1980gg}.
It implies that quantum scale invariance is achieved
with
 some of the running  
 couplings taking finite, instead of vanishing, values in the UV.

The field   has taken up some speed recently due to the discovery 
that asymptotic safety is  realized rigorously in models of particle physics \cite{Litim:2014uca,Bond:2018oco,Bond:2016dvk,Bond:2017sem,Bond:2019npq,Bond:2017lnq}. Gauge fields are key for this to happen 
at weak coupling  \cite{Bond:2018oco} alongside Yukawa and scalar interactions subject to certain constraints 
\cite{Bond:2016dvk,Bond:2017sem}. 
A typical asymptotically safe theory contains gauge fields with charged fermions and meson-like scalars, with 
gauge groups being either unitary \cite{Litim:2014uca}, orthogonal or symplectic \cite{Bond:2019npq}, or of the product type  \cite{Bond:2017lnq} such as in the Standard Model (SM) \cite{Bond:2017wut}. 
Results also 
cover aspects of the quantum vacuum  \cite{Litim:2015iea}, higher order self-interactions   \cite{Buyukbese:2017ehm}, abelian factors \cite{Kowalska:2017fzw}, proofs with supersymmetry \cite{Bond:2017suy},  conformal windows of parameters \cite{Bond:2017tbw}, and radiative symmetry breaking \cite{Abel:2017ujy}.
In a related vein, the proposal that gauge-fermion theories with many flavors may also   realize UV fixed points  \cite{PalanquesMestre:1983zy,Gracey:1996he} has  received  renewed interest as of late
\cite{Litim:2014uca,Pelaggi:2017abg,Mann:2017wzh,Kowalska:2017pkt,Antipin:2018zdg,Abel:2018fls,
 Alanne:2019vuk,Leino:2019qwk}.
For further studies of ultraviolet stable fixed points 
  in particle physics, see \cite{Martin:2000cr,Gies:2003dp,Shaposhnikov:2008xi,Gies:2013pma,Tavares:2013dga,Abel:2013mya,Intriligator:2015xxa,Barducci:2018ysr,McDowall:2018ulq,Schuh:2018hig,Heinemeyer:2019vbc,Gies:2020xuh}.

Asymptotically safe models of particle physics
share   many features of the SM such as non-abelian gauge interactions, a flavorful  fermion sector with Yukawa 
interactions, and a scalar sector. It is therefore natural to ask whether the SM can be extended into an asymptotically 
safe version of itself, 
and if so, what type of  phenomenological signatures this would entail. 
First proposals   \cite{Bond:2017wut,Kowalska:2017fzw}  
have featured
$N_F$ vector-like fermions $\psi$ 
in general representations of the SM gauge groups and hypercharge,
and a $N_F \times N_F$ meson-like complex scalar singlet $S$.
The new matter fields  couple to the SM through the gauge 
interactions and a Higgs portal, while the  BSM Yukawa term 
\begin{equation} \label{eq:Ly}
{\cal L}_y=- y\,\text{Tr}\big[\,\overline{\psi}_LS\psi_R + h.c.\big] \,,
\end{equation}
inspired from exact models \cite{Litim:2014uca,Bond:2017lnq,Bond:2017tbw},
 helps generate interacting UV fixed points for moderate or large $N_F$ \cite{Bond:2017wut,Kowalska:2017fzw,Barducci:2018ysr}. 
Phenomenological signatures at colliders include long-lived particles, R-hadrons, and Drell-Yan production, with 
a scale of new physics potentially as low as a few TeV and  ``just around the corner" \cite{Bond:2017wut}.

\aboverulesep = 0mm
\belowrulesep = 0mm

\begin{table*}[t]
	\centering
	\renewcommand{\arraystretch}{1.2}
	\rowcolors{2}{lightgray}{}
	\begin{tabular}{`cccc`}
		\toprule
\rowcolor{LightBlue}		$\ $ \bf Model $\ $& $\ $ $(R_3,\,R_2,\, Y) \ \ $   & $ \quad $ Yukawa interactions in  $\mathcal{L_{\text{mix}}}$$\quad $  &  $\quad $  $Q_F$   $\quad $ \\ \midrule
		\bf A  &	$(\bm 1,\bm 1,-1)$  	&  $\kappa \,\overline{L} H \psi_R + \kappa'\, \overline{E} S^{\dagger}\psi_L \ $  &  $-1$ \\  
		\bf B  &	$(\bm 1,\bm 3,-1)$  	&  $\kappa \,\overline{L} H \psi_R $     & $-2,-1,0\ \ $\\ 
		\bf C  &	$(\bm 1,\bm 2,-\frac12)$ 	&  $\kappa \,\overline{E} {H}^{\dagger} \psi_L + \kappa'\, \overline{L}S\psi_R \ $  & $\ \ \; -1,0$ \\ 
		\bf D  &	$(\bm 1,\bm 2,-\frac32)$  	&  $\kappa \,\overline{E} \tilde{H}^{\dagger} \psi_L$ & $ -2,-1\ \ \ \ \ $   \\ 
		\bf E  &$(\bm 1,\bm 1,0)$ 	 	&  $\kappa \,\overline{L} \tilde{H}\psi_R$  & $\ \ \ \ \ \ \ \ \, 0$  \\ 
		\bf F  &	$(\bm 1,\bm 3,0)$  		&  $\kappa \,\overline{L} \tilde{H} \psi_R $  & $\ \ \ \ \ \ \ \ \ \ \,-1,0, +1\ \ $  \\ \toprule
	\end{tabular}
	\caption{Shown are the gauge representations $R_3$, $R_2$ and the hypercharges $Y$ of the new vector-like leptons $\psi$ with respect to the SM gauge group $SU(3)_C \cross SU(2)_L \cross U(1)_Y$ for the six basic  models A -- F. Also indicated are  the mixed Yukawa terms involving SM leptons, BSM leptons  and either the   complex gauge singlet BSM scalar $S$, or  the SM Higgs $H$ or its charged conjugate  $\tilde{H} = i\sigma^2H^*$; Yukawa couplings with SM scalars (BSM scalars)
	are denoted by $\kappa$ $(\kappa')$, respectively. The last column
		$Q_F=T_3+Y$ denotes the electric charge of the $\psi$ states. }
	\label{tab:reps}
\end{table*}

In this paper, we 
 put forward a new set of models which,  in addition to \eq{eq:Ly}, are characterized by direct Yukawa interactions  
between SM and BSM matter fields 
\cite{Hiller:2019tvg,Hiller:2019mou}.
 We are particularly interested in the relevance of  flavor portals for the high energy behavior of SM extensions, 
 in the new phenomena which arise from them, and in their interplay with the Higgs portal.
We  focus on those settings where the new fermions $\psi$ are vector-like and colorless.
Moreover, to connect to SM flavor, we use $N_F=3$, that is, three generations of SM and BSM matter.
These choices  restrict the  mixed  Yukawa interactions to the leptons and leave us with   a small number of viable $SU(2)$ 
gauge representations and hypercharges for the new fermions $\psi$  (see Tab.~\ref{tab:reps}), whose features and phenomenology are studied in depth.

\subsection{Setup for Models with Flavor Portals}
In the remainder of the introduction, we detail the basic setup and rationale for our choice of models and flavor symmetries. The renormalizable Lagrangeans of the six basic models are given by
\begin{equation}\label{eq:lag}
\begin{aligned}
\mathcal{L} &= \mathcal{L}_{\text{SM}} +    \mathcal{L}_{\text{BSM}} \\
 \mathcal{L}_{\text{BSM}} &= {\rm Tr}\,\overline{\psi}i\slashed{D}\psi + \text{Tr}\big[(\partial_{\mu}S)^{\dagger}(\partial^{\mu}S)\big] + \mathcal{L}_{\text{s}} + \mathcal{L_{\text{Y}}}
\end{aligned}
\end{equation} 
where 
$\mathcal{L}_{\text{SM}}$ denotes the SM Lagrangean, and traces are over flavor indices.
Throughout, we often suppress  the flavor index of  leptons and $\psi$'s,  and  of the  scalar matrix $S$.
The term $\mathcal{L}_{\text{s}}$ contains the BSM scalar self-interactions and the Higgs portal coupling,  and 
\begin{equation} \label{eq:LY}
\mathcal{L_{\text{Y}}}=\mathcal{L}_y+\mathcal{L_{\text{mix}}}\,
\end{equation}
contains  the Yukawa interactions amongst the new matter fields \eq{eq:Ly}, and those between BSM  
and SM matter $\mathcal{L_{\text{mix}}}$. The latter are specified in Tab.~\ref{tab:reps} for the six basic models to which 
we refer to as model A -- F. The SM fermionic content is denoted as  $L, E$ for the 
lepton $SU(2)_L$-doublet and singlet, respectively, while $H$ denotes the SM Higgs doublet.

We can immediately  state  some of the new phenomenological  features  due to  the flavor portal,
with specifics depending on mass hierarchies and the flavor structure of 
Yukawa  couplings mixing SM and BSM fields:
\begin{itemize}
\item[{(i)}] The BSM sector decays to SM particles.
\item[{(ii)}] The BSM sector can be tree-level produced at colliders  in  pairs or singly.
\item[{(iii)}]  An opportunity to address  flavor data shifted  a few standard deviations away from  SM predictions.
For example, the anomalous magnetic moments of the muon and the electron can be explained  simultaneously with the mixed 
Yukawas  in models A and C, without the necessity to manifestly break lepton flavor universality  \cite{Hiller:2019mou}.
\item[{(iv)}]  
Flavor off-diagonal scalars $S_{ij}$, $i\neq j$  couple to different generations of fermions.
Leptons and new fermions mix after electroweak symmetry breaking, and lead to charged lepton flavor violation (LFV)-like
signals from off-diagonal scalar decays $S_{ij} \to  \ell_i^\pm \ell_j^\mp$ ($\ell=e,\mu,\tau$).
\end{itemize}
Below, we give a general discussion of all models regarding  SM tests with leptons, including prospects for magnetic and electric dipole moments.

Another important part of our study is to ensure that models remain finite and well-defined  up to the Planck scale or beyond, for which we perform a complete two-loop renormalization group (RG)  study of all models.  To keep the technical complexity at bay, 
we make a few pragmatic and symmetry-based assumptions for the flavor structure of the new Yukawa interactions.

To that end, we consider the kinetic part of the Lagrangean \eqref{eq:lag}. Its  large flavor symmetry $\mathcal{G}_F$ can be decomposed as 
\begin{equation}\label{symmetry}
\mathcal{G}_F = U(3)^3_q \otimes U(3)^2_{\ell} \otimes U(3)_\psi^2 \otimes U(3)^2_{S}\,,
\end{equation}
with
\begin{equation}\label{eq:flavsyms}
\begin{aligned}
&U(3)^3_q = U(3)_Q\otimes U(3)_U\otimes U(3)_D\,,\\
&U(3)^2_{\ell} = U(3)_L\otimes U(3)_E \, , \\
&U(3)^2_\psi =  U(3)_{\psi_L}\otimes U(3)_{\psi_R}\,,\\
&U(3)^2_S = U(3)_{S_L}\otimes U(3)_{S_R}
\end{aligned}
\end{equation}
corresponding to the quarks, leptons, BSM fermions, and BSM scalars, respectively. The Yukawas, in general, do not respect the global symmetry \eq{symmetry}.
For instance, the SM part  $ U(3)^3_q \otimes U(3)^2_{\ell} $ is broken down 
to baryon number, lepton number, and hypercharge by the SM Yukawas of quarks and leptons.
Assuming that some subgroup of $\mathcal{G}_F$ is left intact then dictates the flavor structure of the Yukawas. 
For example,  without any assumptions on flavor the BSM Yukawa interactions would read
\begin{equation}\label{eq:generaly}
y_{ijk\ell}\, \overline{\psi}_{Li}\, S_{jk}\, \psi_{R\ell}
\end{equation}
with $3^4$ independent Yukawa couplings $y_{ijk\ell}$. However, identifying $U(3)^2_S$ with $U(3)_{\psi_L}\otimes U(3)_{\psi_R}$, the symmetry-preserving Yukawa interaction is given by \eq{eq:Ly} 
with a universal coupling $y$   instead  \cite{Litim:2014uca,Bond:2017wut}.

Similarly, the mixed fermion couplings with the singlet scalars $(\kappa')$  in Tab.~\ref{tab:reps}  also carry four flavor indices in general. To simplify the flavor structure along the lines of  (\ref{eq:generaly}) versus (\ref{eq:Ly})  we identify $U(3)_{E}$ with $U(3)_{\psi_R}$ (model A) or $U(3)_{L}$ with $U(3)_{\psi_L}$ (model C). As a result, the  interactions  are driven by a single Yukawa coupling instead of a tensor, and read
\begin{equation} \label{eq:AC}
\begin{array}{lr}
\kappa'\,\Tr[\overline{E} \,S^\dagger \psi_L + h.c.] & \mbox{(model A)} \, , \\
\kappa'\,\Tr[\overline{L} \,S^{\phantom \dagger} \psi_R + h.c.] & \mbox{(model C)}  \, . 
\end{array}
\end{equation}
	Finally, all models in Tab.~\ref{tab:reps} 
	contain the  mixed Higgs-Yukawa-matrix  ($\kappa$).\footnote{Notice that we keep the SM Higgs unflavored.}  
	In model A, B, E and F we identify $U(3)_{L}$ with $U(3)_{\psi_R}$ and in model C and D we identify 
	$U(3)_{E}$ with $U(3)_{\psi_L}$,  which results in a diagonal and universal   Yukawa coupling
	\begin{align} \label{eq:kappa-diag}
	\kappa_{ij}=\kappa \,\delta_{ij} \, \quad   \mbox{(model A$-$F)}  \,. 
	\end{align}
Incidentally,   the flavor symmetry   for model A and C entails that $\kappa$ is proportional to the  SM lepton Yukawa coupling in   
$Y_{\ell}\,  \bar L\, H\, E +h.c.$ implying that the latter is flavor-diagonal $Y_\ell\sim \1$. However, the  SM lepton Yukawa couplings 
are irrelevant and will be neglected, unless stated otherwise.
Alternatively,  
 we could have fixed the flavor symmetry by identifying  $U(3)_{E} \sim U(3)_{\psi_R}$  (model B, E, and F), or $U(3)_{L} \sim U(3)_{\psi_L}$ (model D),
to find hierarchical Yukawas
\beq\label{Yell}
\kappa\sim Y_\ell\, \quad   \mbox{(model B, D, E, F)}\,,
\eeq
 instead of \eq{eq:kappa-diag}. 
		  Again, we do not pursue this path any further
	 as the lepton Yukawas  are neglected in the RG study, and adopt   (\ref{eq:kappa-diag}) for all models.
	In all scenarios,  BSM fermion mass terms $\bar \psi_L M_F \psi_R +h.c.$  break the respective  remaining symmetries unless $U(3)_{\psi_L}$  $\sim U(3)_{\psi_R}$, which gives universal and diagonal $ M_F$ in all models.
	
	The symmetry language provides guidance for minimal benchmarks with reduced number of parameters (entries in Yukawa tensors). 
	This makes the study manageable and structures the RG equations.
		If  the origin of flavor would  in fact be symmetries,  there is  a fundamental reduction in complexity, and new physics  patterns observed can 
	provide feedback on flavor \cite{Nir:2007xn}.
	In the following we use  the Yukawa interactions (\ref{eq:LY}) together with  (\ref{eq:AC})  and (\ref{eq:kappa-diag}).
	Unless stated otherwise, we also assume that all BSM couplings are real-valued.

\subsection{Outline}
	The remaining parts of the paper are organized as follows.
		In Sec.~\ref{sec:AS} we  recall the  tools for asymptotic safety of weakly coupled gauge theories with matter
	covering interacting fixed points, scaling exponents, vacuum stability, the critical surface of parameters, and the matching to the Standard Model.
		In Sec.~\ref{sec:benchmarks}, a detailed ``top-down" search of fixed points, RG flows, and matching conditions is provided for  all models 
	to the leading non-trivial orders in perturbation theory.

		In Sec.~\ref{sec:scalar}, the impact of the scalar sector and the interplay between the Higgs and flavor portals are  investigated. 
	RG trajectories from the TeV to the Planck scale are studied in a ``bottom-up" search at the complete two-loop accuracy for the  top, bottom, and new Yukawas, and all gauge and quartic couplings. The BSM critical surface of parameters, $i.e.$~the parameter regions of BSM couplings  at the TeV scale which lead to well-defined (stable vacua, no Landau poles) models up to the Planck scale or beyond, is identified.
	 
	 In Sec.~\ref{sec:pheno}, we  concentrate on the  phenomenology of our models covering  
	 production, decay, fermion mixing, and constraints on model parameters from data. 
	 Effects from scalar mixing and chiral enhancement, the prospects for anomalous magnetic moments, 
	 electric dipole moments (EDMs) or LFV-type processes,
	 and 
	 signatures at $pp$ and lepton colliders such as lepton flavor violation and displaced vertices, are also worked out. 
	We summarize in Sec.~\ref{sec:conclusion}.
	Some auxiliary information and formul\ae\ are relegated into appendices (App.~\ref{sec:appBetaf} -- \ref{sec:app-mixing}).

\section{\label{sec:AS} \bf Tools for Asymptotic Safety}

In this section, we recall the principles and  basic tools for asymptotic safety, and adopt them to the models at hand.
Asymptotic safety requires that the couplings of a theory approach renormalization group fixed points in the high energy limit. 
In the language of the renormalization group, fixed points correspond to zeros of  $\beta$-functions 
\begin{equation}\label{zero}
\beta_a(\alpha)\big\vert_{\alpha=\alpha^*} \equiv \frac{d \alpha_a}{d\ln\mu}\bigg\vert_{*} =0
\end{equation} 
for all couplings $\alpha_a$, with $\alpha_a^*$ denoting the fixed point coordinates. 
Fixed points can be fully interacting with all couplings non-zero, or partially interacting whereby  some couplings become free in the UV. 

Thus, the first step is to compute the $\beta$-functions  and determine whether fixed points exist. This will be achieved using \cite{Machacek:1983tz, Machacek:1983fi, Machacek:1984zw, Luo:2002ti, Schienbein:2018fsw, Pickering:2001aq, Mihaila:2012pz}. Then, one must study if the fixed points can be reached from the IR and finally, if the trajectories can be matched to the SM.

\subsection{Renormalization Group}

We are interested in free or interacting ultraviolet (UV) fixed points in extensions of the SM. The three gauge couplings corresponding to the $U(1)_Y$, $SU(2)_L$ and $SU(3)_C$ gauge sectors are introduced as
\begin{equation}\label{couplings-gauge}
\alpha_1=\frac{g_1^2}{(4\pi)^2}\,,\quad
\alpha_2=\frac{g_2^2}{(4\pi)^2}\,,\quad
\alpha_3=\frac{g_3^2}{(4\pi)^2}\,,
\end{equation}
respectively. In our setup, the BSM fermions do not introduce new $SU(3)$ gauge charges meaning that the strong coupling continues to have an asymptotically free UV fixed point.
One may therefore neglect $\alpha_3$  for the fixed point search: we actually do so in the lowest order analysis in Sec.~\ref{sec:benchmarks}, but treat  $\alpha_3$
at the same order as the electroweak couplings
in the SM-RG and in the higher order analysis in Sec.~\ref{sec:scalar}.
On the other hand, the BSM fermions carry hypercharge and/or weak charges, see Tab.~\ref{tab:reps}. Hence, the weak (hypercharge)  coupling is infrared free in some (in all) models, and  requires an interacting UV fixed point to help cure potential Landau poles and  the triviality problem.

At weak coupling, interacting UV fixed points  arise in exactly two manners \cite{Bond:2016dvk,Bond:2018oco}. An infrared free gauge theory can either directly develop an UV fixed point  with the help of Yukawa interactions, or it may become asymptotically free owing to a gauge-Yukawa (GY) fixed point involving other gauge couplings \cite{Bond:2016dvk,Bond:2017suy}. Either way, Yukawa interactions are key for
 a well-behaved UV limit. 
The Yukawa couplings which may take this role in our models 
are those given in~\eqref{eq:LY} and Tab.~\ref{tab:reps}. We write  them as
\begin{equation}\label{couplings-Yukawa}
\alpha_y=\frac{y^2}{(4\pi)^2}\,,\quad
\alpha_\kappa=\frac{\kappa^2}{(4\pi)^2}\,,\quad
\alpha_{\kappa'}=\frac{\kappa'^2}{(4\pi)^2}\,.
\end{equation}
Let us now turn to the renormalization group equations 
for weakly coupled semi-simple gauge  theory with $n_G$ gauge couplings $\alpha_i$ and $n_Y$ Yukawa couplings 
$\alpha_n$ amongst matter fields \cite{Bond:2016dvk}. Our models 
have three gauge couplings $(i=1,2,3)$ and up to three BSM Yukawa couplings $(n=y,\kappa,\kappa'$), plus SM Yukawas and quartics.

Two remarks on notation: unless indicated otherwise 
we use the letters  $i,j$ as indices for gauge couplings, the letters  $n,m$ as indices for Yukawa couplings,
and the letters $a,b,c$ as indices for any of the gauge, Yukawa, or scalar couplings.
Following  \cite{Litim:2014uca,Bond:2017tbw}, we also  introduce the notation  \texttt{klm} to denote a perturbative approximation of beta functions 
which retains \texttt{k}~loop orders in the gauge beta 
function, \texttt{l}~loops in the Yukawa, and \texttt{m}~loops in the scalar beta functions.

With these conventions in mind,  
the gauge beta functions are given by 
\begin{widetext}
\begin{equation}\label{gauge}
\begin{aligned}
  &{\beta_i} \equiv \frac{d{\alpha_i}}{d \ln \mu} = 
  - \alpha_i^2 \left(B_i - \!\!\! \sum_{j=\rm gauge} \!\!\! C_{ij}\,\alpha_j  + \!\!\! \sum_{n=\rm Yukawa} \!\!\! D_{in}\,\alpha_n\right) +{\cal O}\left(\alpha^3\right).
\end{aligned}
\end{equation}
at the leading non-trivial order in perturbation theory which is the  {\tt 210} approximation. The one-loop  coefficients $B_i$ and the  diagonal two-loop  gauge coefficients $C_{ii}$  (no sum) may take either sign depending on the matter content, though for $B_i<0$ the latter are always positive.  The two-loop Yukawa coefficients $D_{in}$ and the off-diagonal elements $C_{ij}$ ($i\neq j)$ are always positive for any quantum field theory. In these conventions, the  gauge coupling $\alpha_i$ is asymptotic free
if $B_i>0$. 
similarly, the Yukawa beta functions take the form
\begin{equation}\label{Yukawas}
\begin{aligned}
  \beta_n &\equiv \frac{d\alpha_{n}}{d \ln \mu} = 
  \alpha_n \left(\sum_{m=\rm Yukawa} \!\!\! E_{nm}\,\alpha_m  - \!\!\! \sum_{i=\rm gauge} \!\!\! F_{ni}\, \alpha_i\right)  +{\cal O}\left(\alpha^2\right).
\end{aligned}
\end{equation}
\end{widetext}
Any of the loop coefficients $E$ and $F$ are positive in any quantum field theory. 
The  loop coefficients in ~\eqref{gauge} and ~\eqref{Yukawas} corresponding to our models can be found in App.~\ref{sec:appBetaf}.

\subsection{Ultraviolet Fixed Points}

Next, we turn to  renormalization group  fixed points. 
Yukawa couplings at a fixed point are either free or interacting, and ultraviolet fixed points require that some (or all) Yukawa couplings are non-zero.
The vanishing of ~\eqref{Yukawas} implies that the non-zero  Yukawa couplings are related to the gauge couplings as  
\begin{equation}\label{nullcline}
\alpha_n=(E^{-1})_{nm}\, F_{mj}\,\alpha_j\,.
\end{equation}
We refer to these relations as the Yukawa nullclines. Notice that the matrix $E$ is inverted over  the set of non-vanishing Yukawa couplings, and the matrix multiplication   in ~\eqref{nullcline} excludes the vanishing Yukawa couplings (if any). In theories with 
$n_Y$ Yukawa couplings this procedure can lead to as many as $2^{n_Y}-1$ different 
nullclines.  Fixed points for the gauge coupling are found by inserting the nullcline  ~\eqref{nullcline} into ~\eqref{gauge}, leading to
\begin{equation}\label{reduced}
\beta_i\big|_{\beta_n=0}= -\alpha_i^2 \left(B_i -C'_{ij}\,\alpha_j\right) \,.
\end{equation}
Hence, every Yukawa nullcline generates shifted two-loop coefficients $C'$  given by
\begin{equation}\label{shifted}
C'_{ij}=C_{ij} -  D_{in}\,(E^{-1})_{nm}\, F_{mj}
\end{equation}
in terms of the perturbative loop coefficients. In particular, the non-zero fixed points for the gauge couplings follow from ~\eqref{reduced} and ~\eqref{shifted} as
\begin{equation}\label{GY}
\alpha_i^*=(C'^{-1})_{ij} \, B_j\,,
\end{equation}
where the sum over $j$ only includes  the non-vanishing gauge couplings. The  Yukawa fixed point follows from inserting ~\eqref{GY} into the corresponding nullcline ~\eqref{nullcline}. Overall, we may find up to $(2^{n_G}-1)(2^{n_Y}-1)$ different gauge-Yukawa fixed points. Also notice that the physicality condition $\alpha_{i\cdots}^*,\alpha_{n\cdots}^*\ge 0$ is not guaranteed automatically and must still be imposed.
Viable gauge-Yukawa fixed points genuinely exist for asymptotically free gauge sectors. Most importantly, thanks to the Yukawa-induced shift  in ~\eqref{shifted}, physical solutions ~\eqref{GY} 
may even exist for infrared free gauge sectors where $B_i<0$. This is the primary mechanism to stabilize infrared free gauge  sectors in the UV.

Gauge-Yukawa fixed points may also indirectly stabilize an otherwise infrared free gauge sector \cite{Bond:2016dvk,Bond:2017wut,Bond:2017suy,Kowalska:2017fzw}, because the one loop coefficient of a gauge theory can be modified 
in the presence of an interacting fixed point. Conditions for this to happen 
for an infrared free gauge coupling $\alpha_i$ can now be read off from ~\eqref{gauge}, 
\begin{equation}\label{Beff}
B^{\rm eff}_i=B_i - C_{ij}\,\alpha^*_j +  D_{in}\,\alpha^*_n\,.
\end{equation}
The sums run over the non-zero gauge and Yukawa couplings $\{\alpha^*_{j\cdots} ,\alpha^*_{n\cdots}\}$ and we recall that 
$\alpha_i^*=0$. Provided that  the effective one-loop coefficient becomes positive, $B^{\rm eff}_i >0>B_i$, the infrared free gauge coupling  becomes free in the ultraviolet. This is the secondary mechanism to stabilize infrared free gauge sectors in the UV. We stress that Yukawa couplings are mandatory for this as they are the only couplings contributing positively to ~\eqref{Beff}. Below, we will see  that both mechanisms are operative in our models.

If all Yukawa couplings vanish, the gauge sector ~\eqref{gauge} may still achieve free or interacting fixed points. The interacting ones are given by
\begin{equation}\label{BZ}
\alpha_i^*=(C^{-1})_{ij}\,B_j\,,
\end{equation}
where the sum runs over the non-zero gauge couplings.
These are the well-known Banks-Zaks (BZ) fixed points \cite{Caswell:1974gg,Banks:1981nn}, which are always  infrared  and can only be physical  ($\alpha_i^*> 0$) for asymptotically free gauge couplings. In theories with $n_G$ asymptotically free gauge couplings, we may find up to $2^{n_G}-1$ of them. Although Banks-Zaks fixed points play no role for the UV completion of theories, they may still be present and influence the RG evolution of couplings on UV-IR connecting trajectories.

\subsection{Scalar Potential and Higgs Portal} 

Here we briefly discuss the scalar sector  and its ground states. As the BSM scalar carries flavor and couples to the SM fermions its
vacuum expectation values (VEVs) have implications for the flavor structure of the model.

The minimal potential involving the SM and BSM scalars $H$ and $S$ included in \eq{eq:lag} and compatible with the symmetries \eq{symmetry} has the form
\begin{widetext}
\begin{equation}\label{eq:scalarV}
\begin{aligned}
V(H,S) = &-\mu^2H^{\dagger}H -\mu_s^2\Tr[S^{\dagger}S]-\mu_{\rm det}\left(\det S+\det S^\dagger\right) \\
& + \lambda (H^{\dagger}H)^2 + \delta H^{\dagger}H\Tr\left[S^{\dagger}S\right] + u\Tr\left[S^{\dagger}SS^{\dagger}S\right] 
+ v \, \left(\Tr \left[S^{\dagger}S\right]\right)^2 
\end{aligned}
\end{equation} 
\end{widetext}
for all models. It consists of the Higgs self-coupling $\lambda$ and mass parameter $\mu$, the BSM scalar quartics $u$, $v$, as well as the BSM mass parameters $\mu_s$ and the trilinear coupling $\mu_{\rm det}$, and a portal coupling $\delta$ which mixes SM and BSM scalars.  Viable UV fixed points for our models require that the Higgs self-coupling, the portal coupling and the self-couplings of the BSM scalar fields  take  fixed points by themselves, compatible with vacuum stability. 
Interestingly though,  the quartics do not couple back into the gauge-Yukawa system at the leading order. Rather, fixed points in the SM and BSM scalar sectors are fueled by the gauge-Yukawa fixed points, and backcoupling occurs starting at two-loop level in the Yukawa sector, and at three-loop level for the Higgs (four-loop for the BSM scalars) in the gauge sectors.  

The classical moduli space for \eq{eq:scalarV} and conditions for the asymptotic stability of the  vacuum are found following 
\cite{Litim:2015iea,Paterson:1980fc}. 
Depending on the sign of $u$, we find two settings $V^\pm$ with stability conditions 
\begin{equation}\label{eq:vstab}
\begin{aligned}
  V^+ : \quad & 
\left\{ \begin{array}{l}
  \lambda > 0,\quad u > 0,\quad  u + 3\,v > 0,  \\
   \delta >  -2 \sqrt{\lambda\left(u/3 + v\right)}\,,
\end{array} \right. \\
V^- :  \quad &
\left\{ \begin{array}{l}
  \lambda > 0,\quad u < 0,\quad  u + v > 0, \\
   \delta >  -2 \sqrt{\lambda\left(u + v\right)}\,.
\end{array} \right. \\
\end{aligned}
\end{equation}
Both settings allow for the Higgs  to break electroweak symmetry. 
For $V^+$,  the BSM scalar vacuum expectation value (VEV) is flavor-diagonal and  upholds some notion of flavor universality in interactions with the SM. On the other hand, $V^-$ has a VEV  only  in one diagonal component of $S$. In the context of our models, this corresponds to a VEV pointing in the direction of one lepton flavor.
We learn that the Lagrangean (\ref{eq:lag}) offers the possibility to violate lepton  flavor universality spontaneously, an interesting feature also in the context of today's flavor anomalies, e.g.~\cite{Bifani:2018zmi}.
Note, if both scalars $S$ and $H$ acquire a VEV, the portal coupling $\delta$ induces mixing between the scalars $H$ and $S$. Details can be seen in App.~\ref{sec:appscalar}.
In the following we  investigate the availability of fixed points, vacuum stability,  and  phenomenological signatures  at various orders in perturbation theory up to the  \texttt{222} approximation using the methodology of \cite{Machacek:1983tz, Machacek:1983fi, Machacek:1984zw, Luo:2002ti, Schienbein:2018fsw, Pickering:2001aq, Mihaila:2012pz}.

\subsection{Scaling Exponents and UV Critical Surface}\label{ScalingExponents}

The renormalization group flow in the vicinity of fixed point provides information on whether the fixed point can be approached in the UV or IR. Denoting by $\alpha_a$ any of the gauge, Yukawa, or scalar couplings, and expanding the $\beta$-functions around a fixed point $\alpha^*_a$ up to second order in $\delta_a = \alpha_a-\alpha_a^*$, we find
\begin{equation}\label{eq:betaf_taylor}
\beta_a =  M_{ab}\delta_b + P_{abc}\,\delta_b\delta_c + \mathcal{O}\left(\delta^3\right) \,,
\end{equation}
where  $M_{ab}={\partial \beta_a}/{\partial \alpha_b}\rvert_{*}$ is the  stability matrix and $P_{abc}=\frac12{\partial^2 \beta_a}/{\partial \alpha_b\partial\alpha_c}\rvert_{*}$. After diagonalizing $M$ the running of couplings at first order may be written as
\begin{equation}\label{eq:alphasol}
\alpha_a(\mu) =  \alpha_a^* + \sum_b V_{a}{}^{b}\,{c_b}\, {(\mu/\Lambda)}^{\vartheta_b}\,,
\end{equation}
where $\mu$ is the RG scale and $\Lambda$ a UV reference scale, while the UV scaling exponents $\vartheta_b$ arise as the eigenvalues of the stability matrix $M$ with $V^{b}$ the corresponding eigenvectors,
and $c_b$ free parameters. An eigenvector is relevant, marginal, or irrelevant  if the corresponding eigenvalue $\vartheta$ is negative, zero, or positive. For all relevant and marginally relevant couplings, the parameters $c_b$ are fundamentally free  and constitute the ``UV critical surface''  of the theory. Its dimension  should be finite to ensure predictivity. For all irrelevant couplings, we must set  $c_b\equiv 0$ or else the UV fixed point cannot be reached in the limit $\mu\to\infty$. UV fixed points require at least one relevant or marginally relevant eigendirection. 

If a fixed point is partially interacting, that is, some but not  all couplings are non-zero, the relevancy of the vanishing couplings can be established as follows.  If a gauge coupling $\alpha_i$ vanishes at a fixed point, it follows from~\eqref{gauge} being at least quadratic in  $\alpha_i$ that the coupling is marginal. Going to second order in perturbations ~\eqref{eq:betaf_taylor} reveals that  $P_{iii}=-B^{\rm eff}_i$. As expected, the sign of ~\eqref{Beff} determines whether the coupling is marginally relevant ($B^{\rm eff}_i>0$) or marginally irrelevant.
If a Yukawa coupling $\alpha_n$ vanishes at a GY fixed point with coordinates $\{\alpha_{i\cdots}^*,\alpha_{m\cdots}^*\}$, it follows from ~\eqref{Yukawas}  that the corresponding scaling exponent is given by
\begin{equation}\label{theta-Yukawa}
\vartheta_n=E_{nm}\,\alpha_m^*  -F_{ni}\, \alpha_i^*\,.
\end{equation}
As this is a difference between two positive numbers, 
its overall sign is not determined by the existence of the fixed point and the coupling could come out as relevant, marginal, or irrelevant.  For BZ fixed points (all $\alpha_m^*=0$), however, the eigenvalue is always negative and the Yukawas are  relevant.

\subsection{Matching and BSM Critical Surface}

Here we consider how an asymptotically safe UV fixed point must be connected to the SM. At  low energies, any extension of the SM must connect to the measured values
of SM couplings.
 For simplicity, and without loss of generality, we assume that all BSM matter fields have identical masses $M_F$. Moreover, the decoupling of heavy modes is approximated by considering the BSM fields either as massless (for $\mu>M_F$) and as infinitely massive (for $\mu<M_F$). Both of these technical assumptions can be lifted to account for a range of BSM matter field masses, and for a smooth decoupling of heavy modes, without altering the main pattern.  In this setting, the fluctuations of BSM fields are absent as soon as $\mu<M_F$, meaning that the running of all SM couplings $\alpha_{\rm SM}(\mu)$ must be identical to the known SM running  for all $\mu\le M_F$. Therefore, we refer to 
\beq
\label{MatchingScale}
\mu=M_F
\eeq 
as the matching scale. 
On the other hand, the values of the BSM couplings  $\alpha_{\rm BSM}(\mu)$  at the matching scale \eq{MatchingScale} are not predicted by the SM and must  be viewed as free parameters. Schematically, we denote this set of free parameters   as
\beq\label{BSM}
S_{\rm free}=\{\alpha_{\rm BSM}\}\,.
\eeq
 Any BSM renormalization group trajectory 
  is uniquely characterized by the matching scale \eq{MatchingScale}, the (known) values of SM couplings at  the matching scale, and the initial values of BSM couplings \eq{BSM}.
   The latter are, in our models,  the values of the three BSM scalar couplings plus the two (or three) BSM Yukawa couplings at the scale $M_F$,
   \beq
   \alpha_{\rm BSM}=(\alpha_y,\alpha_\kappa,\alpha_{\kappa'},\alpha_\delta,\alpha_u,\alpha_v)\,,
   \eeq
  and the parameter space \eq{BSM}  is hence five (or six) dimensional, depending on the model.  
  
  Depending on the BSM initial values \eq{BSM}, renormalization group trajectories may display a variety of different patterns. These  include  power-law approach towards an interacting  fixed point or cross-over through a succession of  fixed points such as in asymptotic safety proper, or logarithmically slow decay towards the free fixed point such as in asymptotic freedom. Either of these behaviours, or, in fact, any combination thereof, corresponds to a  viable high-energy limit in the sense of Wilson's path integral definition of quantum field theory. In turn, couplings may also run into unphysical regimes where the quantum vacuum becomes meta- or unstable,  or where couplings become non-perturbatively large and RG trajectories terminate due to  Landau pole singularities. 
  
From a bottom-up model building perspective, the set of parameter values $S_{\rm BSM}$ 
for which the BSM trajectories remain finite and well-behaved -- at least up to the Planck scale -- is of particular interest. 
First and foremost, this set  includes initial values for all trajectories which terminate at interacting UV fixed points, should they exist. In general, however, it can often  be larger, simply because it may also include trajectories which remain finite and well-defined up to the Planck scale, but  would otherwise not reach an interacting UV fixed point proper in the transplanckian regime. This feature can be referred to as Planck-safety \cite{Hiller:2019mou}, as opposed to and extending the notion of Asymptotic safety.
  The set of viable BSM parameters $S_{\rm BSM}$ is a subset of \eq{BSM},
  and often of a lower dimensionality. 
  The reason for this is  that  interacting UV fixed points have relevant and irrelevant eigenoperators. All   interactions which are irrelevant in the UV impose constraints on the viable values of BSM couplings at the matching scale \eq{BSM}. Therefore, we refer
  to  the set of viable initial values  $S_{\rm BSM}$  
   as the ``BSM critical surface''. We obtain BSM critical surfaces for models A -- F in Sec.~\ref{sec:crit-sum}.

\section{\label{sec:benchmarks}\bf Benchmark Models and  Fixed Points}

In this section we further specify our benchmark models and investigate their RG flows to the leading non-trivial order in perturbation theory. We focus on the gauge and the Yukawa couplings whose beta functions  are given by \eq{gauge} and \eq{Yukawas} with loop coefficients for all models stated in Sec.~\ref{sec:appBetaf}. Our goal is to gain a first understanding of models and  fixed points, and the availability of matchings to the SM. We postpone the study of  quartic scalar couplings and  higher order loop corrections to Sec.~\ref{sec:scalar}.

The leading order approximation --  known as the {\tt 210} approximation -- retains two loop orders in the gauge and one loop in the Yukawa couplings. Scalar couplings are neglected.  Besides the free Gaussian fixed point, we may find interacting Banks-Zaks  or gauge-Yukawa fixed points, though only the latter will qualify as UV fixed points. Already at this order in the approximation, there can be up to a maximum of $(2^{n_G}-1)$ different Banks-Zaks and a maximum of $(2^{n_G}-1)\times (2^{n_Y}-1)$ different GY  fixed points  \cite{Bond:2016dvk,Bond:2017lnq}. Here $n_G$ denotes the number of SM gauge groups under which the BSM fermions are charged 
($n_G=2,1$ or $0$ for our models), and $n_Y$ the number of BSM Yukawa couplings ($n_Y=2$ or 3 for all models). For this reason,  for Banks-Zaks fixed points in semi-simple gauge theories we  specify the non-zero gauge couplings as an index ($e.g.$ BZ${}_2$). Similarly, for gauge-Yukawa fixed points, we also indicate the non-vanishing Yukawa couplings ($e.g.$ GY$_{1\kappa}$).  

Findings of this section are summarized in Sec.~\ref{sec:sum}.

\aboverulesep = 0mm
\belowrulesep = 0mm

\begin{table*}
	\renewcommand{\arraystretch}{1.2}
	\rowcolors{2}{lightgray}{}
	\begin{tabular}{`cccccccclcl`}
		\toprule
\rowcolor{LightBlue}
	{\ \ \bf Model A\ }
	& $\quad \alpha_1^* \quad$ & $\quad \alpha_2^* \quad$ & $\quad \alpha_{\kappa}^*\quad$ & $\quad\alpha_{\kappa'}^*\quad$ & $\quad\alpha_y^*\quad$ & $\:$rel.$\:$ & $\:$ irrel. $\:$&  
	Info & \ Fig.~\ref{fig:fpsAB}\ 
	& \ \ Matching\ \ \\ \midrule	%
FP$_1$ & 0$^{(+)}$ & 0$^{(-)}$ & 0$^{(+)}$ & 0$^{(+)}$ & 0$^{(+)}$ & 1 & 4 & saddle &  & \cellcolor{white}\\ 
FP$_2$ & 0$^{(+)}$ & 0.543 & 0$^-$ & 0$^{(+)}$ & 0$^{(+)}$ & 1 & 4 & BZ$_2$  &  &\\
FP$_3$ & 0$^{(+)}$ & 0.623 & 0.311 & 0$^{(+)}$ & 0$^+$ & 0 & 5 & GY$_{2\kappa}$ & & \cellcolor{white} 
 \\
FP$_4$ & 2.746 & 0$^{(+)}$ & 0$^-$ & $4.120 - \alpha_y^*$ & $\alpha_y^*$ & 2 & 2 & 
{line}&
& \\ 
FP$_5$ & 1.063 & 0$^{(-)}$ & 0.886 & 1.594 & 0$^+$ & 2 & 3 & GY$_{1\kappa\kappa'}$ & $A_1$ & \cmark\ (Fig.~\ref{fig:FPA1})\\
FP$_6$ & 1.105 & 0.569 & 1.205 & 1.657 & 0$^+$ & 1 & 4 & GY$_{12\kappa\kappa'}$ &  $A_2$ & \xmark \, (Fig.~\ref{fig:FPA2}) \\
FP$_7$ & 2.151 & 0$^{(-)}$ & 0.782 & 0$^-$ & 3.032 & 3 & 2 & GY$_{1 y\kappa }$ &$ A_3$ & \cmark \\
FP$_8$ & 2.267 & 0.200 & 0.933 & 0$^-$ & 3.165 & 2 & 3 & GY$_{12y\kappa }$ & $A_4$ & \xmark 
		\\ \bottomrule
	\end{tabular}
	\caption{Fixed points 
	of model A in the \texttt{210} approximation. FP$_{1,2,3}$ are IR or crossover fixed points, FP$_4$ is a line of fixed points, and  FP$_{5,6,7,8}$ are UV fixed point candidates. Also shown are the number of relevant and irrelevant eigendirections, and whether the  fixed point is 
	 of the BZ or GY type, with indices specifying the non-trivial couplings. Free couplings with power-law running are  marked with a superscript $\pm$ if they are irrelevant/relevant, and an additional parenthesis $(\pm)$  indicates that the flow is logarithmic; see Figs.~\ref{fig:fpsAB}, \ref{fig:FPA2} and \ref{fig:FPA1} for the phase diagram and sample trajectories.}%
	\label{tab:fpsA}
\end{table*}

\subsection{Model A (singlets,  $Y=-1$) \label{sec:A}}

Model A consists of the SM, amended by complex singlet BSM scalars $S$ and $N_F=3$ vector-like BSM fermions $\psi$ in the representation $(\bm 1,\bm 1,-1)$, which is identical to the one of the singlet leptons  $E$ present in the SM, with Lagrangean \eq{eq:lag}. The  Yukawa sector \eq{eq:LY} contains three BSM couplings,
\begin{equation}
-\mathcal{L}^{\text{A}}_{\text{Y}} = \kappa \overline{L} H \psi_R  + \kappa'\overline{E}S^{\dagger}\psi_L + y\, \overline{\psi}_L S \psi_R + \mathrm{h.c.}\,.
\end{equation}
Fixed points for model A are summarized 
 in Tab.~\ref{tab:fpsA} and denoted as FP$_1$ -- FP$_8$. Tab.~\ref{tab:fpsA}  also shows the number of relevant and irrelevant eigendirections. Free couplings are  marked with a superscript $+$ if they are irrelevant or with a $-$ if they are relevant, with power-law running.
 An additional parenthesis, that is, $(+)$  or $(-)$  for irrelevant or relevant, respectively, indicates that the flow along its eigendirection  is logarithmically slow instead.
  It is also shown whether a  fixed point is  of the BZ or GY type, in which case an index is added to specify the non-trivial couplings. 
 \begin{figure}[b]
		\includegraphics[scale=.55]{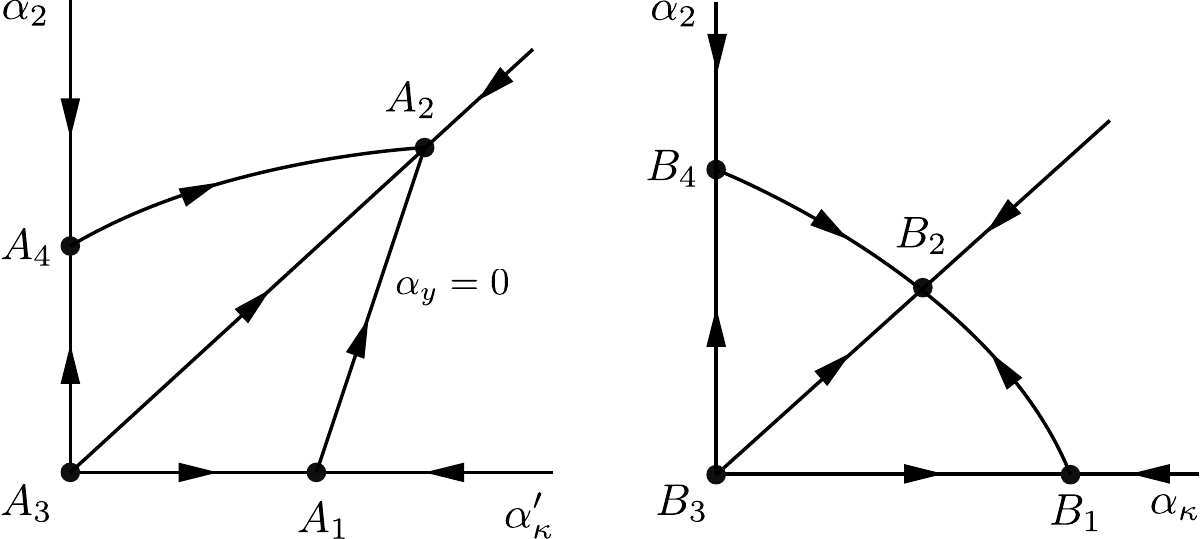}
	\caption{Schematic phase diagram and  various UV fixed points of model A, B and D in the \texttt{210} approximation. Arrows indicating the flow from the UV to the IR. The fixed points of model A  (Tab.~\ref{tab:fpsA}) are projected onto the $(\alpha_2,\alpha_{\kappa'})$ plane (left panel) with $A_2$ denoting the least, and $A_3$ the most ultraviolet attractive fixed point. 
	The fixed points of model B (Tab.~\ref{tab:fpsB}) and model D  (Tab.~\ref{tab:fpsD}) are projected onto the $(\alpha_2,\alpha_\kappa)$ plane  (right panel); results for model D are equivalent to those of model B. Note  that the topology of the projected RG flows in all models is identical.}
	\label{fig:fpsAB}
\end{figure} 

The Gaussian fixed point (FP$_1$)  is a saddle owing to $B_1 < 0 < B_2$ and takes the role of a cross-over fixed point. 
FP$_2$  is an infrared Banks-Zaks fixed point (BZ${}_2$) where  the Yukawa coupling $\alpha_\kappa$ is the sole relevant coupling because the fermions $\psi$, $E$ do not carry weak isospin.  FP$_3$ is an infrared gauge-Yukawa fixed point (GY$_{2\kappa}$) which acts as an infrared sink because it is  fully attractive in all canonically dimensionless couplings. 	 	 	 
FP$_4$ corresponds to a line of fixed points, see Tab.~\ref{tab:fpsA}, which arises from a degeneracy among the   GY$_{12y}$, GY$_{12y\kappa'}$, and GY$_{12\kappa'}$ fixed points. The degeneracy is not protected and lifted by higher loop effects.
The gauge-Yukawa fixed points FP$_5$ -- FP$_8$ are candidates for UV fixed points. They invariably  involve a non-vanishing fixed point for the hypercharge coupling $\alpha^*_1$, with or without a non-vanishing $\alpha^*_2$, and fixed points for the Yukawas. We also note that some of the fixed point couplings are of order unity, in particular the hypercharge coupling. Ultimately, this is a consequence of a low number of BSM fermions and the present approximation. We come back to this aspect in Sec.~\ref{sec:scalar} where the quartic scalar couplings are retained as well.

\begin{figure}[b]
	\begin{minipage}{0.4\textwidth}
		\includegraphics[width=\columnwidth]{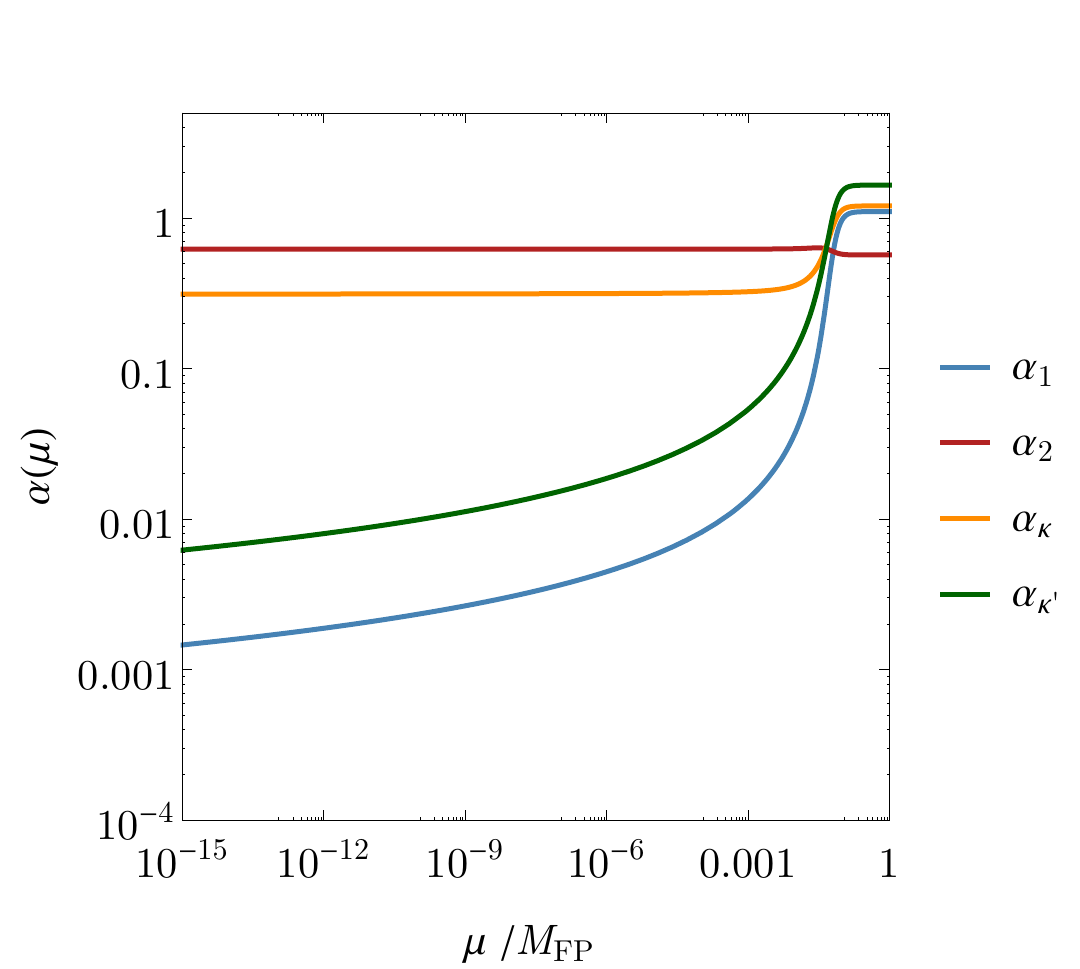}
	\end{minipage}
	\caption{
		Running of couplings of model A in the {\tt 210} approximation from   fixed point $A_2$. Trajectories are invariably attracted by  FP${}_3$ in the infrared,
	and $\alpha_2$ comes out too large compared to the SM value.}
	\label{fig:FPA2}
\end{figure} 

\begin{figure}[b]
	\begin{minipage}{0.4\textwidth}
		\includegraphics[width=\columnwidth]{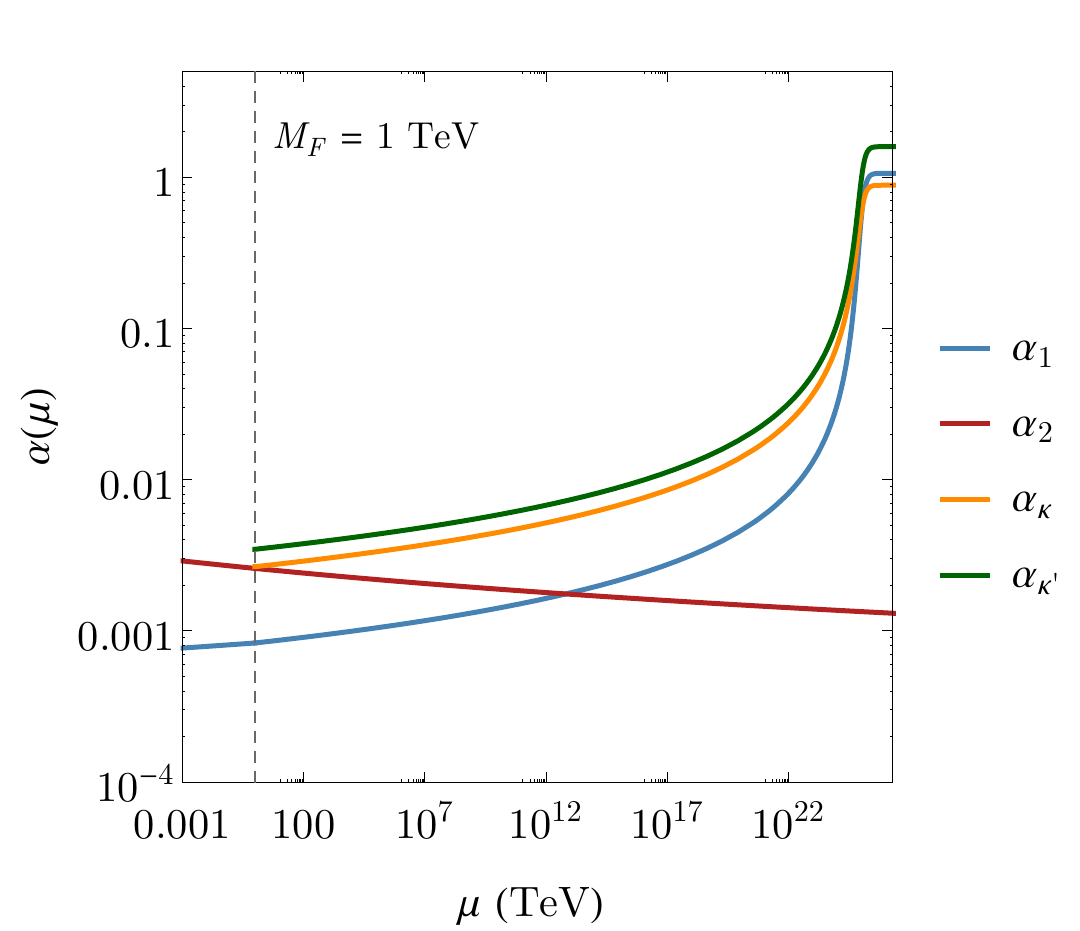}
	\end{minipage}
	\caption{Running of couplings of model A in the {\tt 210} approximation with matching of  the partially-interacting fixed point $A_1$ to the SM at $\mu=1$ TeV (see Tab.~\ref{tab:fpsA}).}
		\label{fig:FPA1}
\end{figure}

Fixed point candidates other than those given in  Tab.~\ref{tab:fpsA} either vanish or come out unphysical. 
For example, the relation $\beta_y / \alpha_y = \beta_{\kappa'}/\alpha_{\kappa'} + 2 \alpha_\kappa$, which holds in model A,  see  \eq{Yukawas}  and Sec.~\ref{sec:appBetaf}  for the RG-coefficients,  implies that at least one of the couplings $\alpha_y^*$ or $\alpha_{\kappa'}^*$ has to vanish provided that $\alpha_\kappa^* \neq 0$. It follows that  fixed points such as GY$_{1\kappa\kappa'y}$ and GY$_{12\kappa\kappa'y}$ cannot arise. For $\alpha_\kappa = 0$, we find a line of fixed points in the  coupling $\alpha_y' = \alpha_y + \alpha_\kappa$. Note also that $\widetilde{\alpha}_y = \alpha_y - c\,\alpha_{\kappa'}$ with $c \neq 1$  a free parameter, is decoupled from the rest of the system.  The fixed points GY$_{2y}$, GY$_{2y\kappa'}$, and GY$_{2\kappa'}$ which are covered by this line of fixed points, are unphysical. As the Yukawa beta functions do not receive vertex corrections, they can be rewritten as $\beta_{y'} = \gamma_{y'}(\alpha_1, \alpha_y')\, \alpha_y'$  and $\beta_{\widetilde{y}} = \gamma_{y'}(\alpha_1, \alpha_y')\, \widetilde{\alpha}_y$ in terms of a single anomalous dimension $\gamma_{y'}$, which, moreover, is independent of $\widetilde{\alpha}_y$. 
Therefore, $\widetilde{\alpha}_y$ becomes exactly marginal for $\gamma_{y'}(\alpha_1^*, \alpha_y'^*) = 0$, and the parameter $c$ remains unspecified. 
Lines of fixed points related to the vanishing of anomalous dimensions are well-known in supersymmetric gauge theories. Here, they are an artifact of the low orders in the loop expansion.
Finally, we note that the fixed points GY$_{1\kappa}$ and GY$_{12\kappa}$ arise with negative $\alpha$ which is  unphysical.

In Fig.~\ref{fig:fpsAB}, we show the schematic phase diagram of model A and the interplay between the UV fixed point FP${}_3$ -- FP${}_7$ (denoted as $A_1$ -- $A_4$) in more detail (see also Tab.~\ref{tab:fpsA}). Trajectories are projected onto the $(\alpha_2,\alpha_{\kappa'})$ plane  and arrows  indicate the flow from the UV to the IR. 
$A_3$ is the most relevant UV fixed point. The separatrices responsible for the cross-over from $A_3$ to $A_1$, from $A_3$ to $A_4$, or from $A_1$ to $A_2$ relate to the lines $\alpha_2 = 0$, $\alpha_{\kappa'} = 0$, or  $\alpha_y = 0$, respectively.  $A_2$ is the  least ultraviolet point only exhibiting $\alpha_1$ as a relevant coupling.

Next, we confirm that some of the UV fixed points  in Tab.~\ref{tab:fpsA} can be matched onto the SM. 
Here, it is worth noting that many renormalization group trajectories  are attracted by the fully  attractive IR fixed point GY$_{2\kappa}$, corresponding to  FP$_3$ in Tab.~\ref{tab:fpsA}. If so,  the gauge coupling $\alpha_2$ remains too large to be matched against the SM. In other words, UV initial conditions within the basin of attraction of FP$_3$ cannot be matched onto the SM.
In concrete terms, this is the case for any trajectory running out of the fixed point $A_2$ or $A_4$ (see Fig.~\ref{fig:FPA2} for an example).
On the other hand, provided that the gauge coupling $\alpha_2$ takes sufficiently  small values in the vicinity of the UV fixed point, trajectories can avoid the  FP$_3$. This is the case for both  UV fixed points $A_1$ and $A_3$. Starting from these, $\alpha_2$  remains sufficiently small throughout the entire RG evolution, and matching against the SM possible at a wide range of matching scales between the TeV and the Planck scale. 
An example for this is shown in Fig.~\ref{fig:FPA1}.
 
 \begin{table*}
	\centering
	\renewcommand{\arraystretch}{1.2}
	\rowcolors{2}{lightgray}{}
	\begin{tabular}{`ccccccclcc`}
		\toprule
\rowcolor{LightBlue}
	{\ \ \bf Model B\ \ }	 & $\quad \alpha_1^* \quad$ & $\quad \alpha_2^* \quad$ &  $\quad\alpha_{\kappa}^*\quad$ & $\quad\alpha_y^*\quad$ & $\:$rel.$\:$ & $\:$ irrel. $\:$&  $\:$Info $\:$& $\:$Fig.~\ref{fig:fpsAB} $\:$ & $\:$Matching$\:$ \\ \midrule	
FP$_1$ & 0$^{(+)}$ & 0$^{(+)}$ & 0$^{(+)}$ & 0$^{(+)}$ & 0 & 4 & G & & \\
FP$_2$ & 1.953 & 0$^{(-)}$ & 1.562 & 1.888 & 2 & 2 & GY$_{1\kappa y}$ & $B_1$ & \cmark \\
FP$_3$ & 1.224 & 0.186 & 1.326 & 1.541 & 1 & 3 & GY$_{12\kappa y}$ & $B_2$ & \cmark \\
FP$_4$ & 2.712 & 0$^{(-)}$ & 0$^-$ & 2.712 & 3 & 1 & GY$_{1y}$ & $B_3$ & \cmark \\
FP$_5$ & 1.732 & 0.216 & 0$^-$ & 2.164 & 2 & 2 & GY$_{12y}$ & $B_4$ & \cmark\\ \bottomrule
	\end{tabular}
	\caption{Partially and fully interacting  fixed points of model B 
	in the \texttt{210} approximation, notation as in Tab.~\ref{tab:fpsA}. 
	Banks-Zaks fixed points are absent since asymptotic freedom is lost in both gauge couplings;
	see Figs.~\ref{fig:fpsAB} and  \ref{fig:FPB1} for the phase diagram and sample trajectories.}
		\label{tab:fpsB}
\end{table*}

Finally, it is noteworthy that, unlike in \cite{Bond:2017wut,Kowalska:2017fzw}, the Yukawa coupling $\alpha_y$ can be switched off as it is not required to generate the fixed points $A_1$ and $A_2$. Instead, the Yukawa couplings $\kappa$ and $\kappa'$ are required to enable a fixed point for $\alpha_1$. Their predicted low energy values 
are $\alpha_{\kappa}(M_F=1 \text{ TeV}) = 2.7\cdot 10^{-3} $ and $\alpha_{\kappa'}(M_F=1 \text{ TeV}) = 3.5\cdot 10^{-3}$ asuming a matching to $A_1$; see Fig.~\ref{fig:FPA1}.

\subsection{Model B (triplets,  $Y=-1$)}

For vector-like fermions $\psi(\bm 1,\bm 3,-1)$ the BSM Yukawa Lagrangean takes the form
\begin{equation}
-\mathcal{L}^{\text{B}}_{\text{Y}} =   + \kappa\overline{L}\psi_R H + y\, \overline{\psi}_L S \psi_R +\mathrm{h.c.} \,.
\end{equation}
The components of $\psi$ can be expressed as $SU(2)_L$ matrix via:
\begin{equation} \label{eq:triplet}
\begin{aligned}
\psi = \left( \begin{array}{cc} {\psi^{-1}}/{\sqrt{2}}&\psi^{0} \\\psi^{-2}&-\psi^{-1}/{\sqrt{2}} \end{array} \right)\,,
\end{aligned}
\end{equation}
in accord with the normalization of the kinetic term in equation \eqref{eq:lag}. The upper indices indicate the $U(1)_{\text{em}}$ charge of each component. 

We have listed all fixed points of model B in Tab.~\ref{tab:fpsB}. In this model,  the one-loop coefficients of both gauge coupling obey $B_{1,2} < 0$, turning the Gaussian into a total IR fixed point, and prohibiting any kind of Banks-Zaks solutions. Moreover, all gauge-Yukawa fixed points only involving $\alpha_2$ (GY$_{2\kappa}$, GY$_{2y}$, GY$_{2\kappa y}$) are unphysical, and for the remaining ones, $\alpha_y^* \neq 0$ is required, additionally excluding GY$_{1\kappa}$, GY$_{12\kappa}$. 

\begin{figure}
  \includegraphics[width=.45\textwidth]{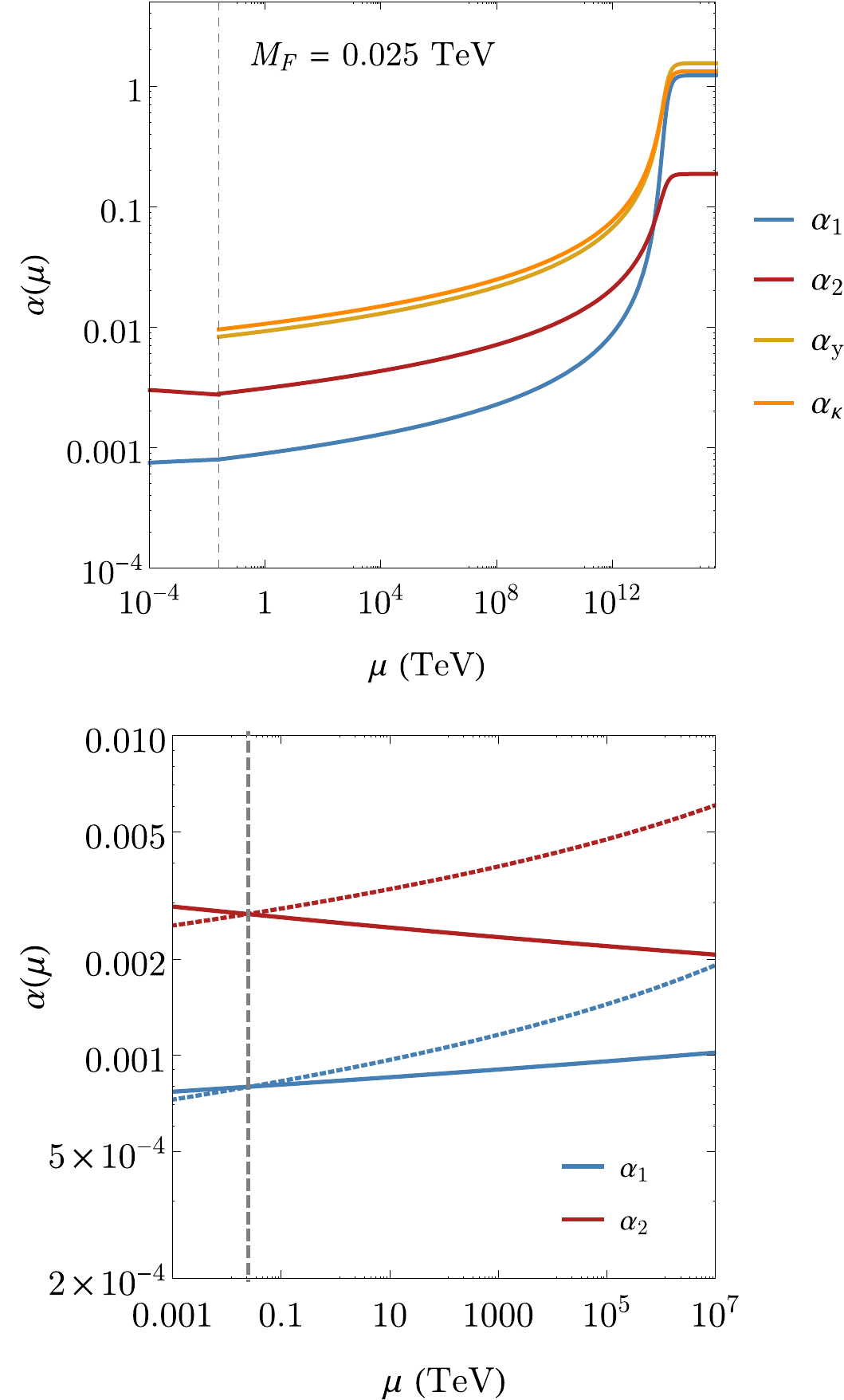}
	\caption{Matching at $M_F=0.025$ TeV  for the fully-interacting fixed point $B_2$ of model B. Top panel: BSM running of the couplings into the fixed point. Bottom: BSM (dotted lines) and SM running (solid lines) of the gauge coupling near the matching scale (dashed vertical  line). }
	\label{fig:FPB1}
\end{figure}

This singles out the fixed points  $B_{1..4}$ as listed in Tab.~\ref{tab:fpsB}. 
Similarly to  the fixed points $A_{1..4}$  of model A,  $B_2$ is the least ultraviolet with $\alpha_1$ being the only relevant coupling, $B_{1,4}$ are  connected to it via a second relevant trajectory, while $B_3$ has  three relevant directions. This is shown schematically on the right hand side of Fig.~\ref{fig:fpsAB}. A crucial difference, however, is that no infrared GY fixed points with $\alpha_2>0$ and $\alpha_1=0$ are realized in model B. Hence, unlike in model A, UV fixed points solutions  with finite $\alpha_2^* \gg \alpha_2^{\text{SM}}( \mu  \gtrsim 0.1 \, \mbox{TeV})$ are not a priori excluded phenomenologically, though constrained, and  the corresponding matching conditions $\alpha_{1,2}^{\text{SM}}(M_F)= \alpha_{1,2}^{\text{BSM}}(M_F)$ can have solutions. 
Integrating the RG trajectories which leave the $B_2$ UV fixed point into  the $\alpha_1$ direction towards lower energies, 
we find  $M_F\sim 0.025$ TeV, as depicted in Fig.~\ref{fig:FPB1}. 
Similarly, for the fixed point $B_4$ we find $M_F = \mathcal{O}(10^{-2} \text{ TeV})$. 
We learn that asymptotic safety can predict the mass scale of new physics. The scale is disfavored phenomenologically, though only narrowly. The impact of higher loop corrections is  studied in the following Sec.~\ref{sec:scalar}.

Fixed point solutions $B_{1,3}$ with $\alpha_2^* = 0$ require more detailed analysis, as asymptotic freedom is absent. Although $\alpha_2$ is relevant at the fixed points $B_{1,3}$ due to the Yukawa interactions, it may turn irrelevant along a trajectory toward the IR, as $\alpha_{\kappa,y}$ become smaller causing $B_2^{\rm eff}$ to become negative.

\subsection{Model C (doublets,  $Y=-\frac12$)}

For model C, the BSM fermions have the representation $\psi \left(1, 2, -\frac12\right)$, which is the same as the one of the SM leptons $L$, leading to the Yukawa interactions 
\begin{equation}
-\mathcal{L}^{\text{C}}_{\text{Y}} = \kappa \,\overline{E} {H}^{\dagger} \psi_L + \kappa'\, \overline{L}S\psi_R + y\, \overline{\psi}_L S \psi_R + \mathrm{h.c.}\,.
\end{equation}
All physical fixed points in the \texttt{210} approximation are listed in Tab.~\ref{tab:fpsC}, and have $\alpha_1$ as an  irrelevant coupling. Besides the Gaussian (FP$_1$), one Banks-Zaks (FP$_2$) and four Gauge-Yukawa fixed points in $\alpha_2$ (FP$_{3..6}$) are realized. Similarly to the arguments used in the discussion of model A, the relation $\beta_y / \alpha_y = \beta_\kappa'/\alpha_{\kappa'} + \alpha_\kappa$, which holds in model C,  see  \eq{Yukawas}  and Sec.~\ref{sec:appBetaf}  for the RG-coefficients,  excludes a solution GY$_{2\kappa\kappa' y}$. In addition, there is a line of fixed points $\alpha_{\kappa'}^* + \alpha_y^* \approx 0.047$ with $\alpha_\kappa^* = 0$ (FP$_6$), that covers three solutions GY$_{2\kappa'}$, GY$_{2y}$ and GY$_{2\kappa' y}$, and give rise to a marginal coupling. However, no physical gauge-Yukawa fixed point involving $\alpha_1$ exists, and hence there is no candidate UV fixed point provided by model C  at lowest loop order.
\begin{table}[b]
	\centering
	\renewcommand{\arraystretch}{1.1}
	\setlength{\tabcolsep}{0pt}
	\rowcolors{2}{lightgray}{}
	\begin{tabular}{`ccccccccl`}
		\toprule
\rowcolor{LightBlue}
	{ \bf Model C$\ $}	 & $\  \alpha_1^* \quad $ & $\quad  \alpha_2^* \quad $ & $\quad  \alpha_{\kappa}^*\ $ & $\ \alpha_{\kappa'}^*\ $ & $\ \alpha_y^*\ $ & $\: $rel.$\, $ & $\,$ ir. $\,$&  $\,$Info \\ \midrule	%
FP$_1$ & 0$^{(+)}$ & 0$^{(-)}$ & 0$^{(+)}$ & 0$^{(+)}$ & 0$^{(+)}$ & 1 & 4 & G \\ 
FP$_2$ & 0$^{(+)}$ & 0.038 & 0$^{-}$ & 0$^{-}$  & 0$^{-}$ & 3 & 2 & BZ$_2$ \\
FP$_3$ & 0$^{(+)}$ & 0.039 & 0.020 & 0$^{-}$  & 0$^{-}$ & 2 & 3 & GY$_{2 \kappa}$ \\
FP$_4$ & 0$^{(+)}$ & 0.054 & 0.027 & 0.049  & 0$^{+}$ & 0 & 5 & GY$_{2 \kappa\kappa'}$ \\
FP$_5$ & 0$^{(+)}$ & 0.053 & 0.011 & 0$^{-}$ & 0.046 & 1 & 4 & GY$_{2 \kappa y}$ \\
FP$_6$ & 0$^{(+)}$ & 0.052 & 0$^{-}$  & $0.047-\alpha_y^*$ & $\alpha_y^*$ & 1 & 3 & GY$_{2  \kappa' y}$
		\\ \bottomrule
	\end{tabular}
	\caption{Partially and fully interacting  fixed points of model C 
	in the \texttt{210} approximation, notation as in Tab.~\ref{tab:fpsA}.
	At this loop order, no viable candidates for UV fixed points exist.} 
	\label{tab:fpsC}
\end{table}

\subsection{Model D  (doublets,  $Y=-\frac32$)}

In model D the BSM Yukawa Lagrangean reads
\begin{equation}
-\mathcal{L}_{\text{Y}}^{\text{D}} =   y\, \overline{\psi}_L S \psi_R + \kappa\bar{E}\tilde{H}^{\dagger}\psi_{L}  + \mathrm{h.c.}\,,
\end{equation}
with $\psi(\bm 1,\bm 2,-3/2)$. Physical fixed points are listed in Tab.~\ref{tab:fpsD}, with remarkable small coupling values $\alpha^* < 1$. All solutions $\alpha_1^* = 0$ suffer from the triviality problem. Besides the Gaussian, and BZ$_2$, all three possible Gauge-Yukawa fixed points involving $\alpha_2$ only are realized (FP$_{3..5}$ in Tab.~\ref{tab:fpsD}), but fall in this category. Viable candidates $D_{1..4}$ for UV fixed points are of the Gauge-Yukawa type involving at least the $\alpha_1$ gauge coupling as well as the BSM Yukawa interaction $\alpha_y$, as only GY$_{1\kappa}$ and GY$_{12\kappa}$ are unphysical. 

\begin{figure}
	\centering
		\includegraphics[width=0.43\textwidth]{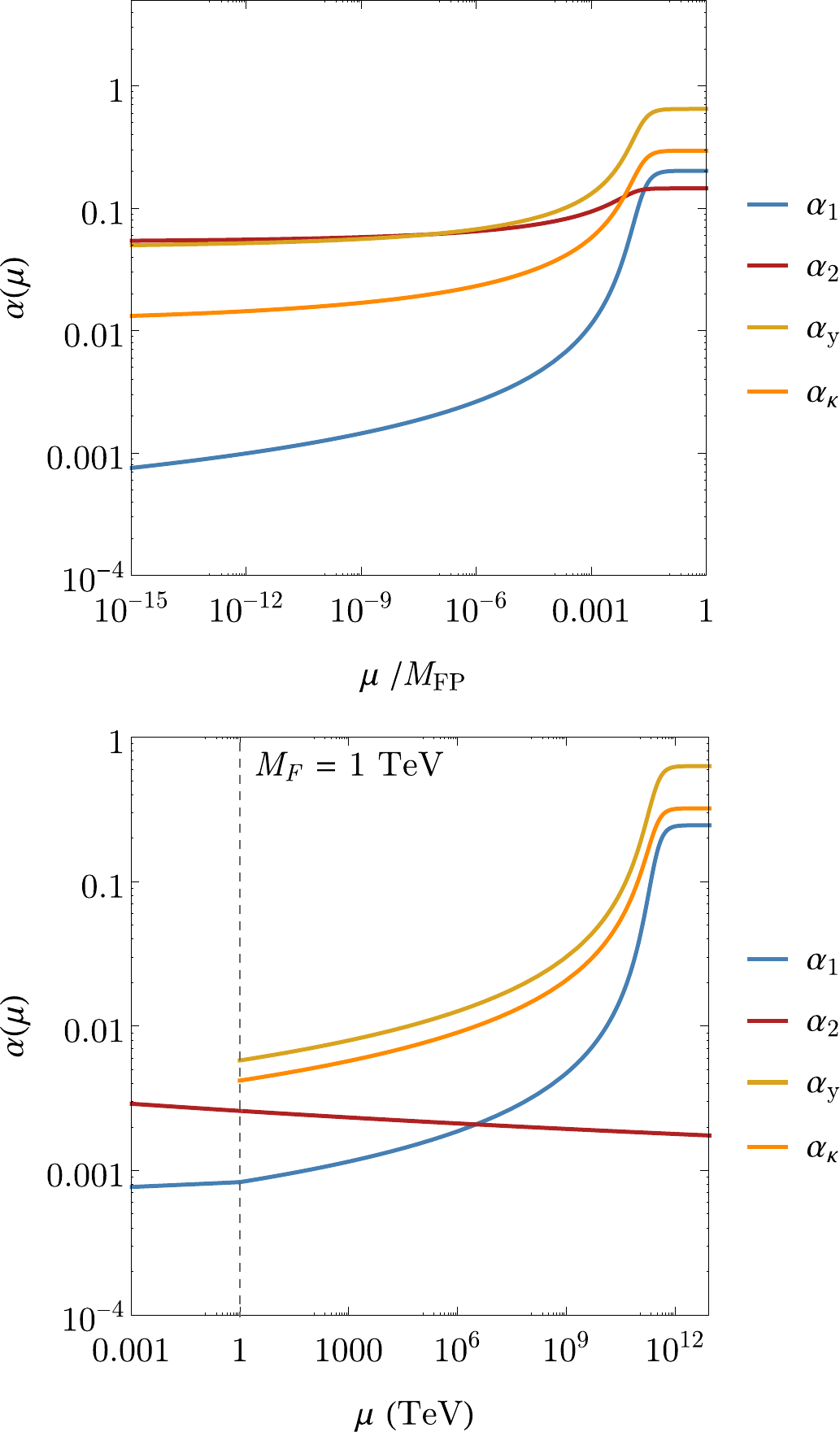}
	\caption{Renormalization group running  of model D.
	Top: BSM running from fixed point $D_2$, where matching is not possible. Bottom: running to the fixed point $D_1$ after matching at $\mu=1$ TeV (dashed vertical line). }
	\label{fig:FPD}
\end{figure}

Projecting onto the $\alpha_2$-$\alpha_\kappa$-plane, the hierarchy is similar to model A,  see Fig.~\ref{fig:fpsAB}, with $D_3$ being the most, and $D_2$ the least ultraviolet fixed points. Moreover, the same argument holds regarding the total IR fixed point GY$_{2\kappa y}$, which attracts trajectories going towards SM coupling values like those following the $\alpha_1$ critical direction from $D_{2,4}$, as depicted on the left hand side in Fig.~\ref{fig:FPD}. Small values of $\alpha_2$ along the trajectory are required, implying solutions $D_{1,3}$ as possible UV fixed points. Matching onto the SM is then possible at a range of scales, for $D_1$ we obtain $\alpha_{\kappa}(M_F=1 \text{ TeV}) = 4.2 \cdot 10^{-3},\alpha_{y}(M_F=1 \text{ TeV}) = 5.8 \cdot 10^{-3}$, which is shown in Fig.~\ref{fig:FPD}. Fixed point $D_3$ has also been studied in \cite{Barducci:2018ysr}, but discarded 
after including higher order contributions. We  retain this fixed point solution, deferring the discussion of higher loop-order effects to Sec.~\ref{sec:scalar}.

\begin{table*}
	\centering
	\renewcommand{\arraystretch}{1.2}
	\rowcolors{2}{lightgray}{}
	\begin{tabular}{`ccccccclcc`}
		\toprule
\rowcolor{LightBlue}
	{\ \ \bf Model D\ \ } 	 & $\quad \alpha_1^* \quad$ & $\quad \alpha_2^* \quad$ &  $\quad\alpha_{\kappa}^*\quad$ & $\quad\alpha_y^*\quad$ & $\:$rel.$\:$ & $\:$ irrel. $\:$&  $\:$Info $\:$& $\:$Name$\:$ & $\:$Matching$\:$ \\ \toprule	%
FP$_1$ & 0$^{(+)}$ & 0$^{(-)}$ & 0$^{(+)}$ & 0$^{(+)}$ & 1 & 3 & G  & &\\ 
FP$_2$ & 0$^{(+)}$ & 0.038 & 0$^-$ & 0$^-$ & 2 & 2 & BZ$_2$ & & \\
FP$_3$ & 0$^{(+)}$ & 0.039 & 0.020 & 0$^-$ & 1 & 3 & GY$_{2\kappa}$  & &\\
FP$_4$ & 0$^{(+)}$ & 0.052 & 0$^-$ & 0.047 & 1 & 3 & GY$_{2y}$  & &\\
FP$_5$ & 0$^{(+)}$ & 0.053 & 0.011 & 0.046 & 0 & 4 & GY$_{2\kappa y}$  & &\\ 
FP$_6$ & 0.246 & 0$^{(-)}$ & 0.322 & 0.631 & 2 & 2 & GY$_{1\kappa y}$ & $D_1$ & \cmark\\
FP$_7$ & 0.202 & 0.145 & 0.295 & 0.647 & 1 & 3 & GY$_{12\kappa y}$ & $D_2$ & \xmark\\ 
FP$_8$ & 0.288 & 0$^{(-)}$ & 0$^-$ & 0.778 & 3 & 1 & GY$_{1 y}$ & $D_3$ & \cmark\\
FP$_9$ & 0.239 & 0.152 & 0$^-$ & 0.782 & 2 & 2 & GY$_{12 y}$ & $D_4$ & \xmark\\
 \bottomrule
	\end{tabular}
	\caption{Partially and fully interacting  fixed points of model D
	in the \texttt{210} approximation, notation as in Tab.~\ref{tab:fpsA};  
	see Figs.~\ref{fig:fpsAB} and  \ref{fig:FPD} for the phase diagram and sample trajectories.}
		\label{tab:fpsD}
\end{table*}
 
\subsection{Model E  (singlets,  $Y=0$)}

The Yukawa interactions in model E read
\begin{equation}
-\mathcal{L}_{\text{Y}}^{\text{E}} =   \kappa\,\bar{L}\tilde{H}\psi_{R} + y\, \overline{\psi}_L S \psi_R + \mathrm{h.c.}\,,
\end{equation}
Since $\psi$ is a singlet under all gauge groups, $\beta_y$ is always positive in the \texttt{210} approximation, requiring $\alpha_y  = 0$ at all scales, as this coupling is irrelevant. This decouples the left-chiral BSM fermion $\psi_L$  and the BSM scalar $S$ from the SM plus $\psi_R$ at this loop order. Only the Gaussian fixed point, the Banks-Zaks in $\alpha_2$ and a gauge-Yukawa GY$_{2\kappa}$ are present, and $\alpha_1$ is irrelevant for all of them. This leaves the model without viable candidates of UV fixed points
at  \texttt{210}  approximation.

\subsection{Model F (triplets,  $Y=0$) \label{sec:F}}

In model F, the BSM fermions $\psi(1,3,0)$ are in the adjoint of $SU(2)_L$ with vanishing  hypercharge. The BSM Yukawa sector can be written as
\begin{equation}
-\mathcal{L}_{\text{Y}}^{\text{F}} =   \kappa\,\bar{L}\tilde{H}\psi_{R} + y\, \overline{\psi}_L S \psi_R + \mathrm{h.c.}\,.
\end{equation}
In this setup, asymptotic freedom is absent for both gauge couplings, making the Gaussian completely IR attractive and excluding any kind of Banks-Zaks fixed points. In the \texttt{210} approximation, $\beta_1$ is independent of $\alpha_y$, and $\beta_y$ is independent of $\alpha_1$, as $\psi$ does not carry hypercharge. Hence the two-loop contributions of $\kappa$ are the only negative terms in $\beta_1$, requiring $\alpha_\kappa^* \neq 0$. Moreover, only $\alpha_2$ contributions are negative in $\beta_y$, which suggests that $\alpha_2^* = 0$ implies $\alpha_y^* = 0$ and irrelevant.
However, none of the remaining gauge-Yukawa solutions GY$_{1\kappa}$, GY$_{2\kappa}$, GY$_{2\kappa y} $, GY$_{12\kappa}$ and GY$_{12\kappa y}$ are realized, as $\kappa$ contributions in $\beta_1$ are too small compared to  one and other two loop terms. This leaves the Gaussian as the only physical fixed point; we conclude that there is no AS fixed point  at \texttt{210} in model F.

\subsection{Summary Top-Down \label{sec:sum}}

In Secs.~\ref{sec:A}-\ref{sec:F} we have gained first insights into the  fixed point structure of models A -- F in a top-down approach of solving the RGEs at leading orders directly, and running towards infrared scales. The results for model A,B and D collected  in Tables \ref{tab:fpsA}, \ref{tab:fpsB} and \ref{tab:fpsD} show several signatures of UV fixed points that can be matched onto the SM, but also indicate that those are borderline perturbative. This suggests that the fixed points are sensitive to contributions from higher loop orders. 
We also found that the models C, E and F do not provide any viable solutions at \texttt{210} and the question arises whether this is just a feature of the approximation. 
In order to address both points,  we go in Sec.~\ref{sec:scalar}  beyond the \texttt{210} approximation. To handle the increased algebraic complexity of higher loop corrections and the quartic sector, a bottom-up approach will be employed, studying the RG running from the IR to the UV instead, mapping out the BSM critical surface.

\section{\bf Running Couplings 
\label{sec:scalar}}

In this section, we  discuss the renormalization group flow of couplings  beyond the leading order approximation which has been employed in the previous Sec.~\ref{sec:benchmarks}. We explore in detail how the running of couplings depends on the values of  BSM couplings $\{\alpha_{\rm BSM}\}$ at the matching scale. The main new technical additions in this section are the quartic scalar and the portal  couplings, and the inclusion of loop effects up to the complete 2-loop order ({\tt 222}  approximation),  or,  if available, the complete 3-loop order ({\tt 333} approximation).  We are particularly interested in the running of couplings from a bottom-up perspective, and study the flow for a given set of BSM initial values $\alpha_{\rm BSM}$  at the matching scale. We then ask whether these values together with the SM input reach Planckian energies without developing poles, exhibit  asymptotic safety, and the stability of the quantum vacuum.

We give our setup and initial conditions in Sec.~\ref{sec:ini}, and briefly review the RG-flow within the SM in Sec.~\ref{sec:sm}.
After identifying relevant correlations between feeble  and weakly-sized BSM couplings in Secs \ref{sec:Feeble} and \ref{sec:Weak}, respectively, we present in Sec.~\ref{sec:crit-sum} the BSM critical surface for each model.

\begin{figure*}
  \centering
  \hskip-.6cm 
  \includegraphics[width=1.7\columnwidth]{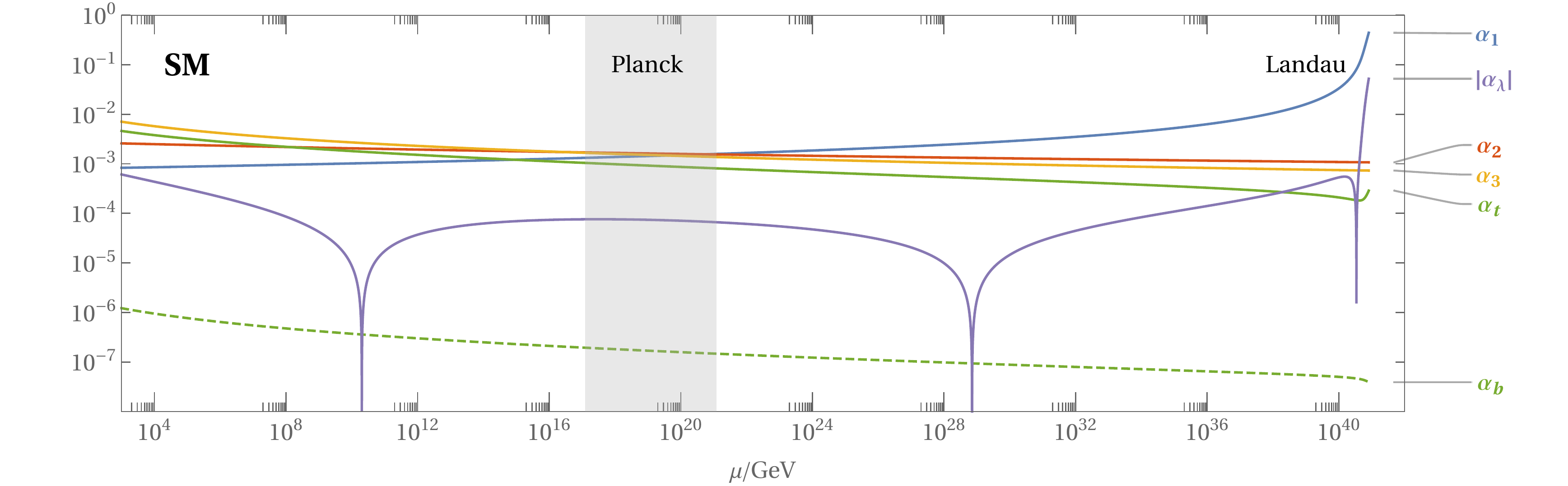}
  \caption{Renormalization group running of the SM. Shown are the gauge, Higgs, top (solid green) and bottom (dashed green) Yukawa couplings at the complete 3-loop order   starting from the 1~TeV regime  up to  the deep UV. The Planck scale is indicated by the gray band. The Higgs self coupling changes sign twice, around $10^{10}$GeV and around $10^{30}$GeV. Inbetween, the SM vacuum is metastable. Ultimately, the hypercharge and the Higgs coupling approach UV Landau pole around $10^{41}$GeV.}
  \label{fig:SM}
\end{figure*}

\subsection{Setup and Boundary Conditions \label{sec:ini}}

We retain the renormalization group running for the three gauge couplings of the SM \eq{couplings-gauge}, and up to three BSM Yukawa couplings \eq{couplings-Yukawa}.  Going beyond the leading order {\tt 210} approximation, we also retain the Higgs quartic self-interaction $\lambda$, the BSM quartics $u,v$, and the quartic portal coupling $\delta$
\begin{equation}\label{couplings-quartics}
\begin{array}{cc}
\di
\alpha_\lambda=\frac{\lambda}{(4\pi)^2}\,, &
\di
\alpha_\delta=\frac{\delta}{(4\pi)^2}\,,\\[2ex]
\di
\alpha_u=\frac{u}{(4\pi)^2}\,,&
\di
\alpha_v=\frac{v}{(4\pi)^2}\,.
\end{array}
\end{equation}
Moreover, it is well-known that the SM top and bottom Yukawa couplings $y_{t,b}$ critically influence the running of the Higgs quartic and, therefore, must be retained as well. We introduce them as
\begin{equation}\label{couplings-tbYukawa}
  \alpha_t=\frac{y_t^2}{(4\pi)^2}\,,\qquad \alpha_b=\frac{y_b^2}{(4\pi)^2}\,.
\end{equation}
Overall, \eq{couplings-gauge}, \eq{couplings-Yukawa}, \eq{couplings-quartics}, and \eq{couplings-tbYukawa} results in 12 (or 11) independent running couplings for models A and C (or models B, D, E, and F).
 
 We also remark that the scalar quartic interactions  
couple back into the Yukawa sectors starting at two loop, and into the gauge sectors starting at three (or four) loop, depending on whether the participating matter fields are charged (uncharged) under the gauge symmetry. Conversely, the Yukawa couplings couple back into the quartic starting at one loop, as do the weak and hypercharge gauge couplings into the Higgs.  We expect therefore a crucial interplay between BSM Yukawas and the portal coupling with Higgs stability.
In addition, the leading order study in Sec.~\ref{sec:benchmarks} showed that some of the fixed point coordinates might come out within the range ${\cal O}(0.1 - 1.0)$, indicating that strict perturbativity cannot be guaranteed. For these reasons, we develop the fixed point search and the study of RG equations up to the highest level of approximation where all couplings are treated on an equal footing, $i.e.$~the complete two loop order ({\tt 222} approximation). The running of SM couplings, which  serves as a reference scenario,  is studied up to the complete three loop order ({\tt 333} approximation). 

All our models require boundary conditions with six SM couplings at the matching scale $\mu_0$, which for all practical purposes corresponds to the mass of the BSM fermions $\psi$. To be specific, we take the matching scale  in this section to be
\beq\label{Matching}
\mu_0=1\,\mathrm{TeV} \, .
\eeq
The initial conditions for the SM couplings then read, using  $M_t \simeq 172.9$~GeV and  \cite{Tanabashi:2018oca,Buttazzo:2013uya},
\begin{equation}\label{matching}
\begin{aligned}
    \alpha_1(\mu_0) &\simeq 8.30 \cdot 10^{-4} , &
     \alpha_\lambda(\mu_0)&\simeq 6.09 \cdot 10^{-4} , \\ 
     \alpha_2(\mu_0) &\simeq 2.58 \cdot 10^{-3} ,&
    \alpha_t(\mu_0) &\simeq 4.61 \cdot 10^{-3} , \\
    \alpha_3(\mu_0) &\simeq 7.08 \cdot 10^{-3} , &
    \alpha_b(\mu_0)&\simeq 1.22 \cdot 10^{-6} .
\end{aligned}
\end{equation}
Hence, in our conventions, initial couplings are within the range  ${\cal O}(10^{-6} - 10^{-2})$. We are now in a position to discuss the running of couplings and 
the ``BSM critical surface'', $i.e.$~the set of values for BSM couplings at the matching scale which lead to viable RG trajectories all the way up to the Planck scale.

\subsection{Standard Model \label{sec:sm}}

We briefly discuss running couplings within  the SM at the complete 3-loop order in perturbation theory \cite{Degrassi:2012ry,Buttazzo:2013uya,Mihaila:2012fm,Bednyakov:2012rb,Bednyakov:2012en,Bednyakov:2013eba,Chetyrkin:2012rz,Chetyrkin:2013wya}, displayed in Fig.~\ref{fig:SM}.
 Overall, the SM running is rather slow with gauge, quartic and Yukawa  couplings mostly below ${\cal O}(10^{-2})$ or  smaller. We also observe that the Higgs potential becomes metastable starting around $10^{10}$~GeV  \cite{Degrassi:2012ry,Buttazzo:2013uya}, an effect which is mostly driven by the quantum corrections from the top Yukawa coupling  $\alpha_t$. 
Further, an imperfect gauge coupling unification is observed around $10^{16}$~GeV. 
Quantum gravity is expected to kick in around the Planck scale,  $M_{\rm Pl}\approx 10^{19}$~GeV, indicated by the gray-shaded area. As an aside, we notice that the Higgs beta function essentially vanishes at Planckian energies
\beq\label{lambda_Planck}
\mu\approx M_{\rm Pl}:\quad\alpha_\lambda\approx 10^{-4}\,,\quad \beta_\lambda\approx 0\,.
\eeq

If quantum gravity can be neglected, hypothetically, we may extend the running of couplings into the transplanckian regime. The hypercharge coupling would then reach a Landau pole  
around $10^{41}$~GeV.  Also, its slow  but steady growth  would eventually  dominate over the slowly decreasing  top Yukawa coupling,  and thereby stabilize the quantum vacuum  starting around $10^{29}$~GeV. Ultimately, however, the Higgs coupling reaches a Landau pole 
alongside the $U(1)_Y$ coupling and the SM stops being predictive. 

\begin{figure*}
  \centering
  \includegraphics[width=1.6\columnwidth]{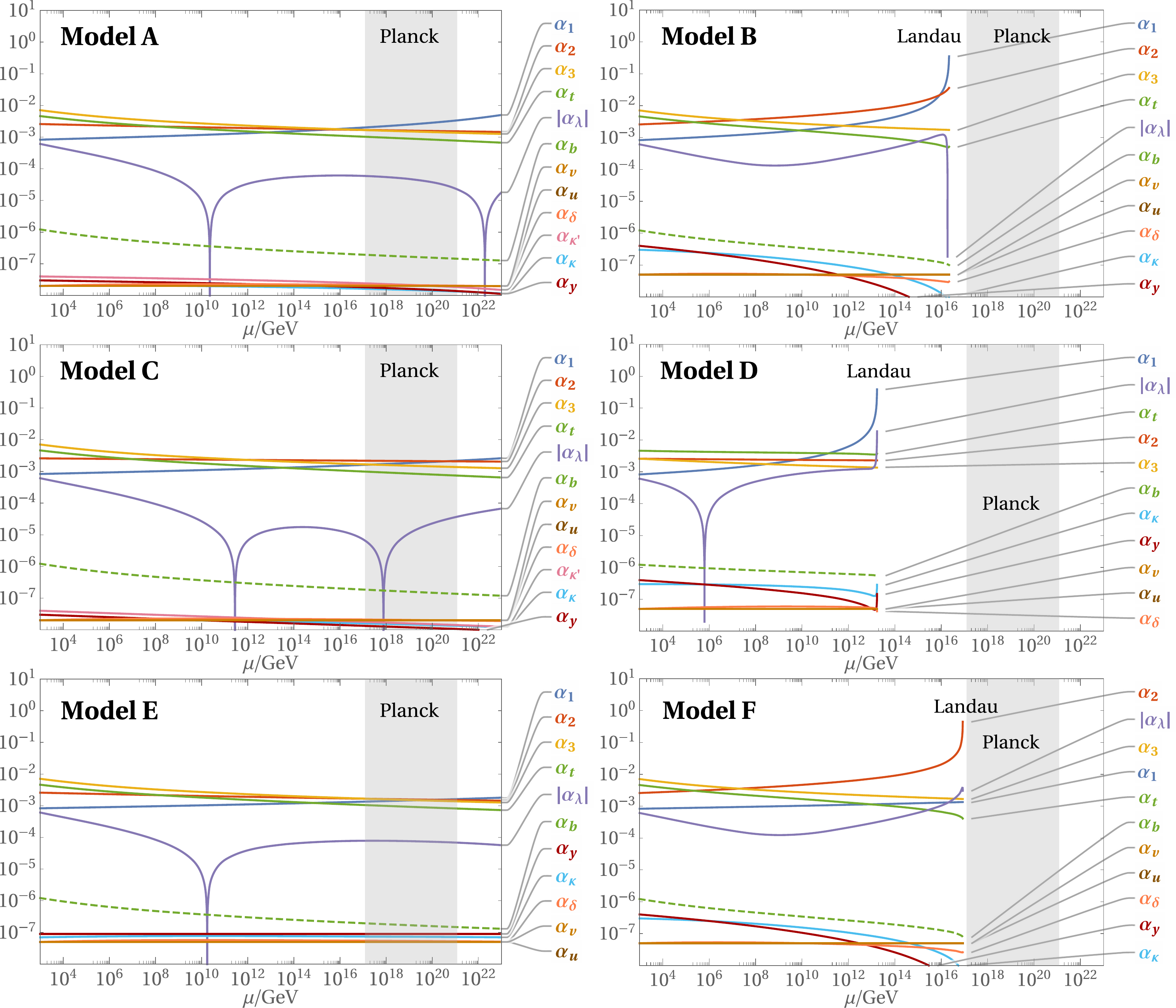}
 \caption{Renormalization group running of models A -- F with  feeble BSM couplings.   The gray-shaded area indicates the Planck scale. Sub-Planckian Landau poles arise in  model B, D (model F)  in the hypercharge (weak) coupling.}
  \label{fig:noBSM}
\end{figure*}

\subsection{Feeble BSM Couplings \label{sec:Feeble}}

Next, we include new matter fields on top of the SM ones and switch on the BSM couplings at the matching scale \eqref{matching}. A minimally invasive  choice  are very small, feeble, BSM  couplings such that they do not significantly influence the renormalization group flow up to the Planck scale. Their own running would then be well encoded already by the leading order  in the perturbative expansion, and models resemble the SM, extended by vector-like fermions. Specifically, we  consider here  initial values of the order of $\alpha_{\rm BSM} \approx 10^{-7}$ or smaller.

\begin{figure*}
\centering
\includegraphics[width=1.6\columnwidth]{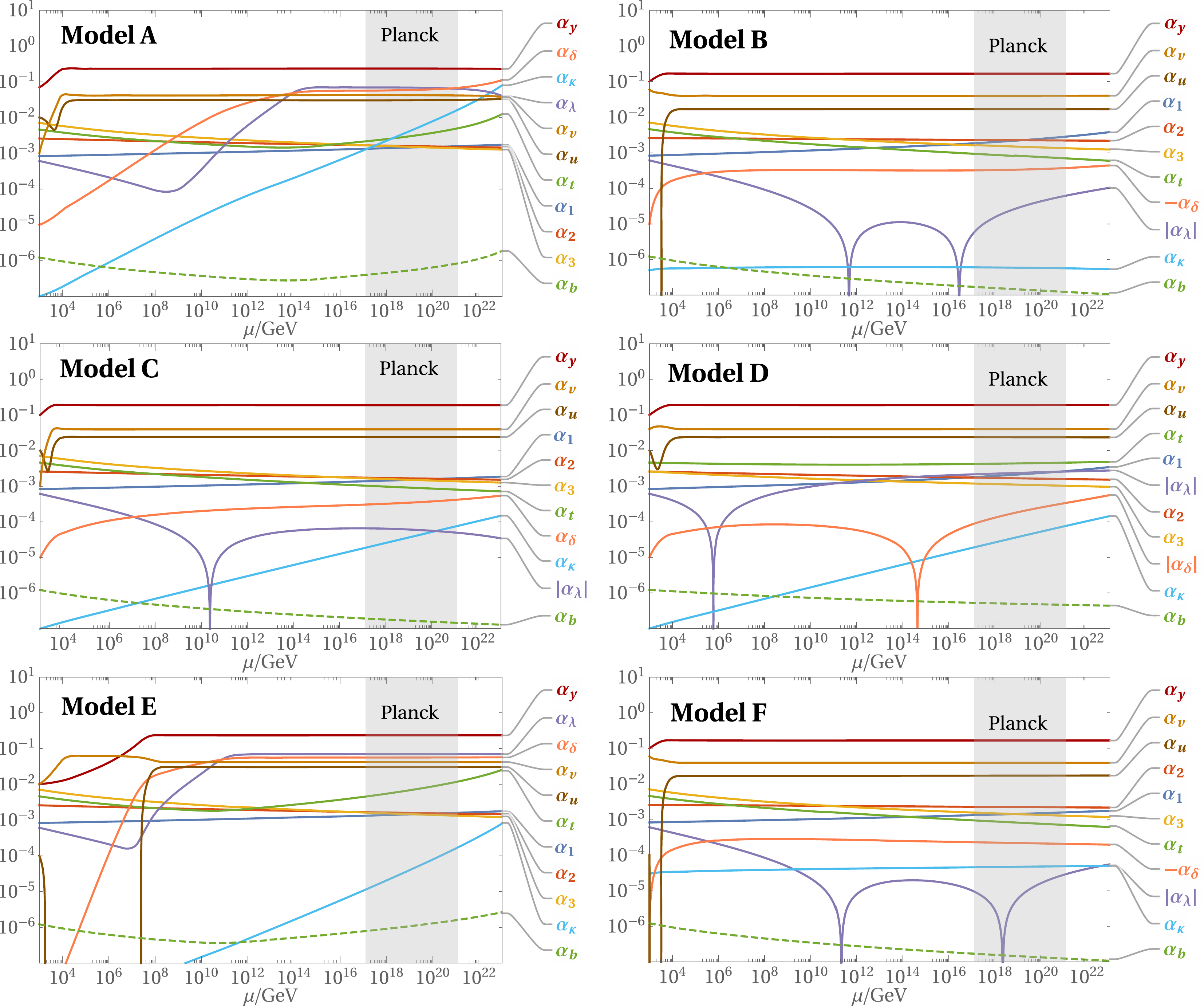} 
  \caption{Renormalization group running of models A -- F and  $\alpha_{\kappa} \approx  0$, $|\alpha_\delta| = 10^{-5}$  (except for model E, where
  $|\alpha_\delta|$  is very feeble),
  as well as $\alpha_{\kappa'} = 0$ for models A and C. In model A, C, D and E, small initial values of $\alpha_\kappa$ (light blue)  blow up in the UV, while for B and F the trajectories remain more stable. The Higgs potential  (lilac) is not stabilized by Yukawa interactions (model B-C,F), but for sufficiently  large initial values of $\alpha_\delta$ (orange)  in the singlet models  (model A and E).}
    \label{fig:cross-y}
\end{figure*}
\begin{figure*}
\includegraphics[width=1.6\columnwidth]{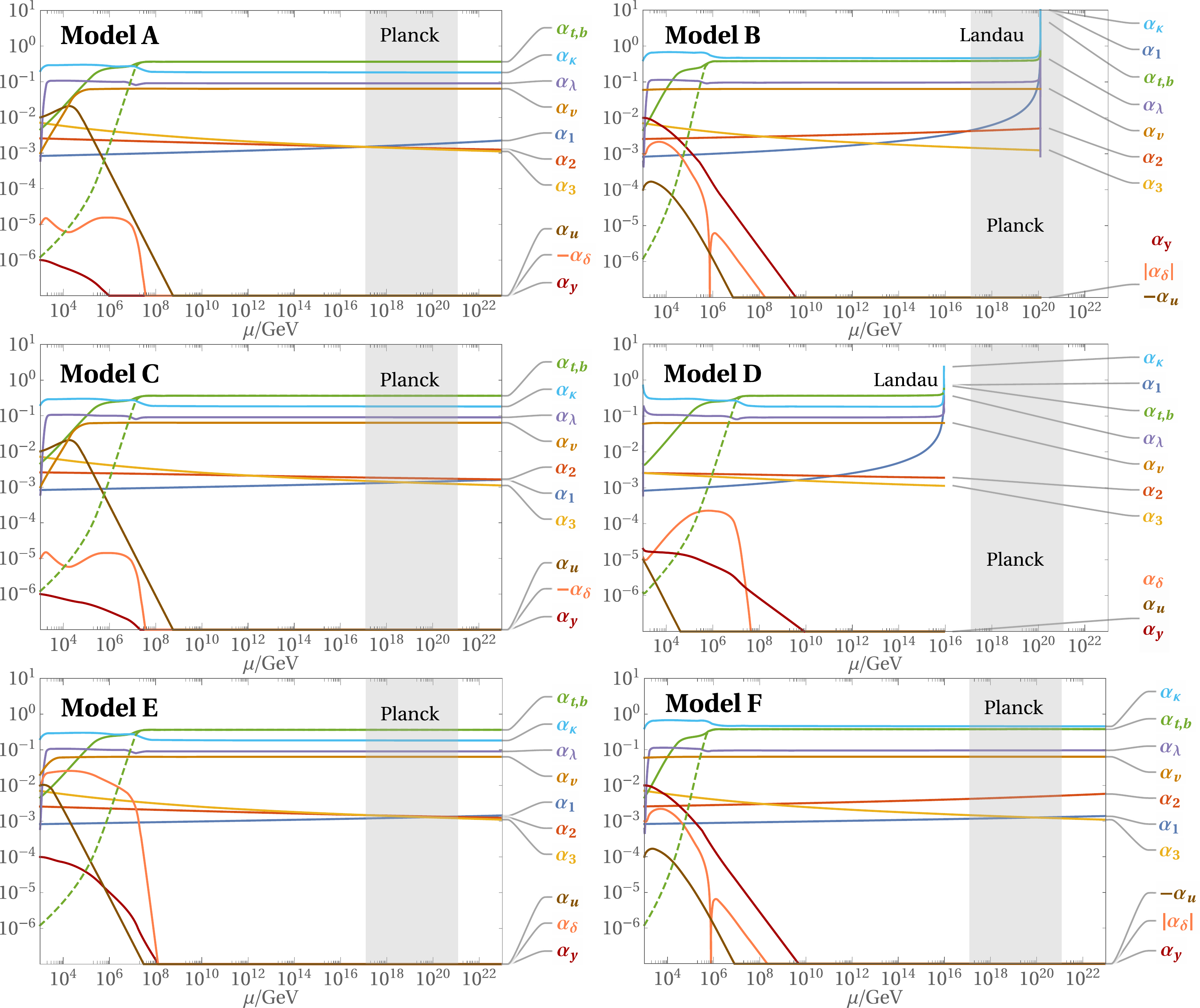}
  \caption{Renormalization group running of models A -- F with $\alpha_{y} \approx  0$ and $|\alpha_\delta| \approx  0$ , and  $\alpha_{\kappa'} = 0$ for models A and C. In all models the couplings $\alpha_{y,u,\delta}$ (red, brown, orange) are driven to zero in the UV. The solid (dashed) green line denotes the flow of the SM top (bottom) Yukawa, which merge
  at the cross-over. }
  \label{fig:cross-kappa}
\end{figure*}
\begin{figure*}
  \includegraphics[width=1.6\columnwidth]{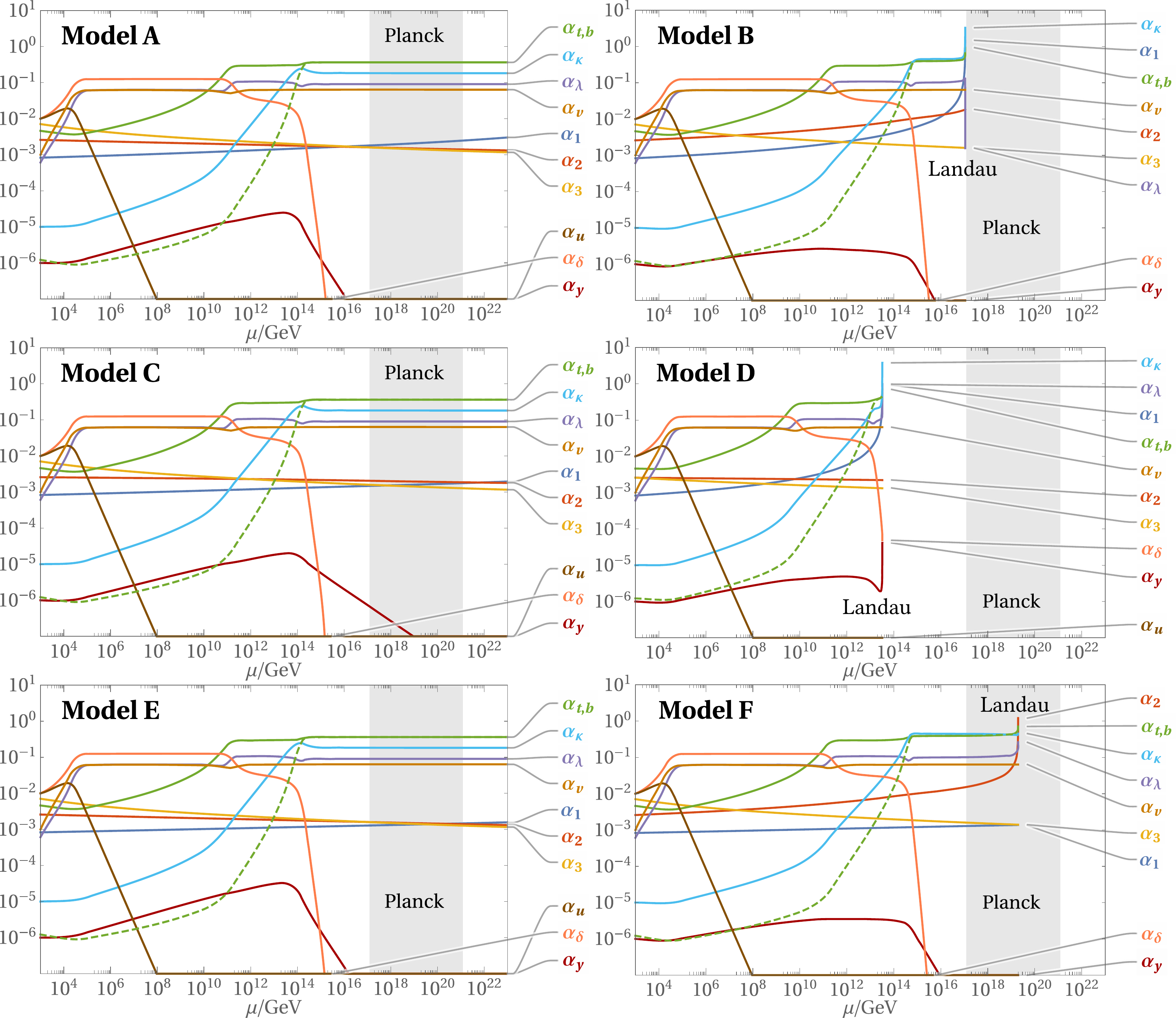}
  \caption{As in  Fig.~\ref{fig:cross-kappa} but for significantly lower values of $\alpha_\kappa$ (light blue) at the matching scale.
  Models B, D and F exhibit Landau poles before or at the Planck scale.  Qualitative features observed in Fig.~\ref{fig:cross-kappa}  for $\alpha_{y,u,\delta}$  (drop towards UV) and  $\alpha_{b,t}$ (merging)   remain.}
    \label{fig:cross-kappa-delta}
\end{figure*}
\begin{figure*}
  \centering
  \includegraphics[width=1.6\columnwidth]{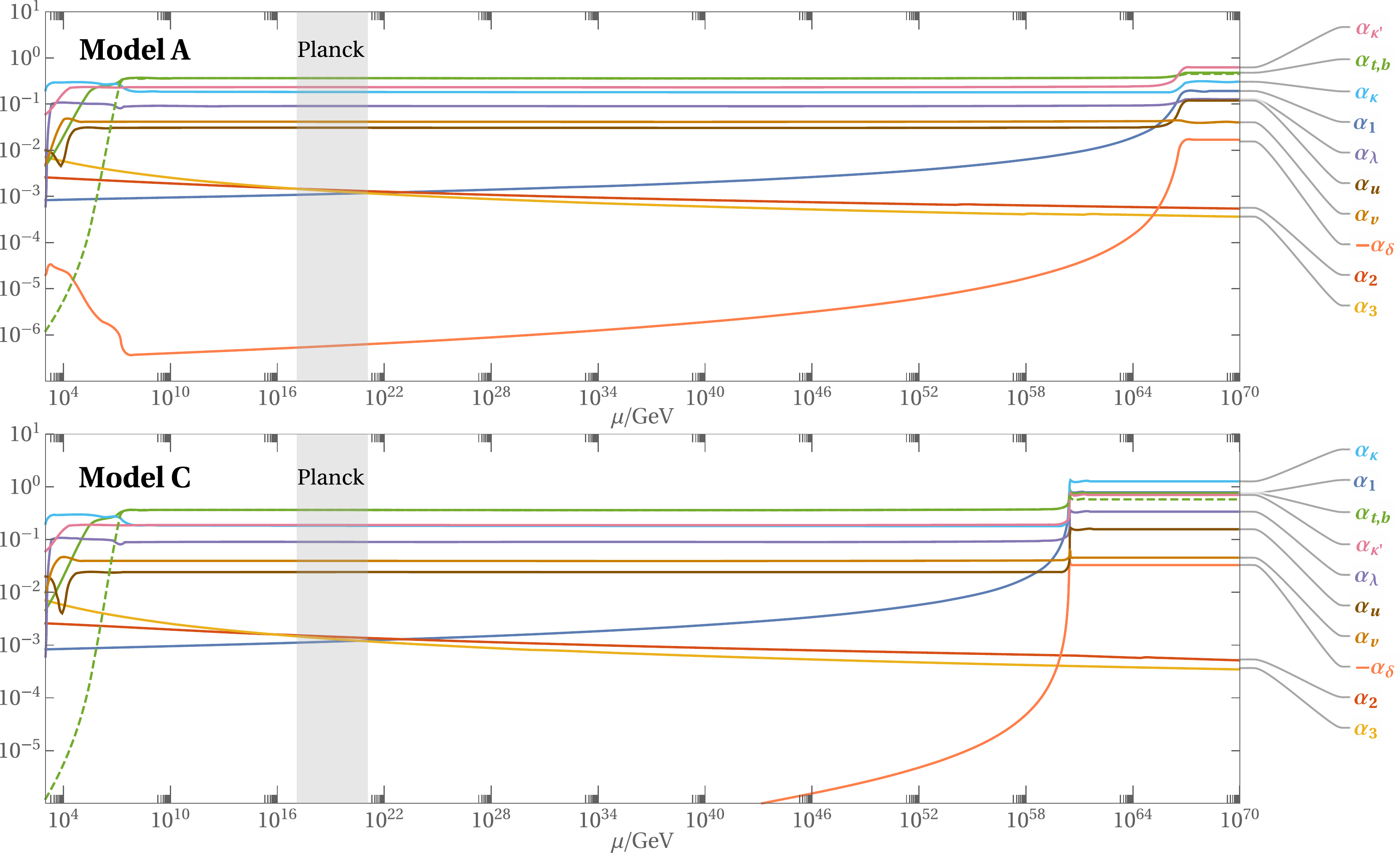}
  \caption{Renormalization group running in models A and C with $\alpha_y = 0$, $\alpha_{\kappa,\kappa'} \neq 0$ and
  $|\alpha_\delta| \lesssim 10^{-5}$. The solid (dashed) green line denotes the SM top (bottom) Yukawa. The flow is stabilized by a cross-over fixed point just before the Planck scale (gray area), see text.
  As $\alpha_{1}$ (steel blue) becomes large, a complete UV fixed point is reached in the far UV. }
  \label{fig:AC_cross}
\end{figure*}
\begin{figure*}
  \centering
 \includegraphics[width=1.6\columnwidth]{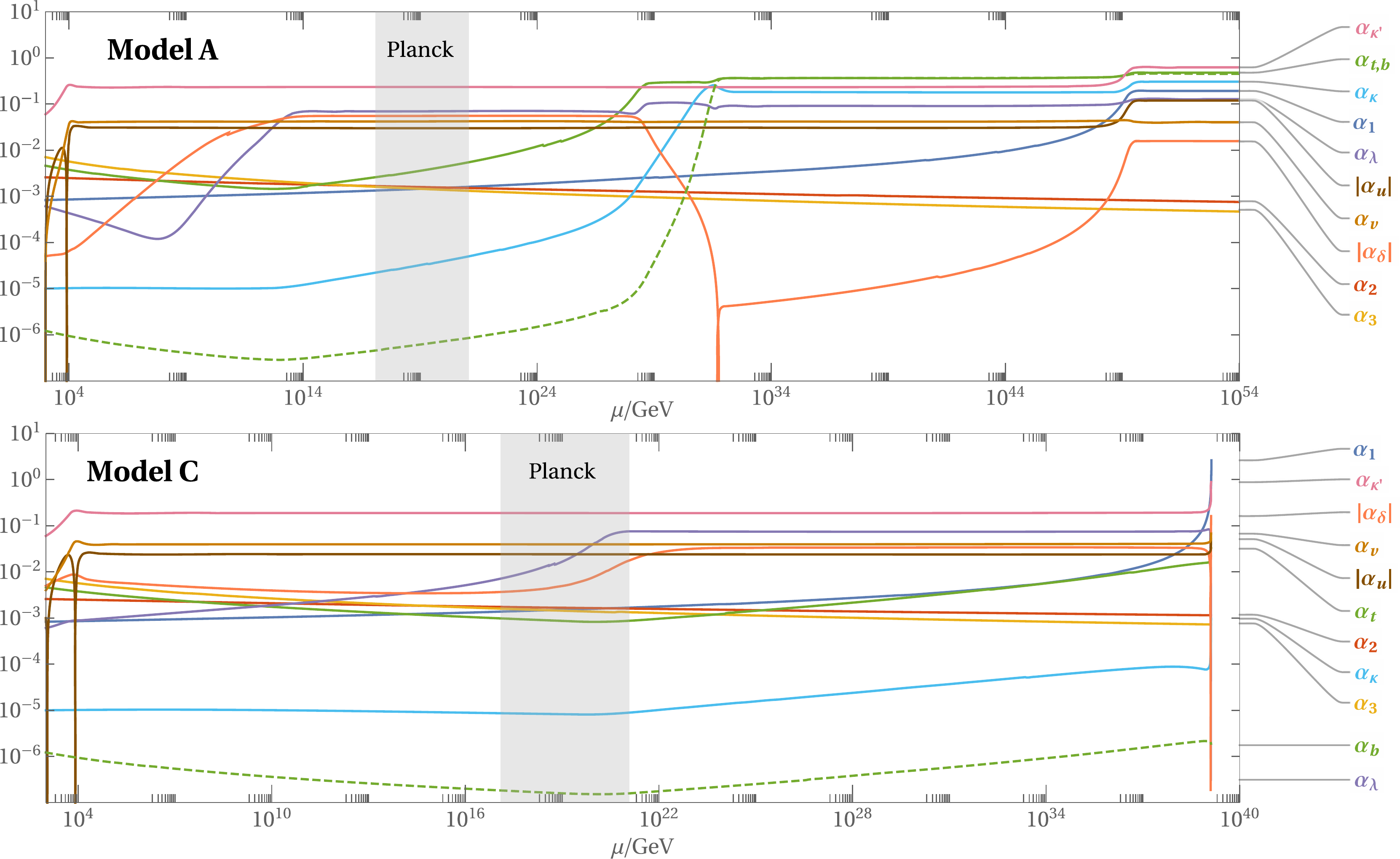}
  \caption{As in  Fig.~\ref{fig:AC_cross} but for smaller $\alpha_{\kappa}$ (light blue) and larger portal coupling $|\alpha_\delta|$ (orange).
The flow is stabilized by a cross-over fixed point involving  $\alpha_\delta$ (orange). In model A, the flow continues into the same walking regime and UV fixed point as in Fig.~\ref{fig:AC_cross}, while model C runs into a pole way beyond the Planck scale.  $\alpha_u$ (brown) changes sign below $M_{\rm Pl}$.}
  \label{fig:AC_cross2}
\end{figure*}
\begin{figure*}[t]
  \centering
  \includegraphics[width=1.6\columnwidth]{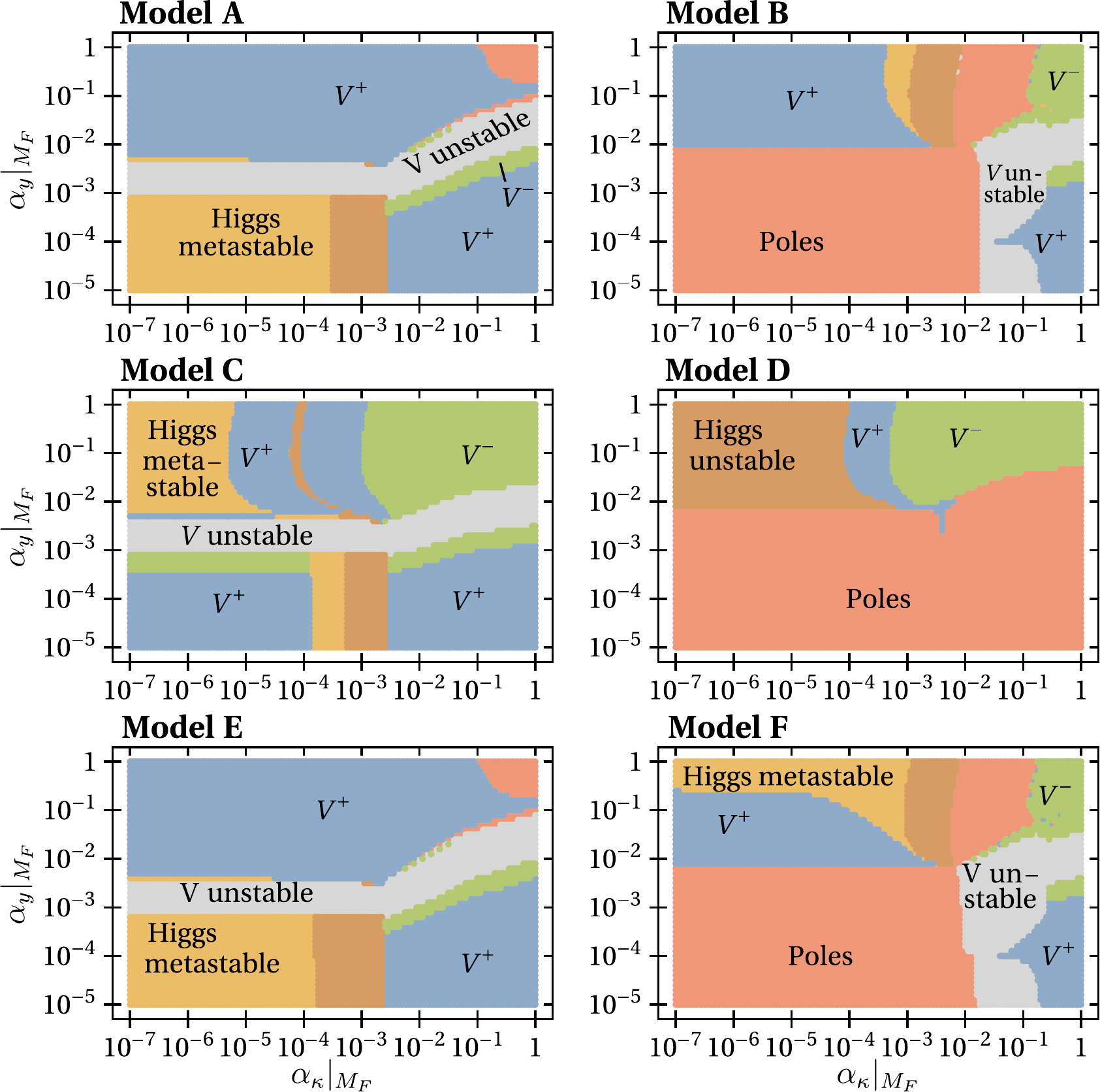}
  \caption{BSM critical surface for models A -- F with $\left\{\alpha_{\kappa'},\, \alpha_\delta,\, \alpha_u,\, \alpha_v \right\}|_{M_F} = \left\{0,\, 5,\, 1,\, 4\right\} \cdot 10^{-5}$ and  values $\left\{\alpha_\kappa,\, \alpha_y \right\}\atMF$ at the matching scale. The colors indicate if the corresponding vacuum  at the Planck scale $M_{\rm Pl}$  is either 
stable  $V^+$ (blue) or $V^-$ (green), \eq{eq:vstab}, an unstable BSM vacuum (gray),  a stable vacuum for $\alpha_{u, v}|_{M_{\rm Pl}}$ but with $\alpha_\lambda|_{M_{\rm Pl}} < 0$ (yellow for $\alpha_\lambda|_{M_{\rm Pl}} > -10^{-4}$, otherwise brown) or if the RG flow runs into a pole (red). Resolution is $141 \times 61$ points per model.}
  \label{fig:surface_kappa_y}
\end{figure*}
\begin{figure}
  \centering
  \includegraphics[width=0.8\columnwidth]{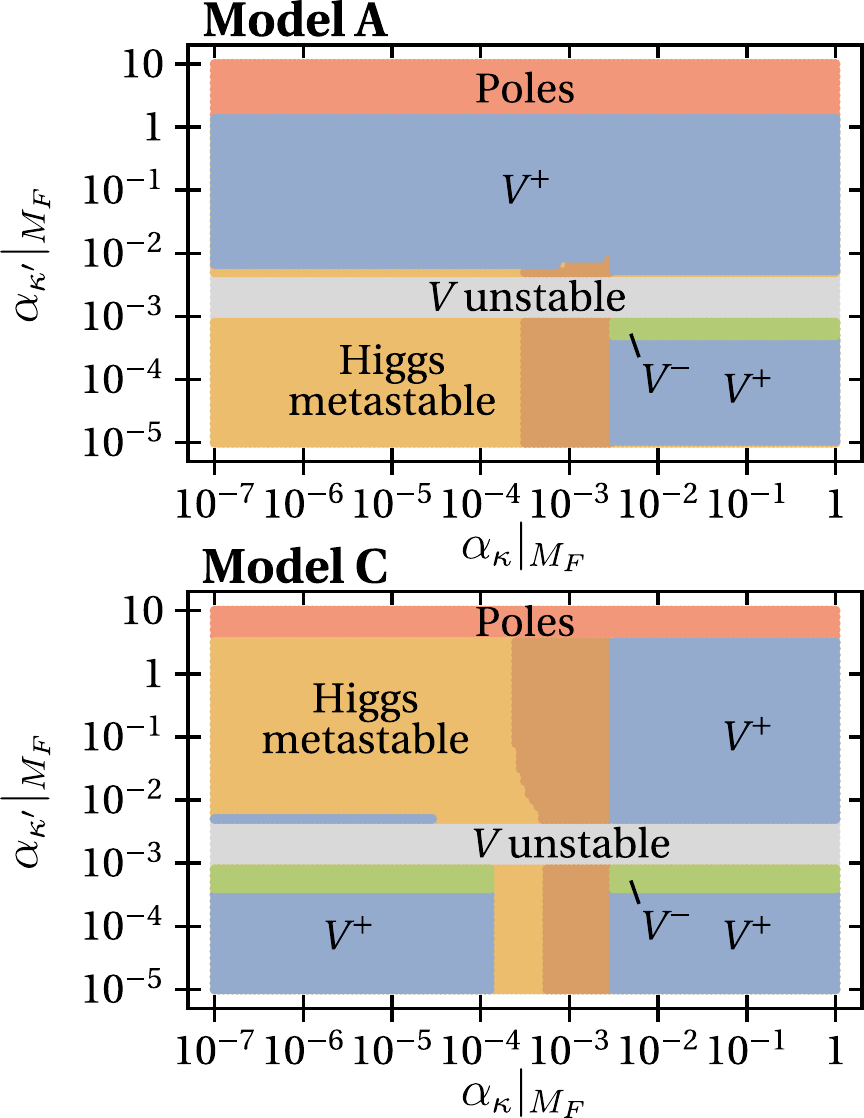}
  \caption{As in Fig.~\ref{fig:surface_kappa_y} for models A and C in the  $\left\{\alpha_\kappa,\, \alpha_{\kappa'} \right\}\atMF$ plane with matching conditions $\left\{\alpha_y,\, \alpha_\delta,\, \alpha_u,\, \alpha_v \right\}\atMF = \left\{0,\, 5,\, 1,\, 4\right\} \cdot 10^{-5}$.  Resolution is $141 \times 61$ points per model. }
  \label{fig:surface_kappa_kappa'}
\end{figure}

\subsubsection{Models A, C, and E}

Sample trajectories with feeble BSM couplings
 are shown in Fig.~\ref{fig:noBSM} (plots to the left) for models A, C and E. In all cases, we observe a SM-like running of couplings. 
The new matter fields modify the running of gauge couplings very mildly. For model A and E, we  find a vanishing beta function for the Higgs quartic coupling, much similar to the SM \eq{lambda_Planck}. For model C, we observe that the regime of Higgs  metastability terminates exactly around the Planck scale,
\beq\label{lambda_Planck}
\mu\approx M_{\rm Pl}:\quad\alpha_\lambda\approx 0\,,\quad \beta_\lambda\approx 0\,.
\eeq
We conclude that  in model A,C and E feeble initial values for the BSM couplings lead to SM-like trajectories including vacuum meta-stability up to the Planck scale. Hence, the BSM critical surface covers the region 
in which all couplings are feeble.

\subsubsection{Models B, D, and F}

The models B, D and F with feeble BSM couplings at $\mu_0$ reach a Landau pole prior to the Planck scale, with sample trajectories shown in Fig.~\ref{fig:noBSM} (plots to the right). Specifically, in model B asymptotic freedom for the weak and hypercharge couplings is lost leading to a Landau pole around $10^{16}$GeV reached first for the hypercharge, going hand-in-hand with the loss of vacuum stability. Similarly, a strong coupling regime with a Landau pole  is reached around $10^{13}$GeV ($10^{16}$GeV) for model D (model F). Hence, none of these models can make it to the Planck scale for feeble BSM couplings, excluding this region from the BSM critical surface. Notice though that the growth of the gauge couplings in model B and F stabilizes the Higgs sector all the way up to close to the pole.

\subsection{Weak  BSM Couplings \label{sec:Weak}}

In the following we explore several matching scenarios for each of the models A -- F with BSM couplings of at least the same order of magnitude as the SM couplings at the matching scale \eqref{matching}. 
In this regime, Yukawa interactions play a crucial role in avoiding Landau poles and stabilizing RG flows, inviting a classification by the couplings involved.
Due to the importance for Higgs stability,  we also distinguish scenarios with or without portal coupling effects.
After identifying relevant correlations between BSM couplings, we obtain in Sec.~\ref{sec:crit-sum} the BSM critical surface for each model.

\subsubsection{Models A -- F with $\alpha_y \neq 0$}

For  $\alpha_{\kappa,\kappa'} \approx  0$, the BSM Yukawa  $\alpha_y \neq 0$ slows down the running of gauge couplings and removes all Landau poles before $M_{\rm Pl}$.
Moreover, it stabilizes the running of the quartics $\alpha_{u,v}$, due to a walking regime $\beta_{y,u,v} \approx 0$, which may extend until after the Planck scale. This is displayed in Fig.~\ref{fig:cross-y}. Due to sizable BSM couplings, the portal $\alpha_\delta$ is being switched on, influencing the running of the Higgs quartic $\alpha_\lambda$. For larger values $\alpha_\delta\atMF$, the Higgs potential can be stabilized, i.e., $\alpha_\lambda > 0$ between $M_F$ and $M_{\rm Pl}$ (model A and E), while smaller values of 
$\alpha_\delta\atMF$ cause the Higgs potential to flip sign twice before the Planck scale (model B and F), or $\alpha_\lambda$ remains negative at $M_{Pl}$ (model  C and D).

In models A, C, D and E, feeble initial values of $\alpha_\kappa$ grow in coupling strength, eventually destabilizing the trajectories in the far UV. For the triplet models B and F, $\beta_\kappa$ remains small for feeble or weakly coupled $\alpha_\kappa$, providing greater windows of stability. In summary, the BSM critical surface covers the parameter space where $\alpha_y$ is weak and $\alpha_{\kappa, \kappa'}$ are feeble at the matching scale.

\subsubsection{Models A -- F with $\alpha_\kappa \neq 0$}

A weakly coupled Yukawa interaction $\alpha_\kappa$ may stabilize the SM scalar sector. The choice 
\begin{equation}\label{eq:cross2}
  \alpha_{y,\kappa'} =  0\,, \quad \alpha_{\kappa} \neq  0\,, \quad |\alpha_\delta| \approx  0\,,
\end{equation}
is depicted in Fig.~\ref{fig:cross-kappa}. 
A common feature of all models A -- F is the stabilization of $\alpha_\lambda$ in a walking region together with $\alpha_\kappa$ and the SM Yukawas, as all of which  couple to the SM Higgs directly. The BSM potential on the other hand lacks a sizable Yukawa interaction, and $\alpha_v$ self-stabilizes around $\alpha_v^* \approx 13/204$.
This phenomenon is not disrupted  by feeble initial values of $|\alpha_{u,y,\delta}|$, which are driven to zero in the UV limit. However, the scenario is not viable for model D as the Landau pole still appears before the Planckian regime. In model B, the pole appears soon after $M_{\rm Pl}$.

The initial value of  $\alpha_\kappa$ can be reduced for  $\alpha_\delta\atMF$ large enough to stabilize the running of the Higgs quartic: 
\begin{equation} \label{eq:cross2_delta}
    \alpha_{y,\kappa'} =  0\,, \quad \alpha_{\kappa} \neq  0\,, \quad |\alpha_\delta| \neq 0.
\end{equation}
 For models A, C and E, this allows for feeble $\alpha_\kappa$ at the matching scale, while in models B, D and F poles arise below or at the Planck regime, as displayed in Fig.~\ref{fig:cross-kappa-delta}.

\subsubsection{Models A and C with $\alpha_{\kappa'} \neq 0$}

Models A and C feature the additional Yukawa interaction $\alpha_{\kappa'}$, giving rise to another walking regime 
\begin{equation}\label{eq:cross1}
  \alpha_y =  0\,, \quad \alpha_{\kappa, \kappa'} \neq 0 \,, \quad |\alpha_\delta| \approx  0\,,
\end{equation}
shown in Fig.~\ref{fig:AC_cross}. Starting from the matching scale $M_F$, these regions are reached before the Planck scale, and at various speeds by different couplings, creating a rich landscape of intermediate pseudo fixed points and scales. Throughout the walking regime, SM and BSM Yukawas and quartics slow down in model A
at
\begin{equation}\label{eq:crossA1}
\begin{aligned}
\alpha_{t,b}^* 		&\simeq 3.61 \cdot 10^{-1} \,, &
\alpha_{\kappa'}^*		&\simeq 2.32 \cdot 10^{-1} \,, \\
\alpha_{\kappa}^* 		&\simeq 1.80 \cdot 10^{-1} \,, &
\alpha_{u}^*		&\simeq 3.07 \cdot 10^{-2}\,, \\
\alpha_{\lambda}^* 		&\simeq 8.95 \cdot 10^{-2} \,, &
\alpha_{v}^*		&\simeq 4.12 \cdot 10^{-2}\,, \\
\end{aligned}
\end{equation}
 and  in model C at
\begin{equation}\label{eq:crossC1}
\begin{aligned}
\alpha_{t,b}^* 		&\simeq 3.61 \cdot 10^{-1} \,, &
\alpha_{\kappa'}^*		&\simeq 1.88 \cdot 10^{-1} \,, \\
\alpha_{\kappa}^* 		&\simeq 1.80 \cdot 10^{-1} \,, &
\alpha_{u}^*		&\simeq 2.44 \cdot 10^{-2}\,, \\
\alpha_{\lambda}^* 		&\simeq 8.95 \cdot 10^{-2} \,, &
\alpha_{v}^*		&\simeq 3.92 \cdot 10^{-2}\,. \\
\end{aligned}
\end{equation} 
On the other hand, the portal $\alpha_\delta$ and gauge couplings continue to run, although the latter is slowed down by the magnitude of the Yukawas. Consequently, Landau poles are avoided even far beyond  the Planck scale. Moreover, the SM [BSM] quartics $\alpha_\lambda$ $\left[\alpha_u,\,\alpha_v\right]$ are stabilized by the $\alpha_\kappa$ $\left[\alpha_{\kappa'}\right]$ Yukawa couplings. All of these phenomena are consequences of the vicinity of a pseudo-fixed point with $\alpha_{1,2,3,y,\delta}^* = 0$, separating the SM and BSM scalar sectors, as well as Yukawa couplings from each other. 
This decoupling is expected to be realized to all loop-orders, because,
in its vicinity, the action decomposes as
\beq
\mathcal{S} =  \mathcal{S}_H\left(H,\, L[E],\, \psi_{R[L]}\right) +  \mathcal{S}_S\left(S, \, E[L],\, \psi_{L[R]}\right)
\eeq 
for model A [C],  up to corrections of the order of the SM lepton Yukawas  $Y_\ell$,  \eq{Yell}.
However, this separation can only be realized approximately for small gauge and portal couplings. Hence, the RG flow eventually leaves the walking regime in the far UV due to the slow residual running 
of $\alpha_1$ or $\alpha_\delta$. 
Ultimately, this triggers a cross-over away from the walking regime and into an interacting  UV fixed point regime where all couplings bar the non-abelian gauge and the BSM Yukawa couplings take non-trivial values. 

Specifically, for model A, the interacting UV fixed point is approximately given by
\begin{equation}\label{eq:AfarUV}
\begin{aligned}
 \alpha_1^* 		&\simeq 1.93 \cdot 10^{-1} \,, &
 \alpha_\kappa^*	&\simeq \ \ 3.05 \cdot 10^{-1} \,,
 \\
 \alpha_3^*		&=  \alpha_2^*= \alpha_y^*=0\,, &
 \alpha_{\kappa'}^*	&\simeq \ \ 6.25 \cdot 10^{-1} \,,
\\
 \alpha_\lambda^*	&\simeq 1.27 \cdot 10^{-1}\,, &
  \alpha_\delta^*	&\simeq -1.55 \cdot 10^{-2} \,,
 \\
 \alpha_t^*		&\simeq 4.78 \cdot 10^{-1} \,,&
 \alpha_u^*		&\simeq \ \ 1.19 \cdot 10^{-1} \,,
 \\   
 \alpha_b^*		&\simeq 4.53 \cdot 10^{-1} \,,&
 \alpha_v^*		&\simeq \ \ 4.03 \cdot 10^{-2} \,,
\end{aligned}
\end{equation} Note that the fixed point is rather close to the values of couplings in the walking regime \eq{eq:crossA1}.
Similarly,  in model C we find an approximate  UV fixed point with coordinates
\begin{equation}\label{eq:CfarUV}
\begin{aligned}
 \alpha_1^* 		&\simeq 7.64 \cdot 10^{-1} \,, &
 \alpha_\kappa^*	&\simeq \ \ 3.05 \cdot 10^{-1} \,,
 \\
 \alpha_3^*		&=  \alpha_2^*= \alpha_y^*=0\,, &
 \alpha_{\kappa'}^*	&\simeq \ \ 7.00 \cdot 10^{-1} \,,
\\
 \alpha_\lambda^*	&\simeq 3.38 \cdot 10^{-1}\,, &
  \alpha_\delta^*	&\simeq -3.30 \cdot 10^{-2} \,,
 \\
 \alpha_t^*		&\simeq 7.51 \cdot 10^{-1} \,,&
 \alpha_u^*		&\simeq \ \ 1.57 \cdot 10^{-1} \,,
 \\   
 \alpha_b^*		&\simeq 5.76 \cdot 10^{-1} \,,&
 \alpha_v^*		&\simeq \ \ 4.54 \cdot 10^{-2} \,.
\end{aligned}
\end{equation}
Again, we note that \eq{eq:AfarUV} is numerically close to  the walking regime \eq{eq:crossC1}.

Reducing $\alpha_\kappa\atMF$ destabilizes the running of the Higgs self-coupling $\alpha_\lambda$, which can however be remedied by a non-vanishing portal coupling $\alpha_\delta$: 
\begin{equation}\label{eq:cross1_delta}
  \alpha_y =  0\,, \quad \alpha_{\kappa, \kappa'} \neq 0 \,, \quad |\alpha_\delta| \neq 0\,.
\end{equation}
In model A, this enables trajectories with feeble $\alpha_\kappa\atMF$ to connect to the phenomena \eqref{eq:crossA1} and \eqref{eq:AfarUV}, while for model C, trans-planckian poles arise. This is displayed in Fig.~\ref{fig:AC_cross2}. In both models 
the coupling $\alpha_u$ (brown), whose overall sign separates the vacuum solutions $V^+$ from $V^-$, \eqref{eq:vstab},  
changes sign below $M_{\rm Pl}$.

In summary, the BSM critical surfaces of model A and C  include regions for both $\alpha_\kappa\atMF$ 
and $\alpha_{\kappa'}\atMF$ being perturbartively small.
For even smaller values of $\alpha_\kappa\atMF$, larger values of $\alpha_\delta\atMF$ are required, 
and Higgs stability is  not automatically guaranteed. 
The interplay of BSM  input values
on Planck-scale features is further detailed below (Sec.~\ref{sec:crit-sum}).

We emphasize that our models are  the first templates of asymptotically safe SM extensions 
with physical Higgs, top, and bottom masses,  and which
connect the relevant SM and BSM couplings at TeV energies with an interacting fixed point at highest energies. 
Another  feature of our models is the  low number  $N_F$ of new fermion flavors required for this.
In contrast, earlier attempts towards 
asymptotically safe SM extensions \cite{Bond:2017wut,Kowalska:2017fzw,Mann:2017wzh,Pelaggi:2017abg}
required moderate or large $N_F$, and either neglected the running of quartic and portal couplings   \cite{Bond:2017wut,Kowalska:2017fzw}, or used an  unphysically large mass for the Higgs \cite{Mann:2017wzh} in large-$N_F$ resummations which require further scrutiny
\cite{Pelaggi:2017abg,Alanne:2019vuk}. It will therefore be interesting to test the fixed point at higher loop orders, once available, and non-perturbatively using lattice simulations  \cite{Leino:2019qwk}, or functional renormalization.

\subsection{BSM Critical Surface \label{sec:crit-sum}}

We analyze  the state of the vacuum at the Planck scale  in dependence on the initial conditions of the BSM couplings 
at $M_F$ to determine the BSM critical surface in each model.
In accord with the reasoning in Sec.~\ref{sec:Weak}, the BSM Yukawas $\alpha_{y}$, $\alpha_\kappa$ are 
varied at the matching scale, 
with  the SM couplings 
fixed by \eqref{matching}. 
The remaining BSM couplings are, exemplarily, set to   
\beq
\left\{\alpha_{\kappa'},\, \alpha_\delta,\, \alpha_u,\, \alpha_v \right\}\atMF = \left\{0,\, 5,\, 1,\, 4\right\} \cdot 10^{-5}\,.
\eeq
For each model, we then sample  $141 \times 61$ different initial values  $(\alpha_{\kappa},\alpha_y)\atMF$ 
and integrate the RG flow at two loop accuracy for all couplings from the matching scale to the Planck scale. 
The result for all models is shown  in Fig.~\ref{fig:surface_kappa_y}.
Different parameter regions are color-coded to indicate the type of ground state at the Planck scale, 
or whether poles or instabilities arise prior to $M_{\rm Pl}$. Specifically,
we distinguish regions in $\alpha_{\kappa, y}\atMF$ that yield stable vacua  $V^+$ (blue) or $V^-$ (green), according to \eq{eq:vstab}, evaluated at the Planck scale.
Regions with negative Higgs quartic are called metastable (yellow), if $0 > \alpha_\lambda  |_{M_{\rm Pl}} > -10^{-4}$, and Higgs-unstable if $\alpha_\lambda  | _{M_{\rm Pl}} < -10^{-4}$.
In the remaining regions with unstable vacuum (gray) either BSM quartics $\alpha_u,\alpha_v$ do not comply with  \eq{eq:vstab} (regardless of $\alpha_\lambda$ and $\alpha_\delta$) or $\alpha_u,\alpha_v$ and $\alpha_\lambda$ do  comply with  \eq{eq:vstab}, but $\alpha_\delta$ does not.
Regions with Landau poles below or at the Planck scale are indicated in red. 

Next, we discuss the pattern of results in Fig.~\ref{fig:surface_kappa_y}.
Connecting to the region of feeble couplings Fig.~\ref{fig:noBSM}, Landau poles  are present before the Planck scale within at least $\alpha_{\kappa, y}\atMF \lesssim 10^{-3}$ in models B, D and F.
For models A, C and E on the other hand, within $\alpha_{\kappa, y}\atMF \lesssim 10^{-4}$ no poles arise and the Higgs potential is metastable  or even becomes stable at the Planck scale (model C), just as depicted in Fig.~\ref{fig:noBSM}.
 
Towards larger values of $\alpha_\kappa\atMF$,  models A, C and E exhibit a metastable and then unstable Higgs potential until $\alpha_\kappa\atMF$ is large enough to stabilize the potential as in Fig.~\ref{fig:cross-kappa}. 
The vacuum configuration at $M_{\rm Pl}$ is then the same as at the matching scale, either $V^+$  or $V^-$.
For models B and F, $\alpha_\kappa\atMF > 10^{-2}$ is required to move the Landau pole past the Planck scale, while this is not possible in model D.

If we are increasing  $\alpha_y \atMF$ instead, this  leads eventually to the ground state $V^+$ in the BSM potential, but Higgs stability is not guaranteed automatically, see Fig.~\ref{fig:cross-y}. 
If not obstructed by poles, each model exhibits a narrow "belt" of parameters  around $\alpha_{\kappa'}\atMF\approx {\cal O}(10^{-3})$ and any $\alpha_{\kappa}\atMF$, within which the BSM  potential is unstable due to $ \alpha_u < - \alpha_v$ in the $V^-$ ground state. Here, Coleman-Weinberg resummations  \cite{Litim:2015iea} or higher order scalar selfinteractions  \cite{Buyukbese:2017ehm} should be included before definite conclusions about stability are taken.

Another feature of models A and E is that for $\alpha_{\kappa,y}\atMF \gtrsim 10^{-1}$  simultaneously, Landau poles occur before the Planck scale. For the other models, RG trajectories are stabilized around $M_\text{Pl}$ in the $V^-$ ground state by quartic interactions. However, this region is especially sensitive to corrections from higher loop orders.

For models A and C, the additional Yukawa interaction $\alpha_{\kappa'}$ adds an extra dimension to the BSM critical surface. 
Its impact is  further investigated in Fig.~\ref{fig:surface_kappa_kappa'} (color-coding as in Fig.~\ref{fig:surface_kappa_y}) where we exemplarily explore the vacuum  state at the Planck scale within the $(\alpha_{\kappa},\alpha_{\kappa'})\atMF$ parameter plane, and 
\beq
\left\{\alpha_y,\, \alpha_\delta,\, \alpha_u,\, \alpha_v \right\}\atMF = \left\{0,\, 5,\, 1,\, 4\right\} \cdot 10^{-5}\,.
\eeq 
We find that the region with $\alpha_{\kappa'}\atMF \lesssim 10^{-4}$ is very similar to the region $\alpha_{y}\atMF \lesssim 10^{-4}$ in Fig.~\ref{fig:surface_kappa_y}, featuring a stable ground state for weakly coupled $\alpha_\kappa\atMF$. 
For both $\alpha_{\kappa, \kappa'}\atMF \gtrsim 10^{-2}$ the phenomena illustrated in Fig.~\ref{fig:AC_cross} occur, implying a stable $V^+$ region. 
The fate of the quadrant  with $\alpha_{\kappa'} \atMF \gtrsim 10^{-2}$ and $\alpha_\kappa \atMF \lesssim 10^{-2}$ hinges on the value of $\alpha_\delta \atMF$.
As can be seen from Fig.~\ref{fig:surface_kappa_kappa'}, its flow  can be  stable, as in Fig.~\ref{fig:AC_cross2}, while poles or Higgs metastability are possible as well.

The BSM critical surface  at the matching scale of  each model consists of the combined  $V^-$ plus $V^+$  regions, with slices in the multi-dimensional parameter space   shown in Fig.~\ref{fig:surface_kappa_y} and Fig.~\ref{fig:surface_kappa_kappa'} in green and blue.
All models  A -- F can be stable at least up to the Planck scale.
The yellow (metastability) regions may be 
included as well, as this corresponds to the situation of the SM.
In general, experimental constraints  on the BSM critical surface apply for matching scales around the TeV-scale, a topic further  discussed in
the next Sec.~\ref{sec:pheno}.

\section{\bf Phenomenology}
\label{sec:pheno}

In this section, we investigate the phenomenological implications of our models. Specifically, 
in Sec.~\ref{sec:prod}  we discuss  BSM sector production  at hadron and lepton colliders, and in Sec.~\ref{sec:decay} the decays of the BSM fermions and scalar. An important ingredient for  phenomenology is mixing between
SM and BSM fermions, the technical details for which are relegated to App.~\ref{sec:appscalar}. Resulting phenomenological consequences are worked out in Sec.~\ref{sec:mix} and include dileptonic decays of the scalars.
Constraints from Drell-Yan data on the matching scale are worked out in Sec.~\ref{sec:DY}. Implications for the leptons' anomalous magnetic moments are studied in Sec.~\ref{sec:AMM}.
In Sec.~\ref{sec:gm2} we  show that the portal coupling $\delta$ in  \eq{eq:scalarV}  together with $\kappa$ and $\kappa'$ can provide a chirally enhanced  contribution to the magnetic moments. This mechanism also induces EDMs for CP-violating couplings, discussed in Sec.~\ref{sec:edm}.
In Sec.~\ref{sec:LFV} we discuss constraints from charged lepton flavor violating (LFV) decays.

\subsection{BSM Sector Production   \label{sec:prod}}

Tree-level production channels of the BSM sector at $pp$ or $\ell\ell$ colliders are shown in~Fig.~\ref{fig:production}. Since the  fermions are colorless, pair production in $pp$ collisions is limited to quark-antiquark fusion to electroweak gauge bosons (diagrams (a) and (b)). Single production through Yukawa interactions with $s$-channel Higgs (diagram (c)) is also possible. In $\ell\ell$ colliders, the $\psi$ can also be produced with $t$-channel Higgs or $S$  in pairs (d) and singly (e). The contribution to $\bar \psi \psi$ production from $s$-channel neutral bosons is especially relevant, since it is present in all models in study (except for model E), in both $pp$ and $\ell\ell$ collisions, and all $N_F=3$ flavors of $\psi$ are produced. In the limit $M_F\gg m_f$, where $f$ is a quark or a lepton and $m_f (Q_f)$ denotes its mass (charge), the contribution to pair production via  photon exchange
at center of mass energy-squared $s$ reads
\begin{widetext}
\begin{equation}
\begin{aligned}
&\sigma_{\gamma}(\overline{f}f \rightarrow \overline{\psi}\psi) = 
N_F \frac{4\pi}{3}\frac{\alpha_e^2 Q_f^2}{s} \sum_{SU(2)_L}  Q^2_F\sqrt{1-\frac{4M_F^2}{s}}\left(1+\frac{2 M_F^2}{s}\right) \quad \text{for} \quad s > 4 M_F^2 \, , 
\end{aligned}
\end{equation}
where  we summed over  the $\psi$'s  flavors and $SU(2)_L$-components; $\alpha_e = e^2/4\pi$ denotes the fine structure constant. Corresponding cross sections are of the order 
$N_F Q_f^2  \sum Q_F^2 90  \, \mbox{fb}/(s [ \mbox{TeV}] )$ \cite{Patrignani:2016xqp}.
Note the enhancement in model B and D which contain fermions with $|Q_F|=2$, and result in effective charge-squares of $\sum Q_F^2=5$.
\begin{figure*}
	\centering
	\includegraphics[width=\textwidth]{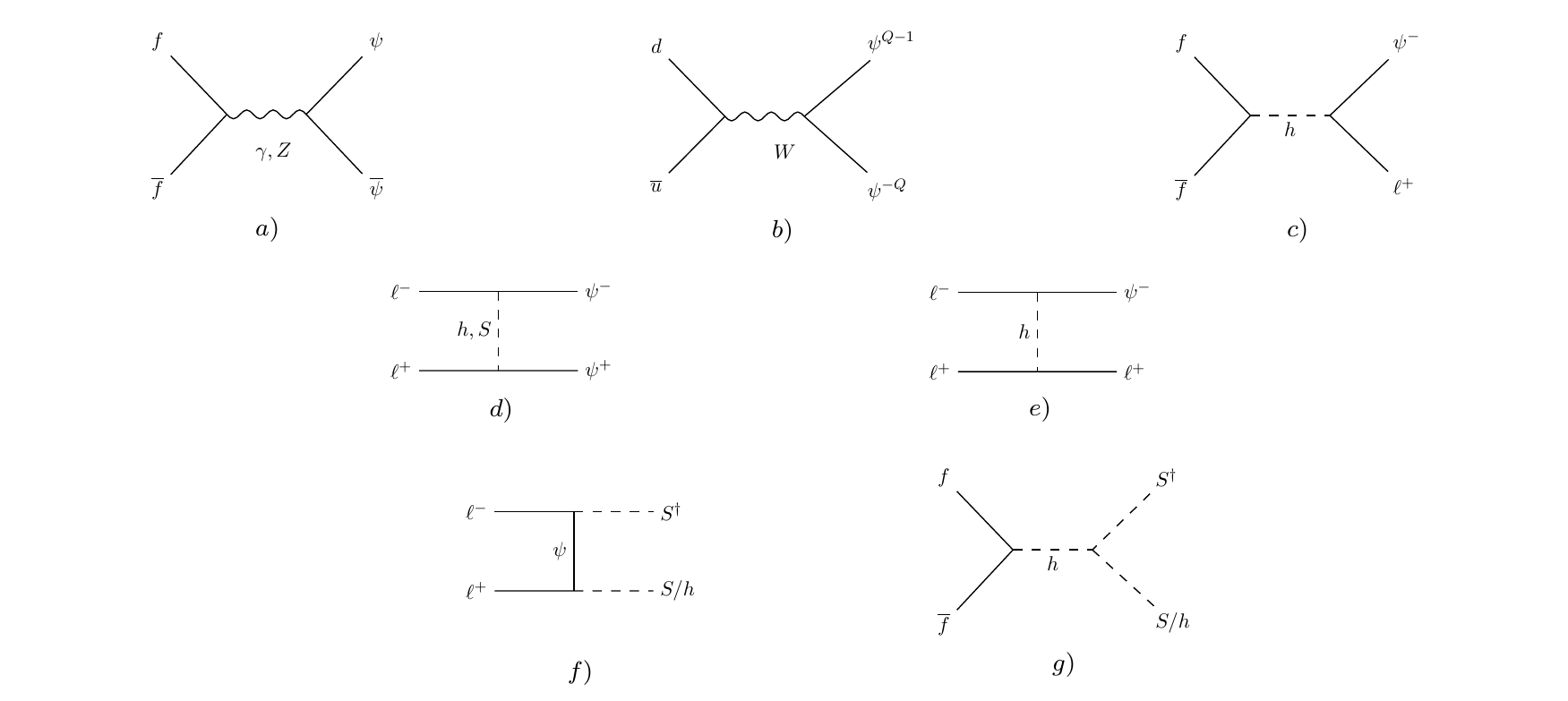}
	\caption{Production channels of the BSM particles at $pp$ and $\ell\ell$ colliders, with $f=\ell,q$. In diagram $f)$
	the $S$ and $S^{\dagger}$ labels are schematic for model A, see 
	text for details.}
		\label{fig:production}
\end{figure*}
The BSM scalars, which are SM singlets, can be pair-produced at lepton colliders in model A and C through the Yukawa interactions $(\kappa')$ with
  $\psi$-exchange (diagram (f)). The cross-section, for $s > 4 M_S^2$, then reads
\begin{equation}
\begin{aligned}
&\sigma(\ell^+\ell^-\rightarrow SS^{\dagger}) = 
\frac{N_F}{32\pi }\frac{\kappa'^4}{s}\left(1-\frac{4M_S^2}{s}\right)^{5/2} 
\int_{-1}^{1}\text{d}x\ {x^2(1-x^2)}{\left[\left(\frac{2(M_F^2-M_S^2)}{s}+1\right)^2-(1-\frac{4M_S^2}{s})x^2\right]^{-2}}\,.
\end{aligned}
\end{equation}
\end{widetext}
Denote by ${\rm Re} [S]$  and  ${\rm Im} [S]$  the  real, CP-even and CP-odd physical degrees of freedom of $S$, respectively.
Together the Yukawas  $\kappa$ and  $\kappa'$ induce single $S$-production, ${\rm Re} [S]$ or ${\rm Im} [S]$, in association with a Higgs (diagram (f)).

Another mechanism to probe the scalars is  through $S$-Higgs mixing (diagram (g)), which arises if the  portal coupling $\sim H^\dagger H \Tr [S^\dagger S]$ is switched on. In this diagram, 
the $h {\rm Re} [S] {\rm Re} [S]$ and $h {\rm Im} [S] {\rm Im} [S]$ couplings arise after electroweak symmetry breaking. In addition,
the $hh {\rm Re} [S]$ vertex is possible when the scalar $S$ acquires a VEV.
A detailed study of $\psi,S$ production at colliders is, however, beyond the scope of this work.

\subsection{BSM Sector Decay  \label{sec:decay}}

We discuss, in this order, the decays of the vector-like leptons $\psi$ and the BSM scalar  $S$.
Both subsections contain a brief summary at the beginning.

\subsubsection{Fermions}

Depending on the representation, coupling and mass hierarchies, the BSM fermions can decay  through the Yukawa interactions to Higgs plus lepton or  to $S$ plus lepton
 (only model A and C),
while some members of the $SU(2)_L$-multiplets need to cascade down within the multiplet first through $W$-exchange. 
These  are the states with electric charge $Q_F=-2$ (model B and D) and $Q_F=+1$ (model F). As detailed below, they
allow for macroscopic lifetimes.
Mixing with the SM leptons induces additional  $\psi$-decays to $Z,W$ plus lepton which are discussed in Sec.~\ref{sec:mix}.

The vector-like fermions with $Q_F=0$ and $Q_F=-1$ can decay through the Yukawa interactions $(\kappa)$ to $h \nu$ and $h \ell^-$, respectively, except in model C, in which
the Higgs  couples to $SU(2)_L$-singlet leptons and only the $Q_F=-1$ decay takes place through $\kappa$. 
Neglecting the lepton mass, the decay rate into Higgs plus lepton is
\begin{equation}\label{DRhell}
\Gamma(\psi\rightarrow \ell h) = \frac{\pi}{4}C_{\psi \ell}^2\,\alpha_{\kappa}{ M_F}\left(1-\frac{m_h^2}{M_F^2}\right)^2\,,
\end{equation}
where 
$C_{\psi \ell}=1/\sqrt{2}$ for the $T_3=0$ states in models B,F and $C_{\psi \ell} = 1$  otherwise. 
For $\alpha_{\kappa} \gtrsim 10^{-14}$ and $M_F$ at least a  TeV, one obtains a lifetime $\Gamma^{-1} \lesssim \mathcal{O}(10^{-13})$ s, which leads to a prompt decay.  In models A (C), the decays $\psi_i \to \ell_j S^\dagger_{ji}$ ($\psi_i \to \ell_j S_{ji}$) are also allowed if the BSM scalars are lighter than the vector-like fermions, with rate 
\begin{equation}
\Gamma(\psi \to \ell S) = \frac{\pi}{2}C_{\psi \ell}^2\,\alpha_{\kappa'}{ M_F}\left(1-\frac{M_S^2}{M_F^2}\right)^2\,.
\end{equation}
Models B and D contain $Q_F=-2$ fermions. After electroweak symmetry breaking, these cascade down through the weak interaction as $\psi^{-2} \to \psi^{-1} W^{*-}$, and subsequent decays.
The lifetime is then driven by the mass splitting within the multiplet. In the limit $M_F \gg m_W,m_Z$ one obtains  for $\Delta m =M_{\psi^{-2}} -M_{\psi^{-1}}$  from SM gauge boson loops
\cite{Cirelli:2005uq}$ \Delta m \simeq \frac{\alpha_2^{\text{PDG}}}{2} \left( 3 \sin \theta_W^2 m_Z + k
\right) $ and $ k  =m_W-m_Z ~(\mbox{model B})$, $  k=0 ~(\mbox{model D})$,
which is around a GeV in both models. Corresponding decay rates $\Gamma (\psi^{-2} \to \psi^{-1} \ell \nu )
\sim  G_F^2 \Delta m^5/(15 \pi^3)\simeq 3 \cdot 10^{-13} \mbox{GeV} (\Delta
m/[\mbox{GeV}])^5$ indicate around picosecond lifetimes of the $\psi^{-2}$, with a small, however macroscopic 
$c \tau \simeq 0.3$ mm resulting in displaced vertex signatures that can be searched
for at the LHC \cite{Evans:2016zau}.
In model F, the $Q_F=+1$ fermions decay similarly through $ \psi^{+1} \to
W^{+*}\psi^0 $, with $\Delta m  = M_{\psi^{+1}} - M_{\psi^0} = 
\alpha_2^{\text{PDG}} M_W \sin^2\frac{\theta_W}{2}$.
Numerically, this is an order of magnitude smaller than the splitting in model B and
D and suppresses the decay rate significantly further, allowing for
striking long-lived charged particle signatures. Note that the presence of fermion mixing,
discussed in the following, can induce more frequent decays unless
couplings are very suppressed.

Note the upper limit on general mass splittings $\delta M$ within the fermion $SU(2)_L$-multiplets by the $\rho$-parameter \cite{Patrignani:2016xqp}
\begin{equation}
N_F  S(R_2) \,  \delta M^2 \lesssim ( 40\, \mbox{GeV})^2 \, ,
\end{equation}
where $S_2(R_2)$ is the Dynkin index of the representation $R_2$ of $SU(2)_L$ (see \cite{Bond:2017wut} for details). 
Specifically, $S_2=0,1/2,2$ for models A and E, models C and D, and models B and F, respectively.
The allowed splitting is hence  about a few percent  for TeV-ish fermion masses.

\subsubsection{Scalars}

If kinematically allowed,  the scalars $S_{ij}$ decay in all models  through Yukawa couplings to $\psi \bar \psi$, and in model A and C to $\psi$ plus lepton.
Only the flavor diagonal components can, except in the SM-singlet model E,  in addition decay to electroweak gauge bosons through the $y$-Yukawa and a triangle loop with $\psi$'s, $S\rightarrow GG'$, with $G,G' = {\gamma, W, Z}$.  
Mixing of the vector-like fermions with the SM leptons induces BSM scalar decays to dileptons, further  discussed in Sec.~\ref{sec:mix}.

For $M_S>M_F + m_{\ell}$ decays to vector-like fermions and leptons through the mixed Yukawas $(\kappa')$, {\it i.e.,} in model A and C, are kinematically open. In all models, the decay to $\overline{\psi} \psi$ is possible for  $M_S>2M_F$ through the Yukawa coupling $y$. 
Only the flavor diagonal components of $S$ can decay in this manner. 
The tree-level decay rates 
for a given flavor-specific  component $S_{ij}$ can be written as
\begin{widetext}
\begin{equation}
\label{eq:Sdecays}
\begin{aligned}
&\Gamma(S_{ij}\rightarrow \overline{\psi}_i \,l_j)+ \Gamma(S_{ij}\rightarrow \overline{l_j}\,\psi_i) =  2\pi \alpha_{\kappa'}  M_S\left(1-\frac{M_F^2}{M_S^2}\right)^2,\\
&\Gamma(S_{ij}\rightarrow \overline{\psi}_i \,\psi_j)+ \Gamma(S_{ij}\rightarrow \overline{\psi_j}\,\psi_i)   =2\pi\alpha_{y} M_S\left(1-\frac{4M_F^2}{M_S^2}\right)^{1/2+\xi},
\end{aligned}
\end{equation}
\end{widetext}
where model-dependent $SU(2)_L$ multiplicities in the final states are not spelled out explicitly. For instance, in model B, $S_{ij}$ decays to  $\overline{\psi}^{-2}_i \psi_j^{-2} + \overline{\psi}^{-1}_i \psi_j^{-1} + \overline{\psi}^{0}_i \psi_j^{0} $ plus CP conjugate ones. The loop-induced decays to gauge bosons read

\begin{equation}
\label{eq:Sdecaysloop}
\Gamma(S_{ii}\rightarrow GG') = \frac{\alpha_e^2 \alpha_y}{16\pi} \frac{M_S^3}{M_F^2}\,\lvert C_{GG'} A_{1/2}(\tau)\rvert^2\,,
\end{equation}
where the coefficients $C_{GG'}$ depend on the representation of $\psi$ and in the limit $M_S\gg M_W$ can be expressed as
\begin{equation}
\begin{aligned}
\label{eq:CGG}
&C_{\gamma\gamma} = S_2(R_2) +d(R_2)Y^2\,, \\
&C_{ZZ} =S_2(R_2)\tan^{-2}\theta_W +d(R_2)Y^2\tan^2\theta_W\,,\\
&C_{WW} = \frac{\sqrt{2}}{\cos^2\theta_W}S_2(R_2)\,,\\ 
&C_{Z\gamma} =\sqrt{2}\left(S_2(R_2)\tan^{-1}\theta_W -d(R_2)Y^2\tan\theta_W\right)\,.
\end{aligned}
\end{equation}
In~\eqref{eq:Sdecays}, \eqref{eq:Sdecaysloop}, $\xi=1$ and $\xi=0$ correspond to the scalar and pseudoscalar parts of $S$, respectively, and $A_{1/2}(\tau) = \frac{2}{\tau^2}\left(\xi \tau + (\tau-\xi)f(\tau)\right)$ with 
\begin{equation}
\begin{aligned}
f(\tau) = \begin{cases}
\arcsin[2](\sqrt{\tau}) \quad &\text{for } \tau\leq 1\,,\\
-\frac{1}{4} \left(\ln \frac{1+\sqrt{1-\tau^{-1}}}{1-\sqrt{1-\tau^{-1}}}-i\pi\right)^2
\quad &\text{for } \tau >  1\,,
\end{cases}
\end{aligned}
\end{equation}
and $\tau = M_S^2/4M_F^2$ \cite{Gunion:1989we}. In the case of one of the $S_{ij}$ mixing with angle $\beta$ with the Higgs, the real part of $S_{ij}$ can decay through mixing with rate $\Gamma_{\text{mix}} = \sin^2\beta\, \Gamma_h^{\text{SM}}$, where $\Gamma_h^{\text{SM}}$ is the decay rate of the Higgs in the SM.

\begin{figure}[b]
	\centering
	\includegraphics[width=.78\columnwidth]{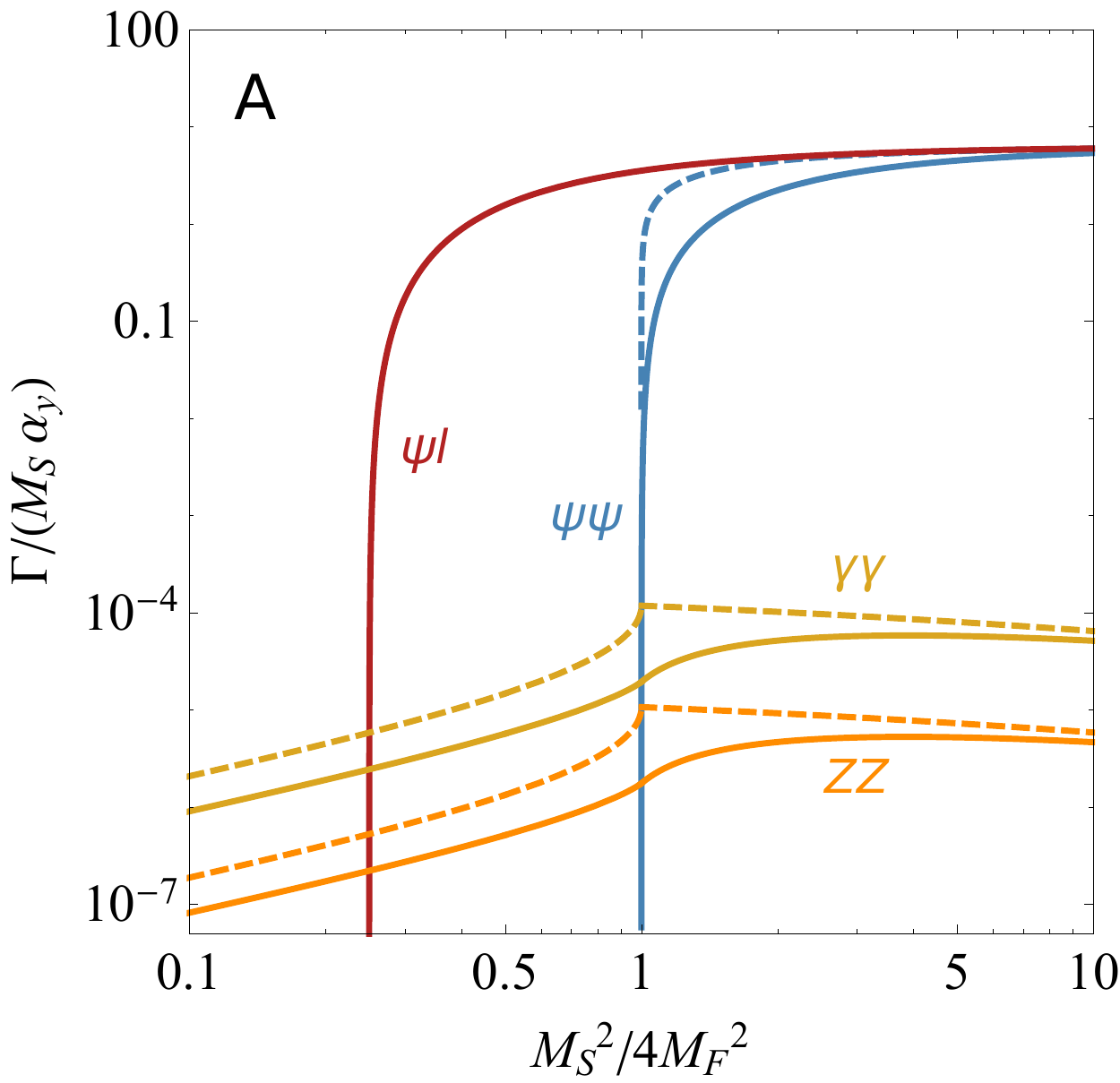}
	\caption{Reduced decay rates $\Gamma/(M_S \alpha_y)$ of the flavor-specific components of the  BSM scalar $S$~\eqref{eq:Sdecays}, \eqref{eq:Sdecaysloop} in model A for $\alpha_y = \alpha_{\kappa'}$. Full (dashed) lines correspond to the scalar (pseudoscalar) decays; for $S\rightarrow \psi \ell$ they coincide. The decay rate into $Z\gamma$ lies between the $ZZ$ and $\gamma\gamma$ curves. }
	\label{fig:decaysA}
\end{figure} 
\begin{figure}[b]
	\centering
	\includegraphics[width=.8\columnwidth]{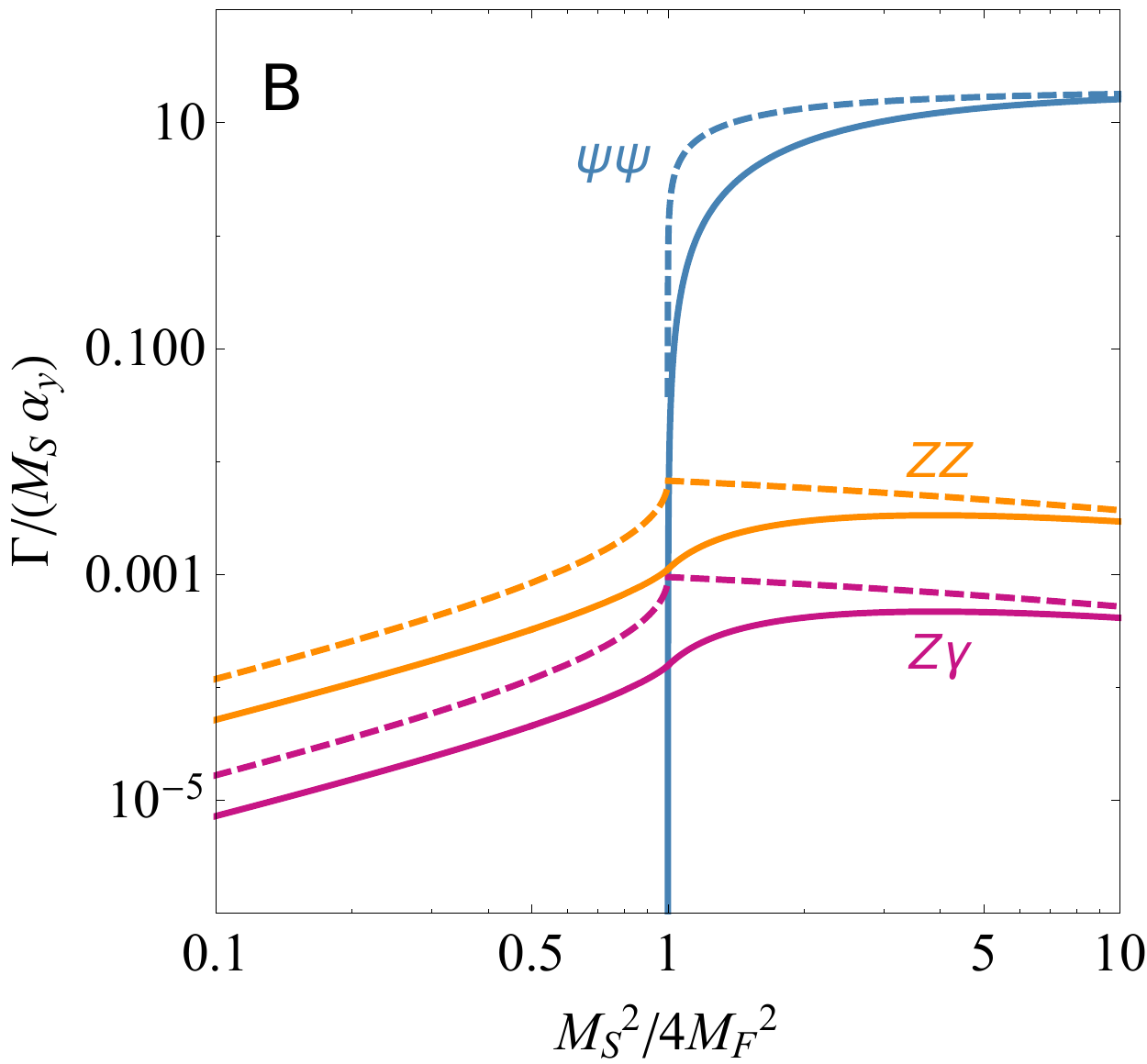}
	\caption{Reduced decay rates $\Gamma/(M_S \alpha_y)$ of the flavor-specific components of the  BSM scalar $S$~\eqref{eq:Sdecays}, \eqref{eq:Sdecaysloop} in model B. Full (dashed) lines correspond to the scalar (pseudoscalar) decays. The decay rates in model B into $WW$ and $\gamma\gamma$ satisfy $\Gamma^{\text{B}}_{ZZ}>\Gamma^{\text{B}}_{\gamma\gamma}>\Gamma^{\text{B}}_{WW}>\Gamma^{\text{B}}_{Z\gamma}$.}
		\label{fig:decaysB}
	\end{figure}

In model A the main $S$ decay channels are $\overline{\psi}\psi$ and $\psi \ell$, followed by  the decay to photons. Other gauge boson modes are further suppressed, as for $T_3=0$ holds  $1 > (C_{Z\gamma} )^2=2 \tan \theta_W^2 >(C_{ZZ} )^2= \tan \theta_W^4$. 
The reduced rates $\Gamma/(M_S \alpha_y)$ as a function of $\tau$ for model A are shown  in Fig.~\ref{fig:decaysA} for $\alpha_y=\alpha_{\kappa'}$.

 In model B,C, D and F the vector-like fermions are charged under $SU(2)_L$, and allow for decays to  $W^+W^-$. When kinematically allowed, the tree-level decays  into 
 $\bar \psi \psi$ are dominant. For model B this is shown in Fig.~\ref{fig:decaysB}. The hierarchy between the gauge boson decay rates in model B
 reads  $\Gamma^{\text{B}}_{ZZ}>\Gamma^{\text{B}}_{\gamma\gamma}>\Gamma^{\text{B}}_{WW}>\Gamma^{\text{B}}_{Z\gamma}$,
 and  in model C 
 $\Gamma^{\text{C}}_{ZZ}>\Gamma^{\text{C}}_{\gamma\gamma}>\Gamma^{\text{C}}_{Z \gamma} \approx  \Gamma^{\text{C}}_{WW}$.
  In model D, the $S_{ii} \to GG'$ hierarchies are $\Gamma^{\text{D}}_{\gamma\gamma}>\Gamma^{\text{D}}_{ZZ}>\Gamma^{\text{D}}_{Z\gamma}>\Gamma^{\text{D}}_{WW}$,
  whereas in model F 
 $\Gamma^{\text{F}}_{ZZ}>\Gamma^{\text{F}}_{Z\gamma}>\Gamma^{\text{F}}_{WW}>\Gamma^{\text{F}}_{ \gamma \gamma}$.

 For $M_S< M_F$  and negligible  $\alpha_y$ 
one may wonder whether 
  $S$ can decay at all. However, fermion mixing induces decays to SM leptons or neutrinos, discussed  next.

\begin{table*}
	\centering
	\setlength{\extrarowheight}{2pt}
	\renewcommand{\arraystretch}{1.2}
	\rowcolors{2}{lightgray}{}
	\begin{tabular}{`ccccccc`}
		\toprule
		\rowcolor{LightBlue}
		{\ \ \bf Vacuum\ \ }	 & $\qquad \theta_L^A \qquad$&  $\qquad\theta_R^A\qquad$ & $\qquad\theta_L^B\qquad$ &  $\qquad\theta_R^B\qquad$ &  $\qquad\theta_L^{0,B}\qquad$ &  $\qquad\theta_L^{0,C}\qquad$ \\ \midrule	%
		$\qquad \bm{V^+} \qquad$& $\frac{\kappa v_h}{\sqrt{2}\,m_2}$ & $\frac{\kappa' v_s}{\sqrt{2}\,m_2}$& $\frac{\kappa v_h}{{2}\,m_2}$ & $\frac{\kappa v_h^2 y_\ell}{{2}\sqrt{2}\,M_2^2}$&  $\frac{\kappa v_h}{\sqrt{2}\,m_2}$ &$\frac{\kappa' v_s}{\sqrt{2}\,m_2}$ \\ 
		$\bm{V^- (\psi_2-\ell_2)}$  & $\frac{\kappa v_h}{\sqrt{2}\,M_F}$ & $\frac{\kappa'v_s}{\sqrt{2}\,m_2}$  & $\frac{\kappa v_h}{{2}\,M_F}$ & $\frac{\kappa v_h^2 y_\ell}{{2}\sqrt{2}\,M_F^2}$ & $\frac{\kappa v_h}{\sqrt{2}\,M_F}$ & $\frac{\kappa' v_s}{\sqrt{2}\,m_2}$ \\
		$\ \bm{V^- (\psi_{1,3}-\ell_{1,3})}\ $  & $\frac{\kappa v_h}{\sqrt{2}\,M_F}$ & $\frac{\kappa v_h^2 y_\ell}{2\,M_F^2}$   & $\frac{\kappa v_h}{{2}\,M_F}$ &  $\frac{\kappa v_h^2 y_\ell}{{2}\sqrt{2}\,M_F^2}$ & $\frac{\kappa v_h}{\sqrt{2}\,M_F}$ &0 \\
		\bottomrule
	\end{tabular}
	\caption{Mixing angles of the $Q_F=-1$ fermions ($\theta^M$) and the $Q_F=0$ fermions ($\theta^{0,M}$) (see Tab.~\ref{tab:reps}), with $m_2=M_F+ y v_s/\sqrt{2}$. In $V^-$, the direction which is aligned with the vacuum (second generation) presents different mixing angles. We have  $\theta^C_{L,R}=\theta^A_{R,L}$ for model C, and $\theta_{L,R}^D=\sqrt{2}\theta_{R,L}^B$ for model D. We also find $\theta^{0,E}_L = \sqrt{2}\theta^{0,F}_L =\theta^{0,B}_L  $  and $\theta^{F}_{L,R} = \sqrt{2}\theta^B_{L,R} $ in models E and F. The additional factor of $1/\sqrt{2}$ in $\theta^B$ and $\theta^{0,F}$ originates from Clebsch-Gordan coefficients \eq{eq:triplet}; see App.~\ref{sec:appscalar} for details.}
	\label{tab:fangles}
\end{table*}

\subsection{Fermion Mixing \label{sec:mix}}

Mixing between SM leptons and BSM fermions provides relevant phenomenology. 
 Mixing angles -- in the small angle approximation to make the parametric dependence explicit --
  for the left-handed $(\theta_L^M)$ and right-handed $(\theta_R^M)$ fermions, with  the model M  indicated as superscript, are given in Table \ref{tab:fangles}.
 Details are given in App.~\ref{sec:appscalar}.
We discuss, in this order, the impact of mixing on scalar decays, 
modified electroweak and Higgs couplings and decays of vector-like leptons to $Z,W+$ SM lepton.
The results are important for experimental searches because they imply that all $S_{ij}$ and $\psi_i$ eventually decay  to SM leptons, charged ones and neutrinos, with the
only exceptions being the diagonal $S_{ii} \to G G'$ decays.

 \subsubsection{LFV-like Scalar Decay}
 
 In models A and C, mixing induces tree-level decays $S_{ij} \to  \ell_i^\pm \ell_j^\mp$  at the order $\kappa^\prime \theta_L^A, \kappa^\prime \theta_R^C \simeq\kappa^\prime \frac{\kappa v_h}{ \sqrt{2} M_F}  $, using the angles of Tab.~\ref{tab:Zint1}. These can be competitive with decays to electroweak bosons: for instance, taking $\kappa'\sim y$ and for $\theta_L^A$ of order $10^{-3}$ or larger they dominate over $S\to \gamma\gamma$ in model A.
Unless the mixing is  strongly suppressed, $\kappa^\prime \theta^A_L, \kappa^\prime \theta_R^C  \lesssim 10^{-7}$ for $M_S$ at the TeV scale, the $S$ lifetime is below picoseconds, and too short for a macroscopic decay length.

In  models B, D and F, for $M_S >M_F$,  fermion mixing induces the decays $S_{ij} \to \psi_j \ell_i$ (models B and F) and $S_{ij} \to \psi_i \ell_j$ (model D) at the order $y \frac{\kappa v_h}{ \sqrt{2} M_F}$. For $M_S<M_F$, the decays $S_{ij} \to  \ell_i^\pm \ell_j^\mp$  at the order   $y\, \theta_L^M \theta_R^M \simeq y (\frac{\kappa v_h}{ \sqrt{2} M_F})^2 \frac{y_\ell v_h}{\sqrt{2} M_F}$ are the leading ones. Using~\eqref{eq:Sdecays} again, one obtains a lifetime of picoseconds or above for a suppression factor $y\, \theta_L^M \theta_R^M \lesssim 10^{-7}$. Due to its flavor dependence, the suppression of the mixing is stronger for tau-less final states. 
This could allow for displaced  decays into dielectrons, dimuons and  $e^ \pm \mu^{\mp}$ , while at the same time, those into ditaus, 
$e^ \pm \tau^{\mp}$ and $\mu^ \pm \tau^{\mp}$ could remain prompt.

Lastly, for models with $Q_F=0$ fermion decays $S\to \overline{\psi}^0_j \nu_i$ are also allowed for $M_S>M_F$, occurring at order $y\,\theta^{0,M}_L$ for models B, E and F  and at order $\kappa'$ for model C. In the case of model E, this is the only available decay mode of the off-diagonal $S_{ij}$ (apart from $S\to \overline{\psi} \psi$ if allowed), leading to below-picosecond lifetimes for  $y\,  \theta_L^{0,E} \simeq y \kappa v_h/\sqrt{2}M_F\lesssim 10^{-7}$.
Study of the   different $S$ decay modes into various  gauge bosons or fermions  can be used for experimental discrimination of models.
The patterns of final state leptons in LFV-like \footnote{Despite the different lepton flavors in the final state processes such as $S_{ij} \to  \ell_i^\pm \ell_j^\mp$ are, strictly speaking, LFV-like only because flavor is conserved in the decay.  \label{foot:like}} decays, $e, \mu$, or $\tau$ can help to understand hierarchies.

\subsubsection{Impact on $Z,W$ and Higgs Couplings}

Fermion mixing gives rise to tree-level effects in the couplings of leptons and vector-like fermions to the massive electroweak bosons. In the case of the $Z$ couplings to two leptons, the Lagrangean in the fermion mass basis $\mathcal{L}_Z = \frac{g_2}{2 \cos\theta_w}\left[ \overline{\ell}\gamma^{\mu}(g_V^\ell - \gamma^5 g_A^\ell)\ell + g^\nu\overline{\nu}\gamma^{\mu}(1 - \gamma^5)\nu \right]Z_\mu  $ acquires couplings
\begin{equation}\label{gagvLeptxt}
g_{_{A}^{V}}^\ell 
= g_{_{A}^{V}}^{\ell, \, \rm SM} + s^2_{\theta_{L}}(T^{3 }_{\psi^{-1}} +1/2 ) \pm  s^2_{\theta_{R}}T^{3 }_{\psi^{-1}}\,,
\end{equation}
 with respect to their SM values $g_V^\ell = - 1/2 + 2s^2_w $ and $g_A^\ell = - 1/2$, and where $T^{3 }_{\psi^{-1}}$ is the isospin of the $Q_F=-1$ component of the vector-like fermions in each model. The rotation angles are to be taken from Tab.~\eqref{tab:fangles} according to  the chosen vacuum structure and the lepton flavor $\ell$. In the case of model A (C), one finds $T^{3 }_{\psi^{-1}}=0$ ($T^{3 }_{\psi^{-1}}=-1/2$), yielding modifications purely proportional to $s^2_{\theta_L}$ ($s^2_{\theta_R}$). In models B, E and F one finds $\theta_R\ll \theta_L$, while model D presents $\theta_L\ll \theta_R$, so that in all models the $g^\ell$ present modifications proportional to $\kappa v_h/M_F$.  In models with $Q_F = 0$ fermions (B, C, E and F), the $Z$ couplings to two neutrinos become
 \begin{equation}\label{gagvNutxt}
 g^\nu =g^{\nu, \, \rm SM} + \Delta g^\nu 
 = g^{\nu, \, \rm SM} + s^2_{\theta_L^{ 0}}\left[ T^{3 }_{\psi^{0}} -1/2\right]
 \end{equation}
 with $g^{\nu, \, \rm SM} = 1/2$. In model C, for which $T^{3 }_{\psi^{0}} = 1/2$, $g^\nu$ remains unaffected. Therefore, in all models $Z$ data mainly constrains the mixing angles proportional to $\kappa v_h/M_F$.  Measurements of the $Z$ couplings to charged leptons and the electron-flavored neutrinos demand $\Delta g \lesssim 10^{-3}$ or smaller \cite{Tanabashi:2018oca},  which implies
 \begin{equation}\label{eq:Zdecay}
 \alpha_\kappa\lesssim 4\cdot 10^{-4}\, \left(M_F/{\rm TeV}\right)^{2}\,.
 \end{equation}
 Modifications of the $W$ couplings remain also in agreement with $W$ decay measurement if~\eqref{eq:Zdecay} is fulfilled (see appendix~\ref{sec:app-mixing} for details).
Additionally, Higgs couplings are modified by mixing as well. Since charged leptons acquire mass from several Yukawa interactions, the couplings of $\mathcal{L}_{h\ell\ell} = \frac{y_\ell}{\sqrt{2}} \overline{\ell} \ell h $ in the mass basis fulfil

\begin{equation}\label{hll}
y_\ell = y_\ell^{\rm SM} + \sin\theta^{\ell}_L \bigg(\kappa' \frac{v_s}{v_h} \cos \theta_R^\ell - \sqrt{2}\frac{ M_F}{v_h} \sin \theta_R^\ell\bigg)
\end{equation}
for model A, while replacing $L\leftrightarrow R$ gives the expression for model C. In all other models, the $\kappa'$ term is absent. For angles fulfilling $Z$ vertex constraints according to Eq.~\eqref{eq:Zdecay}, Higgs signal strength bounds are avoidable for all leptons \cite{Tanabashi:2018oca, Altmannshofer:2015qra}.

 \subsubsection{Electroweak Decays of Vector-like Leptons}
 
  Finally, mixing induces decays of the vector-like fermions to weak bosons and leptons at tree-level, with rates
  \begin{widetext}
 \begin{equation}
\begin{aligned}
\Gamma(\psi^{Q}_i\to  Z f^Q_i) & = \frac{M_F}{64 \pi}\frac{g_2^2}{\cos^2\theta_w}\left(g_V^2 + g_A^2\right)(1-r_Z)^2 (2+1/r_Z)\,,\\
\Gamma(\psi^{Q}_i\to  W^- f^{Q+1}_i) & = \frac{M_F}{64 \pi}{g_2^2}\,\left[(c_L^W)^2 + (c_R^W)^2\right](1-r_W)^2 (2+1/r_W)\,,
\end{aligned}
\end{equation}
 \end{widetext}
 where $r_i = M_i^2/M_F^2$, $f^{-1}=\ell$, $f^0 = \nu$, and the coefficients $c^W_{L,R}$ and $g_{V,A}$ are collected in Tab.~\ref{tab:Wint} and Tab.~\ref{tab:Zint1} respectively for all models.
Let us discuss the decays of the chargeless $\psi^0$ in model C, which occur exclusively through its mixing unless $\psi^0 \to S \nu$ via $\kappa^\prime$ is allowed. For the
universal vacuum $V^+$ and for  the flavor in which the flavor-specific vacuum $V^-$
points, 
it is important to note that the $\psi^0$ is   lighter than the $\psi^{-1}$ by
$\Delta m_C =  M_{\psi^{-1}} -M_{\psi^{0}}=\alpha_2^{\text{PDG}}  \sin \theta_W^2
m_Z/2    \simeq 0.4$ GeV.
This difference causes isospin-breaking in  the mixing angles given in  Table
\ref{tab:fangles}, which induces a CKM-like misalignment between up $T_3=1/2$ and
down $T_3=-1/2$ sectors $\theta_L^{0,C}- \theta_L^C \simeq \theta_L^C( \Delta
m_C/M_F)$, 
such that the decay
$\psi^0 \to \ell^{\prime -} W^{+*} \to \ell^{\prime -} \ell^+ \nu$ can take place.
Assuming $\theta_R^C\ll \theta_L^C$, we estimate $\Gamma (\psi^{0} \to  \ell^{\prime -} \ell^+ \nu) \sim  G_F^2   |
\theta_L^C |^2 \Delta m^2_C M_F^3/(192 \pi^3)\simeq 4 \cdot 10^{-6} \mbox{GeV}  |
\theta_L^C |^2 ( M_F/[\mbox{TeV}])^3$.
Unless $ \theta_L^C \lesssim 10^{-3}$, the $\psi^0$ decays faster than picoseconds. 

For  the flavors $k$ in the lepton-specific vacuum $V^-$ which do not get a
	corresponding VEV in $S$,
	the left- and right-handed angles have the opposite hierarchy, fulfilling  $\theta_L^C\ll \theta_R^C$. Since $\theta_R^{0,\, C}=0$, the $\psi^0_k$ decay promptly through $\psi^0_k \to W^- \ell^+_k$ with $\rvert c_R^W\rvert = \sin \theta_R^C \simeq \kappa v_h/\sqrt{2} M_F$.

\subsection{Drell-Yan \label{sec:DY}}

\begin{figure}[t]
	\centering
	\includegraphics[width=.9\columnwidth]{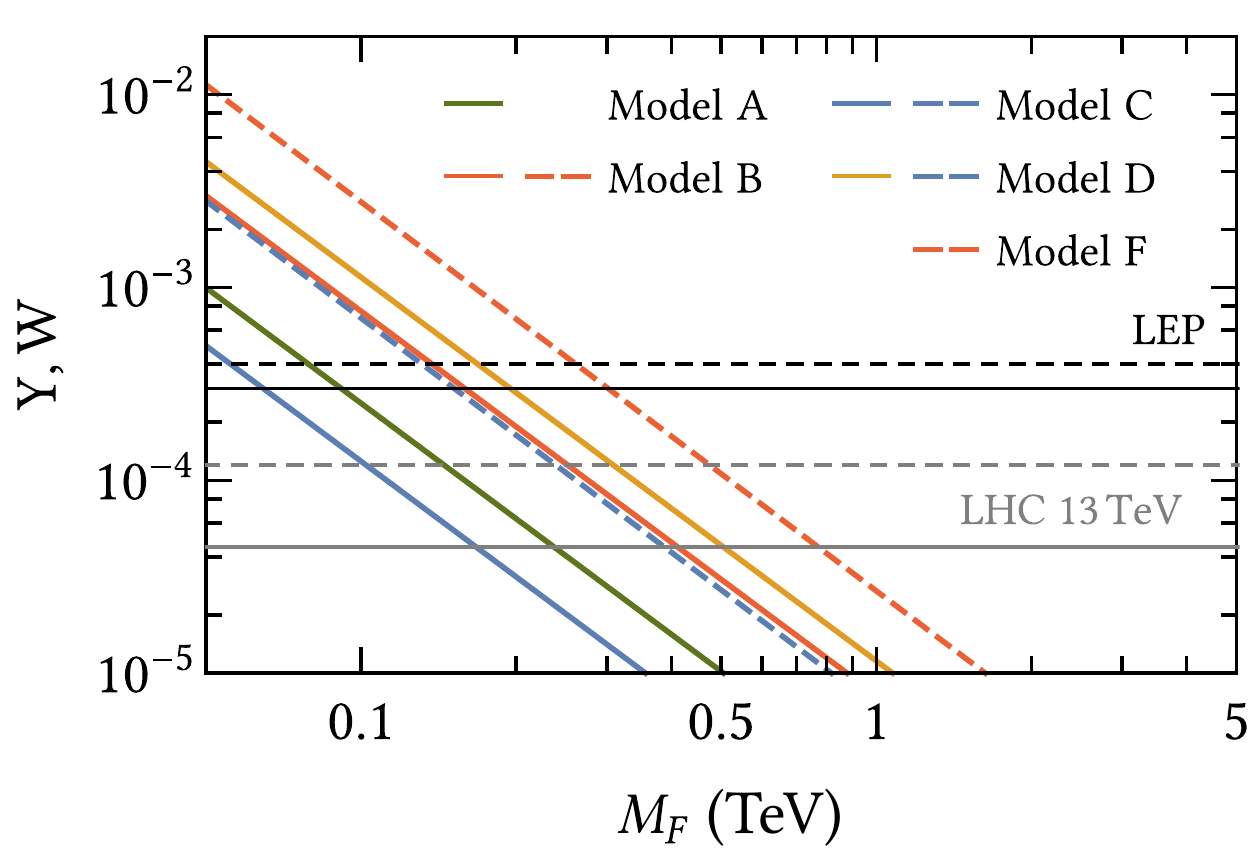}
	\caption{The electroweak parameters $Y$ (full lines) and $W$ (dashed lines) for model A-D, F as functions of the BSM fermion mass, and 
	in comparison with the most stringent constraints 
	from either LHC 8 TeV or LEP (black), and the projected sensitivity of LHC 13 TeV (gray)  taken from \cite{Farina:2016rws}.
	$W$-constraints in model B,F (orange dashed) and C,D (blue dashed) are identical.}
	\label{fig:WY}
\end{figure}

Modifications of the running of the electroweak couplings can be constrained directly from charged and neutral current Drell-Yan processes. Of particular interest are the electroweak precision parameters $W$ and $Y$, which are linearly dependent on the BSM contribution to the running of $\alpha_2$ and $\alpha_1$ respectively as \cite{Alves:2014cda}
\begin{equation}\label{eq:WY}
\begin{aligned}
W,Y = \alpha_{2,1}\,\frac{C_{2,1}}{10}\,\frac{M_W^2}{M_F^2}\,(B_{2,1}^{\text{SM}}-B_{2,1})\,,
\end{aligned}
\end{equation}
where $C_2=1$ and $C_1=3/5$. A lower limit on the mass of the vector-like fermions can  be directly extracted from experimental bounds on $W,Y$ \cite{Farina:2016rws}. As shown in Fig.~\ref{fig:WY}, these require $M_F \gtrsim 0.1$ TeV for model A and $M_F \gtrsim 0.3,0.2$ TeV for models B, C respectively. In models D, F one obtains $M_F \gtrsim 0.2,0.3$ respectively, while in model E one cannot extract bounds due to the BSM sector being uncharged under the SM gauge symmetries. The bound for model B excludes fixed points $B_2$ and $B_4$, which can only be matched at $M_F \simeq 0.02$ TeV. Remarkably, the fixed points that remain viable in terms of matching are only those which present a free $\alpha_2$.
The effect of two-loop corrections in $W,Y$ may be estimated by taking the effective coefficients $B_i^{\text{eff}}$ instead of $B_{2,1}$ in \eqref{eq:WY}. In our matching scenarios, this typically induces relative changes of order $1\%$ or less in $W,Y$ with respect to the one-loop values, and $W,Y$ remain positive. The smallness of these corrections is due to the fact that all couplings at low scales present values of order $10^{-2}-10^{-3}$, which are suppressing the two-loop effects, while  $B_{2,1}$ are typically of order 1 or larger.

\subsection{Anomalous Magnetic Moments \label{sec:AMM}}

The measurements of the electron and muon anomalous magnetic moments 
 are  in tension with SM predictions, offering hints for new physics. 
In the case of the muon, the long-standing discrepancy  amounts to \cite{Tanabashi:2018oca}
\begin{equation}\label{eq:deltagminus2}
\Delta a_{\mu} = a_{\mu}^{\text{exp}} - a_{\mu}^{\text{SM}} = 268(63)(43)\cdot 10^{-11}\,.
\end{equation}
Adding uncertainties in quadrature, this  represents a $3.5 \, \sigma$ deviation from the SM, while recent theory predictions find up to $4.1 \, \sigma$ \cite{Jegerlehner:2017lbd,Davier:2016iru}.\footnote{The  possibility of rendering $ \Delta a_\mu$  insignificant has recently been suggested by a  lattice determination of the hadronic vacuum polarization \cite{Borsanyi:2020mff}. Further scrutiny is required \cite{Aoyama:2020ynm} due to  tensions with electroweak data \cite{Crivellin:2020zul,Keshavarzi:2020bfy} and earlier lattice studies.} 
For the magnetic moment of the electron, recent measurements lead to \begin{equation}\label{eq:deltagminus2e}
\Delta a_{e} = a_{e}^{\text{exp}} - a_{e}^{\text{SM}} = -88(28)(23)\cdot 10^{-14}\,,
\end{equation}
corresponding to a pull of $- 2.4\,\sigma$ from the SM prediction \cite{Hanneke:2008tm,Parker:2018vye}.

From a model building perspective it is  important to understand which new physics ingredients are required to  explain  the anomalies  \eqref{eq:deltagminus2}, \eqref{eq:deltagminus2e} simultaneously. Given that the electron and muon deviations point into opposite directions, it is commonly assumed that an explanation  requires the manifest breaking of lepton flavor universality.
BSM models which explain  both anomalies by giving up on lepton flavor universality have used   either 
new light scalar fields \cite{Davoudiasl:2018fbb,Liu:2018xkx,Gardner:2019mcl,Cornella:2019uxs,Bauer:2019gfk,Dutta:2020scq}, 
supersymmetry \cite{Dutta:2018fge,Endo:2019bcj,Badziak:2019gaf,Yang:2020bmh}, 
bottom-up models \cite{Crivellin:2018qmi,Crivellin:2019mvj}, 
leptoquarks \cite{Bigaran:2020jil,Dorsner:2020aaz}, 
two-Higgs doublet models \cite{Botella:2020xzf,Jana:2020pxx}, 
or other BSM mechanisms which treat electrons and muons manifestly differently 
\cite{Han:2018znu,Abdullah:2019ofw,CarcamoHernandez:2020pxw,Haba:2020gkr,
Calibbi:2020emz,Arbelaez:2020rbq,Chen:2020jvl,Hati:2020fzp,Jana:2020joi}.
In   the spirit of Occam's razor, however, we have shown recently that the data can very well be explained without any manifest breaking of lepton  universality  \cite{Hiller:2019mou}, 
which  is in marked contrast to any of the alternative explanations offered  by \cite{Davoudiasl:2018fbb,Liu:2018xkx,Gardner:2019mcl,Cornella:2019uxs,Bauer:2019gfk,Dutta:2020scq,Dutta:2018fge,Endo:2019bcj,Badziak:2019gaf,Yang:2020bmh,Crivellin:2018qmi,Crivellin:2019mvj,Bigaran:2020jil,Dorsner:2020aaz,Botella:2020xzf,Jana:2020pxx,Han:2018znu,Abdullah:2019ofw,CarcamoHernandez:2020pxw,Haba:2020gkr,Calibbi:2020emz,Arbelaez:2020rbq,Chen:2020jvl,Hati:2020fzp}.

 In this and the following subsection, we detail how the models A, B, C, D, and F induce anomalous magnetic moments at one-loop, and why, ultimately, only models A and C can explain the present data. Note that model E does not appear in the list, the reason being that the charged SM leptons do no longer couple to BSM fermions after electroweak symmetry breaking.  The setting previously put forward by us in \cite{Hiller:2019mou} corresponds to  model A and model C of the present paper.

\begin{figure}
	\centering
	\includegraphics[width=\columnwidth]{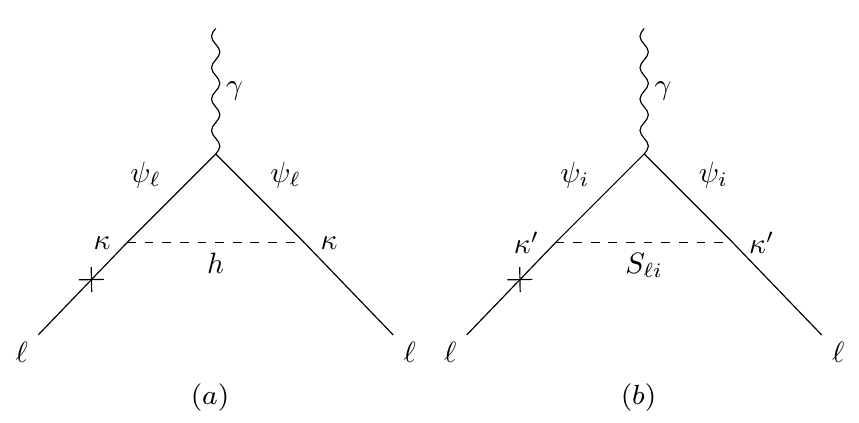}\\
	\caption{Contributions to  $\Delta a_\ell$ $(\ell=e,\mu,\tau)$  with a lepton chiral flip (cross on solid line) via $h$ (a) or $S_{i \ell}$ exchange, with $i=1,2,3$, only present in models A, C (b).}
	\label{fig:gminus2feyn2}
\end{figure}

\begin{figure}[b]
\begin{center}	
	\begin{minipage}{0.47\textwidth}
		\includegraphics[width=.9\columnwidth]{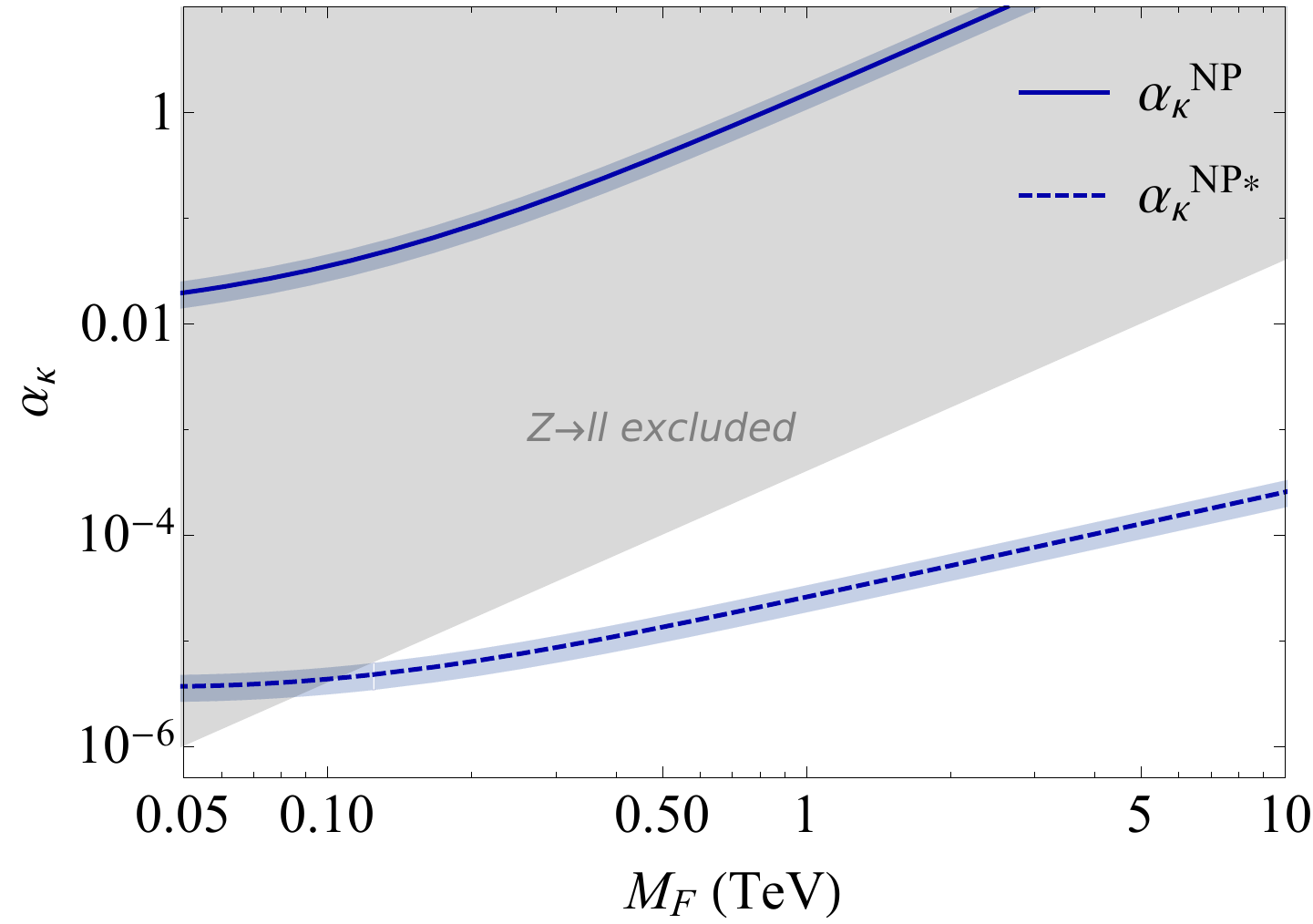}
	\end{minipage}
\end{center}
	\caption{Requisite values of $\alpha_{\kappa}$ to  account for $\Delta a_{\mu}$~\eqref{eq:deltagminus2} for new physics contributions scaling as \eqref{eq:BSM_g-2} (full line) and~\eqref{eq:NP2} (dotted line).  The shaded region is excluded by $Z$-data (\ref{eq:Zdecay}).}
		\label{fig:g-2}
\end{figure}

 Specifically, new physics contributions to  $\Delta a_{\ell}$ arise through the 1-loop diagrams shown in Fig.~\ref{fig:gminus2feyn2}.
 In the limit where $M_F$ is much larger than the mass of the lepton and the scalar propagating in the loop, the NP contribution typically scales as
\begin{equation}\label{eq:BSM_g-2}
a_{\ell}^{\text{NP}} \sim \alpha_{\eta} \, \frac{m_{\ell}^2}{M_{F}^2}\,,
\end{equation}
where $m_{\ell}$ denotes the lepton mass and $\eta=\kappa, \kappa^\prime$ is one of the mixed Yukawa couplings; see appendix \ref{sec:appg-2} for details. For  couplings 
$\kappa', \kappa$ of comparable order, the largest contribution comes from the latter, which couples the vector-like fermions to the lighter scalar (the Higgs). The parameter space $\alpha_{\kappa},\,M_F$ compatible with  \eqref{eq:deltagminus2} is shown in Fig.~\ref{fig:g-2}. 
As obvious from \eqref{eq:g-2full}, \eqref{eq:BSM_g-2} is manifestly positive, and cannot account for $\Delta a_{e}$.
For the muon anomaly \eqref{eq:deltagminus2}, the coupling $\alpha_\kappa\,M_F^{-2} \approx \left(1.4 \pm 0.4\right) \text{TeV}^{-2}$ in model A, C and D as well as $\alpha_\kappa\,M_F^{-2} \approx \left(4.2 \pm 1.2\right) \text{TeV}^{-2}$ for model B and F is required. 
This is however ruled out by the constraint \eqref{eq:Zdecay}. We learn that the models B, D, E and F cannot accommodate either  of the present data \eqref{eq:deltagminus2}, \eqref{eq:deltagminus2e}. 
Models A and C on the other hand have an additional diagram from $S$ exchange, Fig.~\ref{fig:gminus2feyn2}b). In fact, since the $S$ field is a matrix in flavor space the unobserved flavor index of the BSM fermion $\psi_i$ in the loop makes this in total $N_F=3$ contributions. The external chirality flip again induces a contribution quadratic in lepton mass \eqref{eq:BSM_g-2} which can account for $(g-2)_\mu$, since the coupling to the scalar singlet $\kappa'$ is much less constrained than the one to the Higgs \cite{Hiller:2019mou}.

\begin{figure}
	\centering
		\includegraphics[scale=.9]{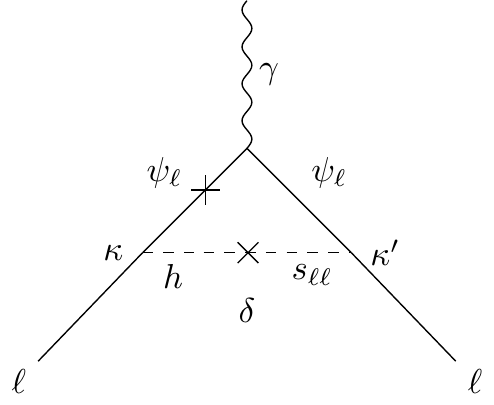}\\
	\caption{Chirally enhanced contribution to the anomalous magnetic moment of a lepton $\ell$ through scalar mixing (cross on dashed line) and
	a $\psi_\ell$ chiral flip (cross on solid line). }
	\label{fig:gminus2feyn}
\end{figure}

Certain NP scenarios, notably supersymmetric ones, can evade one power of  lepton mass suppression  in (\ref{eq:BSM_g-2}) by having instead  the requisite chiral flip on the
 heavy fermion line in the loop, as in  Fig~\ref{fig:gminus2feyn}, such that
\begin{equation}\label{eq:NP2}
a_{\ell}^{\text{NP*}} \sim \alpha_{\eta} \, \frac{m_{\ell}}{M_{F}}\,, 
\end{equation}
opening up the possibility for larger  contributions to $g-2$, and dipole operators in general. For $g-2$ we explore this further for  models A and C in Sec.~\ref{sec:gm2}. 
Another application are electric dipole moments, discussed in Sec.~\ref{sec:edm}.

\subsection{Scalar Mixing and   
Chiral Enhancement} \label{sec:gm2}

The  scalar potential involving the SM and BSM scalars $H$ and $S$ and the various quartic couplings and scalar mass terms has been given in \eq{eq:scalarV}. To investigate the prospect of chiral enhancement for dipole operators, such as those responsible for $(g-2)$, EDMs, or $\mu \to e \gamma$,  we need to investigate the ground states.
Using the methods of \cite{Litim:2015iea,Paterson:1980fc} two ground states $V^\pm$  have been identified in  \eq{eq:vstab}, including the 
conditions for couplings.
The ground state  $V^+$  respects flavor universality in interactions with the SM  because it breaks $SU(3)_{\psi_L} \cross SU(3)_{\psi_R} \rightarrow SU(3)_{diag}$ due to the diagonal VEV $\braket{S_{ij}} = \frac{v_s}{\sqrt{2}}\, \delta_{ij}$. Conversely, $V^-$   spontaneously violates flavor universality because it breaks 
$SU(3)_{\psi_L} \cross SU(3)_{\psi_R}$ to  $SU(2)_{\psi_L} \cross SU(2)_{\psi_R} \times U(1)$ by only allowing a single diagonal component $k$ to pick up a non-vanishing VEV $\braket{S_{ij}} = \frac{v_s}{\sqrt{2}}\, \delta_{i\dot{k}} \delta_{j\dot{k}}$.

If both scalars $S$ and $H$ acquire a VEV, the portal coupling $\delta$ induces a non-diagonal mass term in the potential which allows  the scalars to mix. 
Together with both BSM Yukawas $\kappa$, $\kappa'$ chiral enhancement can occur  in models A and C.  A corresponding  contribution to $g-2$ is shown  
in Fig.~\ref{fig:gminus2feyn}. 
Firstly, we study the case $V^-$, where a single diagonal component of $S$ generates a VEV. 
The $S_{ii}$ component is chosen in order to target the  generation $i$ of leptons in the term $\kappa'\,\Tr[\overline{\psi}_LSE]$.  
We define
\begin{equation}\label{eq:vevsinscalars}
\begin{aligned}
&H = \left( \begin{array}{c} h^+ \\\frac{1}{\sqrt{2}} (h + ih^c +v_h) \end{array} \right)\,,
& \ S_{ii} =\frac{1}{\sqrt{2}}(s_{ii} + i s_{ii}^c + v_s) \,.
\end{aligned}
\end{equation}

The mass matrix of the entire scalar sector is diagonal except for the mixing of $s_{ii}$ and $h$. 
Concentrating on this sub-system, the mass eigenstates $h_1,h_2$ can be expressed in terms of the mixing angle $\beta$ as
\begin{equation}\label{eq:rotation}
\left( \begin{array}{c}h_1\\h_2 \end{array} \right) = \left( \begin{array}{cc} \cos\beta & \sin\beta  \\ -\sin\beta & \cos\beta \end{array} \right) \left( \begin{array}{c}s_{ii}\\h \end{array} \right)\,,
\end{equation}
where
\begin{equation} \label{eq:s2b}
\begin{aligned}
\tan 2\beta = \frac{\delta}{\sqrt{\lambda(u+v)}} \frac{m_h}{m_s}\bigg(1+  \mathcal{O}( m_h^2/m_s^2)\bigg)\, , 
\end{aligned}
\end{equation}
 see App.~\ref{sec:appscalar} for details. 
 Neglecting for the sake of this discussion the mixing induced by the scalar VEVs in the fermion system,
 the BSM Yukawa Lagrangean in the scalar mass basis reads
\begin{equation}\label{eq:scalarportalLag}
\begin{aligned}
\mathcal{L}_{\beta} &= -\overline{\psi}_j \Big[\big(\kappa \sin \beta \, \delta_{jk} P_L + \kappa'  \cos \beta \, \delta_{ij} \,\delta_{ik} P_R\big) h_1 \\
&+\big(\kappa\cos \beta \,\delta_{jk} P_L -\kappa' \sin \beta \,\delta_{ij}\delta_{ik} P_R \big)  h_2  \Big]\ell_k + h.c.\\
\end{aligned}
\end{equation}
where we have again assumed $\kappa,\kappa'$ real and $\kappa_{jk} =\kappa \delta_{jk} $. Provided that the mass eigenstate $h_1$ is much heavier than $h_2$ and $\psi$, and in the limit $M_F\gg m_{h_2},$ the leading contribution to $(g-2)_{\ell}$ reads, 
for $\ell=i$ with $\langle S_{ii} \rangle \neq 0$,
\begin{equation}\label{eq:BSM_g-2_scalar}
a_{\ell}^{V^-} 
=  -\frac{m_{\ell}}{2M_F}\, \frac{\kappa\kappa'}{16\pi^2}\, \sin 2\beta \,,
\end{equation}
see App.~\ref{sec:appg-2} for details. This contribution is switched on only when both left and right chiral couplings of the lepton to the 
same scalar are present, a condition which is met by scalar mixing, and which comes with 
an enhancement factor $\frac{M_F}{m_{\ell}}(\frac{\alpha_{\kappa'}}{\alpha_{\kappa}})^{1/2} |\sin 2\beta |$ with respect to NP contributions such as~\eqref{eq:BSM_g-2}. 
$a_\ell^{V^-}$ can have either sign.

\begin{figure}
	\centering
	\begin{minipage}{0.472\textwidth}
		\includegraphics[width=.9\columnwidth]{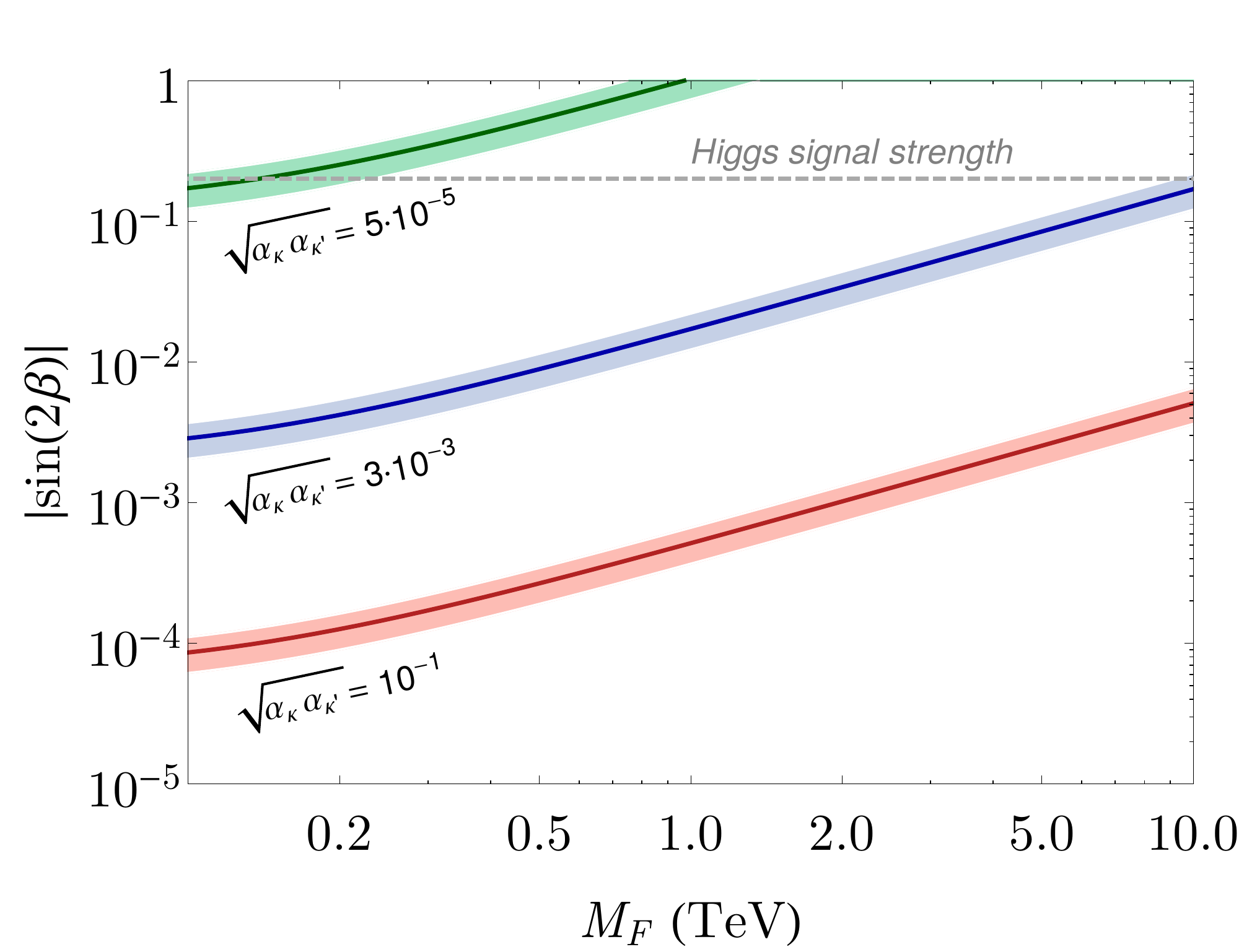}
	\end{minipage}\qquad
	\begin{minipage}{0.472\textwidth}
	\end{minipage}
	\caption{The mixing angle $|\sin 2 \beta|$ as a function of $M_F$ that explains  the $(g-2)_{\mu}$ anomaly within its  $1 \sigma$ uncertainty  (\ref{eq:deltagminus2})  for a muon-aligned vacuum $V^-$ 
	(\ref{eq:BSM_g-2_scalar}) and for different values of $\sqrt{\alpha_{\kappa}\alpha_{\kappa'}}$, with upper bound from Higgs signal strength measurements \cite{Parker:2018vye}. For $V^+$ the corresponding, requisite  value of  $|\sin 2 \beta'|$ is a factor of $\sqrt{3}$ larger (\ref{eq:g-2V-}).}
		\label{fig:beta}
\end{figure}

If  the vacuum is aligned in the muon direction, $(g-2)_\mu$ benefits from  chiral enhancement~\eqref{eq:BSM_g-2_scalar}. 
Fig.~\ref{fig:beta} shows for which values of $M_F, |\sin 2\beta |$ the contribution to $(g-2)_{\mu}$ equals $\Delta a_{\mu}$  \eqref{eq:deltagminus2}  for some benchmark values of $\sqrt{\alpha_\kappa \alpha_{\kappa'}}$. Also shown is an upper limit on the mixing angle $\sin\,2\beta < 0.2$ from Higgs signal strength measurements \cite{Patrignani:2016xqp}.

Next we consider the case $V^+$, where the BSM VEV is universal in all flavors. Here mixing occurs between the $h$ and the three $s_{ii}$ states acquiring the VEV. However, two of the mass eigenstates contain no component in the $h$ direction, and thus only mix the $s_{ii}$ states within themselves.  The two normalized mass eigenstates which have non-vanishing components in the $h$ direction are
\begin{equation}
\begin{aligned}
h_1' &= \frac{1}{\sqrt{3}}\left(\cos\beta's_{11},\ \cos\beta's_{22},\ \cos\beta's_{33},\,\sqrt{3}\sin\beta'h \right)\,,\\ 
h_2' &= \frac{-1}{\sqrt{3}}\left(\sin\beta's_{11},\ \sin\beta's_{22},\ \sin\beta's_{33},\ -\sqrt{3} \cos\beta'h \right)\,.\\
\end{aligned}
\end{equation} 
Hence, the mixing pattern with $h$ is identical for all $s_{ii}$. The enhanced contribution to the anomalous magnetic moments affects all lepton generations, and reads  
\begin{equation}\label{eq:g-2V-}
a_{\ell}^{V^+}
=  -\frac{m_{\ell}}{2\sqrt{3}M_F}\, \frac{\kappa\kappa'}{16\pi^2}\, \sin 2\beta' \,.
\end{equation}
As $a_\ell^{V^-}$, it can a priori have either sign,  and can accommodate future $(g-2)_\mu$ data  by adjusting  $\kappa \kappa^\prime \delta/M_F$ together with the quartics.
The parameter space $\beta',M_F$ that fits $\Delta a_{\mu}$ is, up to a factor $\sqrt{3}$, the same as for $V^-$, shown in Fig.~\ref{fig:beta}, and we note that  this factor  cancels
with  the one for  the angle $\beta'$, which obeys  
\begin{equation} \label{eq:betprime}
\begin{aligned}
\tan 2\beta' = \frac{ \sqrt{3} \delta}{\sqrt{\lambda u}} \frac{m_h'}{m_s'}\bigg(1+  \mathcal{O}( m_h'^2/m_s'^2)\bigg)\,.
\end{aligned}
\end{equation}
for $v=0$ and $\mu_{\rm det}=0$,  see App.~\ref{sec:appscalar} for details.  

Owing to \eq{eq:g-2V-}, we emphasize that fixing the parameter $\sin 2\beta' $ in $V^+$ to explain $\Delta a_{\mu}$ in \eq{eq:deltagminus2} also induces  a contribution to the anomalous magnetic moments of the electron  and the tau,
\begin{eqnarray}
a_{e}^{ V^+}&=& (m_e/m_\mu)  a_{\mu}^{ V^+} \simeq  1.4\cdot 10^{-11}\\
a_{\tau}^{ V^+}&=& (m_\tau/m_\mu)  a_{\mu}^{ V^+} \simeq  4.5\cdot 10^{-8}\,.
\end{eqnarray} 
The former, however, is in conflict with  the data for $\Delta a_e$  in \eq{eq:deltagminus2e}, both in magnitude and in sign, 
while the latter is four orders of magnitude away  from present limits on $\Delta a_\tau \equiv a^{\rm exp}_\tau-a^{\rm SM}_\tau$ \cite{Tanabashi:2018oca}.

On the other hand, larger couplings $\kappa'/M_F [\rm TeV] \sim \mathcal{O}(10) $ allow for a simultaneous explanation  of both data points \eq{eq:deltagminus2}  and \eq{eq:deltagminus2e}.
This mechanism uses the diagrams in Fig.~\ref{fig:gminus2feyn2}b)  to generate $\Delta a_\mu$,  and the chirally enhanced diagram of Fig.~\ref{fig:gminus2feyn} to generate $\Delta a_e$, without introducing  flavor structure  explicitly, and irrespective of the vacuum being  flavor blind $(V^+)$ or  electron-aligned $(V^-$ with $\ell=e$). 
Moreover, the underlying mechanism is not fine-tuned and could, in principle,  accommodate a wide range of deviations $\Delta a_\mu$ and $\Delta a_e$ different from present data. 

Since the underlying Lagrangean does not break lepton flavor, this mechanism leads additionally to a prediction for  the deviations of the tau anomalous magnetic moment $\Delta a_\tau$. Using  the  data \eq{eq:deltagminus2}  and \eq{eq:deltagminus2e}, our models predict
\beq
\Delta a_\tau ^{ V^+}\simeq (7.5\pm 2.1)\cdot 10^{-7}\,,
\eeq
if the vacuum is flavor-blind, or
\beq
\Delta a_\tau^{ V^-}\simeq (8.1\pm  2.2)\cdot 10^{-7}\,,
\eeq
if the ground state is electron-aligned, respectively. Further details of this scenario  can be found in  \cite{Hiller:2019mou}. 

Within our set of models, we conclude  that the muon anomaly \eq{eq:deltagminus2}  alone, or the electron anomaly \eq{eq:deltagminus2e} alone, or both anomalies together, can only be explained by models A and C.

\subsection{EDMs  \label{sec:edm}}

Unlike in the remainder of this work, here we allow the  BSM Yukawas to be complex-valued. If the portal interaction $\delta$ is present, in models A and C a
relative phase between $\kappa$ and $\kappa'$ induces an Electric Dipole Moment (EDM) of the SM leptons through the chirally enhanced  1-loop diagram Fig.~\ref{fig:gminus2feyn}.
The EDM-Lagrangean can be written as 
\beq
\mathcal{L}_{\rm EDM} = d_\ell (-i/2)\overline{\ell} \sigma_{\mu\nu}\gamma_5F^{\mu\nu}\ell\,,
\eeq
 where $F^{\mu \nu}$ denotes the electromagnetic field strength tensor and $d_\ell$ the lepton electric dipole moment with mass dimension $-1$.
 
 For model A, and in the large-$M_F$ limit,  we find
\begin{equation} \label{eq:edm}
	\begin{aligned}
		\frac{d_\ell^{V^-}}{e} = -\frac{\sin 2\beta}{4 M_F}\,\frac{{\rm Im}[ \kappa^* \kappa']}{16\pi^2}\,
	\end{aligned}
\end{equation}
where the flavor-specific vacuum $V^-$ is assumed with $\ell$ denoting the flavor  distinguished by the ground state  $(\langle S_{\ell \ell} \rangle \neq 0)$. 
Here, an EDM arises solely for the lepton flavor selected spontaneously by the vacuum. 
In turn, assuming 
the vacuum $V^+$ and provided that the CP-phases are lepton-universal, we find instead
\begin{equation} \label{eq:edm+}
	\begin{aligned}
		\frac{d_\ell^{V^+}}{e} = -\frac{\sin 2\beta'}{4 \sqrt{3} M_F}\,\frac{{\rm Im}[ \kappa^* \kappa']}{16\pi^2}\,,
	\end{aligned}
\end{equation}
for any flavor, and  all EDMs are predicted to be equal. The same expressions \eq{eq:edm}, \eq{eq:edm+} and results hold true for model C except for the replacement $ \kappa^* \kappa'\rightarrow \kappa \kappa'^*$.

The current experimental  bounds on $d_e$ and $d_\mu$
\begin{equation}
\begin{aligned}
\lvert d_e\rvert&<1.1\cdot10^{-29}\,e \,{\rm cm }\,,\\
\lvert d_\mu\rvert&<1.5\cdot10^{-19}\,e \,{\rm cm }\, ,
\end{aligned}
\end{equation}
 by the ACME and Muon g-2 collaborations  @90 \% CL \cite{Andreev:2018ayy,Bennett:2008dy} respectively, 
imply the bound
\begin{equation}
	\begin{aligned}
		\big\vert \sin 2\beta\,{{\rm Im}[ \kappa^* \kappa']}  \big\vert /{16\pi^2} < 2.2 \cdot 10^{-12} (M_F/\mbox{TeV})  \,,
	\end{aligned}
\end{equation}
from the electron data, while the bound from muons is ten orders of magnitude weaker.
Comparing this  to $\Delta a_\mu$ \eqref{eq:deltagminus2}  induced by the same mechanism, see Fig.~\ref{fig:beta}, the CP-phases must be suppressed at the order $10^{-7}$ ($d_e$-bound) and are unsuppressed by the muon EDM data. If the lepton EDMs are induced by a  lepton flavor nonuniversal mechanism,
flavor-dependent CP-phases or in a vacuum $V^-$ pointing in the muon direction, the electron EDM bound could be bypassed and
the muon EDM could be as large as $d_\mu \sim 2.5  \cdot 10^{-22} \, e \, \mbox{cm}$  given \eqref{eq:deltagminus2} with order one phases.
Interestingly, this is in reach of future experiments $\lvert d_\mu\rvert  \sim 5\cdot10^{-23}\,e \,{\rm cm }\,$ \cite{Crivellin:2018qmi}.

\subsection{Charged LFV Processes \label{sec:LFV}}

In the setup  with Yukawa interactions (\ref{eq:LY}),  (\ref{eq:AC})  and (\ref{eq:kappa-diag})  flavor is conserved. While there is  intergenerational mixing  in Yukawas with $S$, 
no charged LFV proper occurs, see footnote \ref{foot:like}.
Here we envision a situation beyond  (\ref{eq:kappa-diag}) and allow for additional flavor off-diagonal couplings.
Our aim is to see whether and how well such variants can be probed in LFV processes.

\begin{figure}
\begin{center}	
	\begin{minipage}{0.47\textwidth}
		\includegraphics[width=.9\columnwidth]{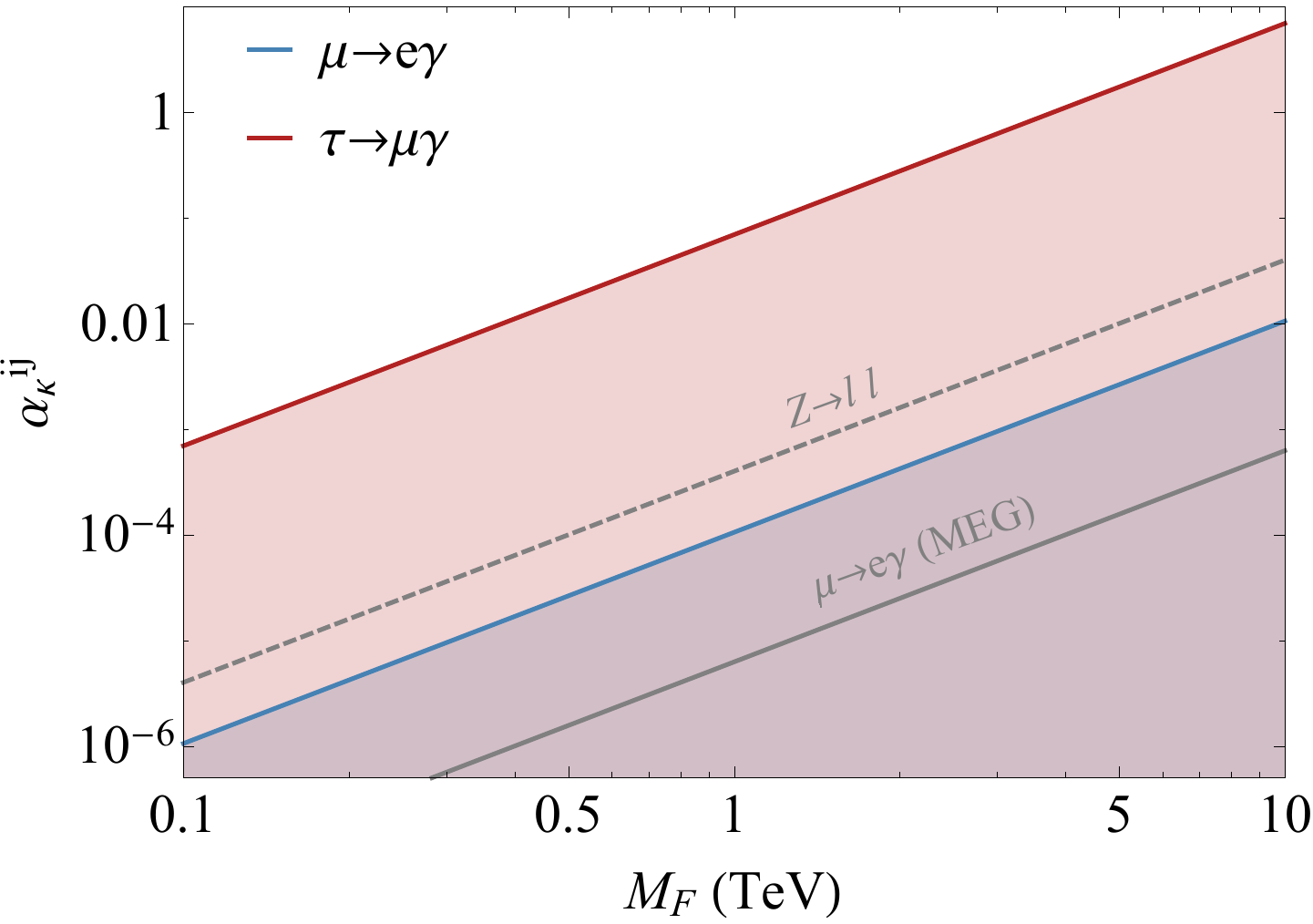} 
	\end{minipage}
	\end{center}
	\caption{
	Allowed regions (shaded) for  $\alpha_{\kappa}^{\mu e}, \alpha_{\kappa}^{\tau\mu}$ and $M_F$ from LFV decays (\ref{eq:LFV-br}). Due to the proximity of upper limits on ${\cal{B}}(\tau \to e \gamma)$ and  ${\cal{B}}(\tau \to \mu \gamma)$
	only the latter is shown.
	The projected sensitivity of the MEG-II experiment \cite{Cavoto:2017kub} is shown by the solid gray line.
	The dashed  gray line denotes the $Z\to \ell \ell$-constraint (\ref{eq:Zdecay}).}
	\label{fig:LFV}
\end{figure}

The $\ell_i\rightarrow \ell_j\gamma$ decay rate induced by a Higgs-fermion loop  in all models except the singlet model E  for  $m_{\ell},m_h\ll M_{F}$ and $m_j\ll m_i$ can be written as \cite{Lavoura:2003xp} 
\begin{equation}\label{eq:LFVrateF}
\Gamma(\ell_i\rightarrow \ell_j\gamma) = 
\frac{\alpha_e}{576}(\alpha_{\kappa}^{ij})^2\ \frac{m_i^5}{M_F^4}\,,
\end{equation}
with 
\begin{equation}\label{eq:alphaij}
\begin{aligned}
\alpha_{\kappa}^{ij} = \frac{1}{(4\pi)^2} \sum_{m}  \kappa_{mi}\kappa_{mj}
\,,
\end{aligned}
\end{equation}
where $m$ corresponds to the  flavors of the  BSM fermion in the loop,
see appendix \ref{sec:appLFV} for details. In~\eqref{eq:alphaij}, a flavor pattern  proportional to $\kappa  \delta_{ij}$ plus small off-diagonal entries of the order $\kappa \epsilon$
is assumed that is responsible for charged LFV. Hence, $\alpha_{\kappa}^{ij}\propto \alpha_\kappa\epsilon$, and  $\Gamma(\ell_i\rightarrow \ell_j\gamma)$ arises at order  $\epsilon^2$.  Fig.~\ref{fig:LFV} shows how present bounds \cite{Tanabashi:2018oca} 
\begin{equation}  \label{eq:LFV-br}
\begin{array}{ll}
{\cal B} \left(\mu\rightarrow e\gamma\right) &< 4.2\cdot 10^{-13}\,,\\
{\cal B} \left(\tau\rightarrow e\gamma\right) &<3.3\cdot 10^{-8}\,,\\
{\cal B} \left(\tau\rightarrow \mu\gamma\right) &<4.4\cdot 10^{-8}\,,
\end{array}
\end{equation}
at 90\% CL and projected bounds ${\cal B}\left(\mu\rightarrow e\gamma\right) \lesssim 2\cdot 10^{-15}$ from the MEG-II experiment \cite{Cavoto:2017kub} constrain  $\alpha_{\kappa}^{\mu e}, \alpha_{\kappa}^{\tau\mu}$ depending on $M_F$. 
Also shown for comparison is the constraint on diagonal couplings from $Z$-data (\ref{eq:Zdecay}).
While present constraints on the off-diagonal entry $\alpha_{\kappa}^{\mu e}$ (blue) are comparable to the diagonal ones from $Z \to \ell \ell$, 
the ones on $ \alpha_{\kappa}^{\tau\mu}$ and $ \alpha_{\kappa}^{\tau e}$  (red) are significantly weaker.

LFV decays into three lepton final states are also possible, receiving contributions from both penguin and box diagrams with $\kappa,\kappa'$. We estimate \cite{Chowdhury:2018nhd} 
\begin{equation}
{\cal B} (\mu \to ee\bar{e}) \sim \frac{3(4\pi)^2 \alpha_e^2}{8 G_F^2}\frac{(\alpha_{\kappa}^{\mu e})^2}{M_F^4}{\cal B} (\mu \to e\bar{\nu}_e\nu_\mu) \,,
\end{equation}
which  is  $\epsilon^2$-suppressed as (\ref{eq:LFVrateF}). Present bounds from the SINDRUM collaboration  ${\cal B} (\mu \to ee\bar{e}) < 10^{-12}$ \cite{Bellgardt:1987du} give
$\alpha_{\kappa}^{\mu e}/(M_F[\text{TeV}])^2< (2-3)\cdot 10^{-4}$. This is indeed comparable with $\mu\to e\gamma$ bounds in Fig.~\ref{fig:LFV}, yet not more excluding. The parameter space will be further probed by the Mu3e experiment, which aims at a reach of  ${\cal B} (\mu \to ee\bar{e}) < 10^{-16}$ \cite{Perrevoort:2018cqi}. For $\tau$ decays to three charged leptons, present bounds pose loose constraints on off-diagonal couplings,  $\alpha_{\kappa}^{\tau \ell}/(M_F[\text{TeV}])^2\lesssim 0.1$.

On the other hand, $\mu$ to $e$ conversion processes have a limit in gold nuclei on the conversion rate (CR)  of 
CR$(\mu-e, {\rm Au})\lesssim7\cdot10^{-13}$ at 90\% CL by the SINDRUM II collaboration \cite{Bertl:2006up}. In our models the conversion process is possible through $Z$ and $\gamma$ penguin contributions which receive $\epsilon^2$ suppression. We estimate CR$(\mu-e, {\rm Au})\sim \mathcal{O}(10^{-12}) (\alpha_{\kappa}^{\mu e}/10^{-4})^2/(M_F[\text{TeV}])^4$ \cite{Kitano:2002mt}, in close competition with $\mu \to e\gamma$ bounds. The future Mu2e experiment \cite{DonghiaonbehalfoftheMu2eCalorimeterGroup:2017aei}, with expected sensitivity CR$(\mu-e, {\rm Au})<6.7\cdot10^{-17}$, can improve the bound from SINDRUM II on $\alpha_{\kappa}^{\mu e}$ by about two orders of magnitude.

Along the lines of the anomalous magnetic moments, scalar mixing induces chirally enhanced contributions to LFV dipole operators  if $\kappa$ contains non-vanishing off-diagonal elements. For instance, the rate for $\mu\rightarrow e\gamma$ becomes  
\begin{equation}
\Gamma(\mu \rightarrow  e\gamma) =
 \frac{\alpha_e}{64} \left(\frac{\kappa_{e\mu}\kappa' \sin 2\beta}{16\pi^2}\right)^2 \frac{m_{\mu}^3}{M_F^2}  \,,
\end{equation}
in the same approximations as in  \eqref{eq:BSM_g-2_scalar} and $V^-$ pointing in the muon direction.
(There is a similar contribution induced by $\kappa_{\mu e}$ which requires a scalar VEV in the electron direction.)
 Constraints on  $\alpha^{\mu e}_{\kappa \kappa^\prime \delta}= \kappa_{e\mu}\kappa' \sin 2\beta/(16\pi^2)$  from the  chirally enhanced amplitude are stronger than on 
$\alpha^{\mu e}_\kappa$ by a factor $m_\mu/3 M_F$.

\section{\bf \label{sec:conclusion}Conclusions}

We have studied SM extensions with three generations of vector-like leptons and a new singlet matrix scalar field, inspired by asymptotic safety.  
The main focus has been on new physics implications for settings where the running couplings  
remain finite and well-defined at least up to the Planck scale, and possibly beyond.
A key novelty over earlier models  
are  Higgs and  flavor portals
(Tab.~\ref{tab:reps}) which are explored in depth.
Within this setup we show that the number of new fermion generations 
required for
asymptotically safe or Planck safe extensions
can be much lower than thought previously. 

Using the renormalization group, we have provided a comprehensive study of six basic models. All of them are found to be well-behaved up to  the Planck scale in 
certain parameter regimes, owing to Yukawa couplings linking SM and BSM fermions with the 
Higgs  (Figs.~\ref {fig:noBSM} -- \ref{fig:cross-kappa-delta}). The  TeV scale initial conditions 
for  BSM couplings (Fig.~\ref{fig:surface_kappa_y}) illustrate parameter regions which do not run into Landau poles and vacuum 
 instabilities or metastabilities. 
 Similar results are   found for  models which admit 
a secondary Yukawa coupling between SM and BSM fermions and the new scalars (Figs.~\ref{fig:AC_cross},~\ref{fig:AC_cross2})
 with  a corresponding critical surface of parameters (Fig.~\ref{fig:surface_kappa_kappa'}).
Very explicitly we  learn that the requirement for safety up to the Planck scale provides a testable selection criterion in the BSM parameter space.

A conceptual novelty is the use of both "top-down" 
and "bottom-up" searches to find fixed points and Planck safe parameter regions. On the technical side, we have retained
 the RGEs for the  gauge, Yukawa, and quartic couplings up to the complete two loop order, extending upon previous studies. 
 New  features are walking regimes, and new patterns for fixed points due to  a competition 
between  Yukawa, portal, and gauge couplings. 
Theories where the running of couplings can be extended to infinite energy are of  interest in their own right.
Our asymptotically safe 
extensions are the first ones which achieve this for the key SM and BSM couplings, and in accord 
with the measured values of the gauge couplings and the Higgs, top, and bottom masses.

Our models  also offer a rich phenomenology due to 
their close ties with the SM through Yukawa and Higgs portals.
 Genuine features are LFV-like signatures in scalar decays
$$S_{ij} \to  \ell_i^\pm\, \ell_j^\mp\,,$$
also with displaced vertices for sufficiently small coupling.
The  vector-like leptons  can have exotic charges which can lead to displaced vertex signatures.
The models can be experimentally probed at  colliders  (Fig.~\ref{fig:production}),  specifically through $\bar\psi \psi $ and  Drell-Yan production, and additionally, at $e^+ e^-$ or $\mu^+  \mu^-$ machines,  through single $\psi$ production \cite{Benedikt:2018qee,Zimmermann:2018wfu}.
The BSM scalars can be pair-produced at lepton colliders, or in $pp$, if portal effects are present.
  It would be interesting to check whether existing new physics searches at the LHC in lepton-rich final states lead to constraints on model parameters.  As no dedicated analysis for the models here has been performed, however, this requires a  re-interpretation of existing searches  which is beyond the scope of this work.

Finally, we comment on outstanding features related to lepton universality and  low energy probes for new 
physics, i.e.~measurements of the  lepton's magnetic or electric dipole moments. 
Except for the breaking by SM Yukawas,  lepton universality is manifest in all our models and  may or may 
not be broken spontaneously by the vacuum. Irrespective of the ground state, however, we find that two of the six basic models
can explain  the electron anomaly alone, the muon anomaly alone,
or both anomalies together.  
The latter is rather  remarkable in that it also entails a
prediction for the tau anomalous magnetic moment   \cite{Hiller:2019mou}, whereas any other BSM explanation of the muon and electron anomalies 
requires a manifest breaking of lepton universality \cite{Davoudiasl:2018fbb,Liu:2018xkx,Gardner:2019mcl,Cornella:2019uxs,Bauer:2019gfk,Dutta:2020scq,Dutta:2018fge,Endo:2019bcj,Badziak:2019gaf,Yang:2020bmh,Crivellin:2018qmi,Crivellin:2019mvj,Bigaran:2020jil,Dorsner:2020aaz,Botella:2020xzf,Jana:2020pxx,Han:2018znu,Abdullah:2019ofw,CarcamoHernandez:2020pxw,Haba:2020gkr,Calibbi:2020emz,Arbelaez:2020rbq,Chen:2020jvl,Hati:2020fzp}.
In addition, provided  the vacuum is flavorful and points into the muon direction,
we find that the electron EDM bound can be bypassed with a sizeable muon EDM at the level of $\sim  10^{-22} \, e \,  \mbox{cm}$. 

We look forward to further exploration of asymptotically safe model building and  searches.

\section*{\bf Acknowledgements}
GH and DL  thank  the Aspen Center for Physics, which is supported by National Science Foundation grant PHY-1607611,  for hospitality while part of this work was performed. Some results have been presented at the Moriond Electroweak conference  2019 and at the workshop "Asymptotic Safety meets Particle Physics" (Dortmund 2019).
This work has been supported  by the DFG Research Unit FOR 1873 ``Quark Flavor Physics and Effective Field Theories''.

\appendix
\renewcommand{\thesubsection}{\Alph{subsection}}
\renewcommand{\theequation}{\thesubsection.\arabic{equation}}

\counterwithin*{equation}{subsection}

\newpage
\section*{\bf Appendices}

The following appendices collect technical details and useful formul\ae\ used within the main manuscript.

\subsection{\label{sec:appBetaf}Two Loop $\beta$-functions}
In this appendix, we detail $\beta$-functions for the models A -- F up to two-loop order. The results are based on \cite{Machacek:1983tz, Machacek:1983fi, Machacek:1984zw, Luo:2002ti, Schienbein:2018fsw, Pickering:2001aq, Mihaila:2012pz}. 

\begin{table*}[t]
	\centering
	\renewcommand{\arraystretch}{1.2}
	\setlength{\tabcolsep}{4pt}
	\rowcolors{2}{lightgray}{}
	\begin{tabular}{`ccccccccccccccccc`}
		\toprule
		\rowcolor{LightBlue}
		\ \ \bf Model\ \ & $D_{1\kappa}$  & $D_{1 \kappa'}$  & $D_{2 \kappa}$ & $D_{2 \kappa'}$ & $E_{t\kappa}$ & $E_{b\kappa}$ & $E_{y\kappa}$ & $E_{y\kappa'}$ & $E_{\kappa \kappa}$ & $F_{\kappa 1}$ &$F_{\kappa 2}$ 
		& $E_{\kappa' \kappa'}$ & $E_{\kappa' y}$ & $E_{\kappa'\kappa}$ & $F_{\kappa' 1}$ &$F_{\kappa' 2}$\\ \midrule	
		\bf A &15 & 36 & 3 & 0 & 6 & 6 & 2 & 8 & 9 &  15/2 & 9/2 &8 & 8 & 0 & 12 & 0\\
		\bf B &45/4 & 0 & 33/4 & 0 & 6 & 6  & 1/2 & 0 & 23/4  & 15/2 & 33/2&&&&&\\
		\bf C &15 & 18 & 3 & 18 &  9/2 & 9/2 & 1 & 10 & 9 &  15/2 & 9/2&10 & 10 & 0 & 3 & 9\\
		\bf D &39 & 0 & 3 & 0 &  6 & 6 & 1 & 0 & 9 &  39/2 & 9/2&&&&& \\
		\bf E &3 &  0 & 3 & 0 & 6 & 6 & 2 & 0 & 9 &  3/2 & 9/2 &&&&&\\
		\bf F &9/4&	0 & 33/4 & 0 &  6 & 6 &  1/2 & 0 & 23/4 & 3/2 & 33/2&&&&& \\ \toprule
	\end{tabular}
	\caption{Model-specific  loop coefficients for the gauge and Yukawa beta functions \eq{A1}.}
		\label{tab:gauge-yuk_1}
\end{table*}

The two-loop gauge and one-loop Yukawa RGEs at can be expressed as
\begin{equation}\label{A1}
\begin{aligned}
  &{\beta_i} =
  - \alpha_i^2 \left(B_i - \!\!\! \sum_{j=\rm gauge} \!\!\! C_{ij}\,\alpha_j  + \!\!\! \sum_{n=\rm Yukawa} \!\!\! D_{in}\,\alpha_n\right),\\
    &\beta_n  = 
  \alpha_n \left(\sum_{m=\rm Yukawa} \!\!\! E_{nm}\,\alpha_m  - \!\!\! \sum_{i=\rm gauge} \!\!\! F_{ni}\, \alpha_i\right)\,,
\end{aligned}
\end{equation}
corresponding to  \eq{gauge} and \eq{Yukawas} in the main text. Some of the loop coefficients are model-specific and  listed in Tab.~\ref{tab:gauge-yuk_1},  while others are universal or can be expressed in a general way in terms of the representation $R_2$ of the vector-like fermions under $SU(2)_L$ and their hypercharge $Y$. In what follows, $C_2(R_2)$ denotes the quadratic Casimir invariant and $S_2(R_2)$ the Dynkin index (see \cite{Bond:2017wut} for details).

For the hypercharge coupling, 
these generic coefficients read 
\begin{equation}\label{eq:a1coeffs}
\begin{aligned}
B_1 &= -\frac{41}{3} - 8\,d(R_2)Y^2\,,\\
C_{11} &=\frac{ 199}{9} + 24\,d(R_2)\,Y^4\,,\\
C_{12} &= 9 + 24 \, C_2(R_2)  d(R_2) Y^2 \,,\\
C_{13} &=\frac{88}{3} \,,\quad 
D_{1t} = \frac{17}{3} \,,\quad
D_{1b} = \frac{5}{3} \,,\\
D_{1y} &= 36\,d(R_2)\,Y^2 \,.
\end{aligned}
\end{equation}
For the weak coupling, one obtains

\begin{equation}\label{eq:a2coeffs}
\begin{aligned}
B_2 &= \frac{19}{3} -8\,S_2(R_2)\,,\\
C_{21} &= 3 + 24\, S_2(R_2) \, Y^2\,,\\
C_{22} &= \frac{35}{3} + 12\,S_2(R_2)\,\left(2C_2(R_2) + 20/3\right)\,,\\
C_{23} &= 24 \,,\ \quad\\
D_{2t} &= D_{2b} = 3\,,\\
D_{2y} &= 36 \, S_2(R_2)  \,.
\end{aligned}
\end{equation}
Finally, for the strong coupling the coefficients are independent of the BSM sector
\begin{equation}\label{eq:a3coeffs}
\begin{aligned}
B_3 &= 14\,,
&D_{3t} &= 4 \,,
&D_{3b} &= 4 \,,\\
C_{31} &= \frac{11}{3}\,,
&C_{32} &= 9\,,
&C_{33} &= -52\,.
\end{aligned}
\end{equation}
For the Yukawa coefficients, we note that only $\alpha_\kappa$ couples into the one-loop running of the top and bottom Yukawas $\beta_{t,b}$, see  Tab.~\ref{tab:gauge-yuk_1}. Further loop coefficients for Yukawa couplings at one loop are given by 
\begin{equation}
  \begin{aligned}
    E_{tt} &= E_{bb} = 9\,, &E_{tb} &= E_{bt} = 3\,,\\
    F_{t1} &= \frac{17}{6}\,, &F_{b1} &= \frac{5}{6}\,,\\
    F_{t2} &= F_{b2} = \frac{9}{2}\,, &F_{t3} &= F_{b3} = 16\,.
  \end{aligned}
\end{equation}
In a similar vein, there are no one-loop contributions from $\alpha_{3,t,b}$ to $\beta_y$ and $\beta_{\kappa'}$. 
For $\beta_y$ one finds
\begin{equation}\label{eq:aycoeffs}
\begin{aligned}
E_{yy} &= 2 \left(3 + d(R_2)\right)\,,\\ 
F_{y1} &= 12\, Y^2 \,,\\
F_{y2} &= 12\, C_2(R_2)\,.
\end{aligned}
\end{equation}
Loop coefficients for $\beta_\kappa$  which are universal in all models are given by
\begin{equation}
\begin{aligned}    
E_{\kappa t} &= E_{\kappa b} = 6\,,\quad \\
    E_{\kappa y} &= 3\,,\quad \\
    E_{\kappa \kappa'} &= F_{\kappa 3} = 0\,,
\end{aligned}
\end{equation}
while those which are  model specific  are summarized in 
Tab.~\ref{tab:gauge-yuk_1}.

 For the scalar couplings at one loop, it is convenient to use the definition
 \begin{equation}
   \widetilde{\alpha}_y = \left\{
   \begin{array}{rl}
    \alpha_{\kappa'} +  \alpha_y & \qquad\text{model A, C}\\
    \alpha_y & \qquad\text{model B, D, E, F}
   \end{array}
   \right.\,.
 \end{equation} 
 \begin{widetext}
In this convention, the one loop  RGEs for the scalar quartic couplings read
\begin{equation}\label{eq:qrt-1L}
  \begin{aligned}
    \beta_\lambda^{(1)} &= \beta_\lambda^{\mathrm{SM}(1)} +  9 \alpha_\delta^2  + I_{\kappa}\, \alpha_\kappa \alpha_\lambda  - J^\lambda_{\kappa \kappa}\, \alpha_\kappa^2\,,\\
    \beta_\delta^{(1)} &= \alpha_\delta \Big[ 4 \alpha_\delta + 12 \alpha_\lambda + 24 \alpha_u + 40 \alpha_v + 6 \alpha_t 
    + 6 \alpha_b  +  \frac12 I_{ \kappa} \, \alpha_\kappa  +  2\, d(R_2) \, \widetilde{\alpha}_y - \frac32 \alpha_1 - \frac92 \alpha_2\Big] 
    - \frac13 I_{\kappa} \,\alpha_\kappa \alpha_y\,,\\
    \beta_u^{(1)} &= 24 \alpha_u \left(\alpha_u + \alpha_v\right) + 2\, d(R_2) \widetilde{\alpha}_y \left(2 \alpha_u - \widetilde{\alpha}_y\right),
    \\
    \beta_v^{(1)} &=  52 \alpha_v^2 + 12 \alpha_u \left(\alpha_u + 4 \alpha_v\right)+ 2 \alpha_\delta^2 
    +4\,d(R_2)\, \widetilde{\alpha}_y \alpha_v \,.
  \end{aligned}
\end{equation}
Here, $\beta_\lambda^{\mathrm{SM}(1)}$ denotes the one-loop $\beta$-function of the Higgs quartic in the SM.
The one loop coefficients $I_{\kappa}$ and $J^\lambda_{\kappa \kappa}$ are tabulated in Tab.~\ref{tab:gauge-yuk_2}.

\begin{table*}[b]
	\renewcommand{\arraystretch}{1.2}
	\setlength{\tabcolsep}{4pt}
	\rowcolors{2}{lightgray}{}
	\begin{tabular}{`cccccccccccc`}
		\toprule
		\rowcolor{LightBlue}
		\ \ \bf Model\ \ & $I_{\kappa}$  &  $J^\lambda_{\kappa \kappa}$ & $K^\lambda_{11\kappa}$ & $K^\lambda_{12\kappa}$ & $K^\lambda_{22\kappa}$ & $H^\lambda_{\kappa\kappa\kappa}$ & $H^\delta_{\kappa\kappa y}$ &$L^\lambda_{1\kappa}$ & $L^\lambda_{2\kappa}$ & $L^\delta_{1y}$ & $L^\delta_{2y}$ 
		\\ \midrule	%
		\bf A & 12 &  6 & 75/4 & $-$33/2 & 9/4 & 10 & 14 & 12 & 0 & 12 & 0 \\
		\bf B & 9  &  15/8 & 225/16 & $-$51/8 & $-$21/16& 47/32 & 39/8 & 15/4 & 15/2 & 9 & 18 \\
		\bf C & 12 &  6 & 75/4 & $-$33/2  & 9/4 & 10 & 16 & 12& 0 & 6 & 6 \\
		\bf D & 12 &  6 & 219/4 & 39/2  & 9/4 & 10 & 16 & 36& 0 & 30 & 6 \\
		\bf E & 12 &  6 & 3/4 & 3/2& 9/4 & 10 & 14 & 0 & 0 & 0 & 0 \\
		\bf F & 9  &  15/8 & 9/16 & 57/8 & -21/16 & 47/32 & 39/8 & 0 & 15/2 & 0 & 18 \\ \toprule
	\end{tabular}
	\caption{Model-specific  loop coefficients for the quartic and Yukawa beta functions \eqref{eq:qrt-1L}, \eqref{eq:qrt-2L} and \eqref{eq:BSM-qrt-2L}.}
	\label{tab:gauge-yuk_2}
\end{table*}

At two-loop order, running of the couplings $\alpha_{t,b,\lambda}$ is modified via 
\begin{equation}\label{eq:qrt-2L}
  \begin{aligned}
    \frac{\beta_{t}^{(2)}}{\alpha_{t}} &= \frac{\beta_{t}^{\mathrm{SM}(2)}}{\alpha_{t}} + 9 \alpha_\delta^2 -\frac{9}{4} J^\lambda_{\kappa \kappa}\, \alpha_\kappa^2 - \frac{27}{24} I_{\kappa} \, \alpha_\kappa \left(\alpha_y  + \alpha_{t} - \frac{15}{27} \alpha_{b}\right) + \frac54\left(D_{1\kappa}\,\alpha_1 + 3 D_{2\kappa}\,\alpha_2\right) \alpha_\kappa  \\&\phantom{= \ }  + 6 \,S_2(R_2) \,\alpha_2^2
    + \frac{58}{9} Y^2  \, d(R_2) \,\alpha_1^2\,,\\
\frac{\beta_{b}^{(2)}}{\alpha_{b}} &= \frac{\beta_{b}^{\mathrm{SM}(2)}}{\alpha_{b}} + 9 \alpha_\delta^2 -\frac{9}{4} J^\lambda_{\kappa \kappa} \, \alpha_\kappa^2 - \frac{27}{24} I_{\kappa} \,\alpha_\kappa \left(\alpha_y  + \alpha_{b} - \frac{15}{27} \alpha_{t}\right) + \frac54\left(D_{1\kappa}\,\alpha_1 + 3 D_{2\kappa}\,\alpha_2\right) \alpha_\kappa   \\
&\phantom{=\ } + 6 \,S_2(R_2) \,\alpha_2^2 
 - \frac{2}{9} Y^2  \, d(R_2) \,\alpha_1^2\,,\\
 \beta_{\lambda}^{(2)} &= \beta_{\lambda}^{\mathrm{SM}(2)} - 90 \alpha_\delta^2 \alpha_\lambda -36\alpha_\delta^3
 - 18 \,d(R_2) \,\widetilde{\alpha}_{y} \alpha_\delta^2 
   - 12\, I_\kappa \,\alpha_\kappa \alpha_\lambda^2 - \frac12 J^\lambda_{\kappa \kappa}\, \alpha_\kappa^2 \alpha_\lambda - \frac{27}{12} I_\kappa \,\alpha_\kappa \alpha_y \alpha_\lambda\\
  &\phantom{= \ }  + 3\, J^\lambda_{\kappa \kappa} \alpha_y \alpha_\kappa^2 + 3\, H^\lambda_{\kappa\kappa\kappa} \alpha_\kappa^3 - L^\lambda_{1 \kappa} \alpha_1 \alpha_\kappa^2 - L^\lambda_{2 \kappa} \alpha_2 \alpha_\kappa^2\\
 &\phantom{= \ }  + \frac52\left(D_{1\kappa}\,\alpha_1 +3 D_{2\kappa}\,\alpha_2\right) \alpha_\kappa \alpha_\lambda -K^\lambda_{11\kappa} \alpha_1^2 \alpha_\kappa -K^\lambda_{12\kappa} \alpha_1 \alpha_2 \alpha_\kappa -K^\lambda_{22\kappa} \alpha_2^2 \alpha_\kappa \\
 &\phantom{= \ } + 30 \,S_2(R_2) \,\alpha_2^2 \alpha_\lambda + 10\, d(R_2) Y^2\,\alpha_1^2 \alpha_\lambda 
 - 4 d(R_2) Y^2 \left(\alpha_1 + \alpha_2\right)\alpha_1^2  -4\, S_2(R_2) \left(\alpha_1 + 3 \alpha_2\right) \alpha_2^2
\,, 
  \end{aligned}
\end{equation}
using coefficients in Tab.~\ref{tab:gauge-yuk_1},\ref{tab:gauge-yuk_2}, and $\beta_{t,b,\lambda}^{\mathrm{SM}(2)}$ denote the two loop beta functions of the SM.

Two loop RGEs of $\alpha_{y,\kappa,\kappa'}$ read
\begin{equation}\label{eq:BSM-yuk-2L}
  \begin{aligned}
      \frac{\beta_{\kappa}^{(2)}}{\alpha_{\kappa}} = &-P^\kappa_{\kappa\kappa} \alpha_\kappa^2 -\frac{9}{4} \left(1+ 2 d(R_2)\right)  \widetilde{\alpha}_y \alpha_y - \frac{3}{32} \left[90 + d(R_2)\,\left(89 - 27\, d(R_2)\right) \right] \alpha_y \alpha_\kappa + R^\kappa_{11} \,\alpha_1^2 \\
      & - \frac{1}{6}\left[49 + d(R_2) \, \left(39 - 19 \,d(R_2)\right)\right] \alpha_2^2 + Q^\kappa_{1y}\, \alpha_1 \alpha_y + Q^\kappa_{2\kappa} \, \alpha_2 \alpha_\kappa 
       + Q^\kappa_{12}\, \alpha_1\alpha_2 + Q^\kappa_{1\kappa}\, \alpha_1\alpha_\kappa
      \\
      & + \frac{5}{12}\,\alpha_1 \left[17\,\alpha_t + 5\,\alpha_b\right] + \frac{45}{4}\,\alpha_2 \left[\alpha_t + \alpha_b\right] + 40\,\alpha_3 \left[\alpha_t + \alpha_b\right] - 6 S_2(R_2)\left[1 - 2 d(R_2)\right]\,\alpha_2 \alpha_y  \\
      &- \frac{27}{2} \left[\alpha_t^2 + \alpha_b^2\right] + 3\,\alpha_t\,\alpha_b  - J^\lambda_{\kappa \kappa}\,I_\kappa^{-1} \,\left(27\,\alpha_t + 27\,\alpha_b + 48\, \alpha_\lambda\right) \alpha_\kappa -12\, \alpha_y \alpha_\delta + 9 \alpha_\delta^2 + 12 \alpha_\lambda^2
      \,.
      \,.\\
      \frac{\beta_{\kappa'}^{(2)}}{\alpha_{\kappa'}} = &\left[ \frac{211}{3} + 2\,Y^2\left(20\, d(R_2) - 3\right)\right]Y^2 \,\alpha_1^2-  \left[\frac{257}{3} + 6\,C_2(R_2) - 40\,S_2(R_2) \right] C_2(R_2)\,\alpha_2^2\\
    &- 12\, C_2(R_2) Y^2\,\alpha_1 \, \alpha_2  + \left[48 + 10\, d(R_2)\right]\, Y^2\, \alpha_1 \, \widetilde{\alpha}_y  + \left[48 + 10\, d(R_2)\right]\,C_2(R_2)\, \alpha_2\widetilde{\alpha}_y  \\
    &+ 8\left[5\, \alpha_u^2 + 5\, \alpha_v^2 + 6\, \alpha_u \alpha_v\right] + 2 \alpha_\delta^2 - 16\, \left(5\,\alpha_u + 3\, \alpha_v\right) \widetilde{\alpha}_y  - \left[\frac12 + 18 \, d(R_2)\right] \widetilde{\alpha}_y^2\\
    &- P^y_{\kappa\kappa} \, \alpha_\kappa^2 + Q^y_{1\kappa}\,\alpha_1 \alpha_\kappa + Q^y_{2\kappa}\,\alpha_2 \alpha_\kappa  - \frac{3}{2^{d(R_2)}} \left(2\, d(R_2) + 1\right) \alpha_y \alpha_\kappa\,,\\
    \frac{\beta_y^{(2)}}{\alpha_y} = &\left[ \frac{211}{3} + 2\,Y^2\left(20\, d(R_2) - 3\right)\right]Y^2 \,\alpha_1^2-  \left[\frac{257}{3} + 6\,C_2(R_2) - 40\,S_2(R_2) \right] C_2(R_2)\,\alpha_2^2\\
    &- 12\, C_2(R_2) Y^2\,\alpha_1 \, \alpha_2  + \left[48 + 10\, d(R_2)\right]\, Y^2\, \alpha_1 \, \widetilde{\alpha}_y  + \left[48 + 10\, d(R_2)\right]\,C_2(R_2)\, \alpha_2\widetilde{\alpha}_y  \\
    &+ 8\left[5\, \alpha_u^2 + 5\, \alpha_v^2 + 6\, \alpha_u \alpha_v\right] + 2 \alpha_\delta^2 - 16\, \left(5\,\alpha_u + 3\, \alpha_v\right) \widetilde{\alpha}_y  - \left[\frac12 + 18 \, d(R_2)\right] \widetilde{\alpha}_y^2\\
    &- P^y_{\kappa\kappa} \, \alpha_\kappa^2 + Q^y_{1\kappa}\,\alpha_1 \alpha_\kappa + Q^y_{2\kappa}\,\alpha_2 \alpha_\kappa  - 2^{-d(R_2)}\left[18 \,\alpha_t + 18\,\alpha_b  + 3\left(2\, d(R_2) + 1\right) \alpha_y +16\, \alpha_\delta\right] \alpha_\kappa\,,
  \end{aligned}
\end{equation}
also using the loop coefficients tabulated in Tab.~\ref{tab:gauge-yuk_3}.

Finally, the two-loop contributions for the BSM scalar quartics are
\begin{equation}\label{eq:BSM-qrt-2L}
  \begin{aligned}
    \beta_u^{(2)} = &- 336\, \alpha_u^3 - 1056\, \alpha_u^2 \alpha_v - 688\, \alpha_u \alpha_v^2 + \left[Y^2 d(R_2)  \alpha_1 + 3\, S_2(R_2) \alpha_2 \right] \left[20 \alpha_u - 8 \widetilde{\alpha}_y\right] \widetilde{\alpha}_y  \\ 
    &- 48 d(R_2) \widetilde{\alpha}_y \left( \alpha_u + \alpha_v\right) \alpha_u +2 d(R_2) \, \left[6 \widetilde{\alpha}_y -9 \alpha_u + 4 \alpha_v\right] \widetilde{\alpha}_y - 20\, \alpha_\delta^2 \alpha_u + I_\kappa\, \alpha_\kappa \alpha_y\left[\frac13 \widetilde{\alpha}_y - \frac12 \alpha_u\right]\,,\\
    \beta_v^{(2)} = &- 288\, \alpha_u^3 - 688 \, \alpha_u^2 \alpha_v  - 1056\, \alpha_u \alpha_v^2 - 816\,\alpha_v^3  + 20 \left[ Y^2 d(R_2) \,\alpha_1 + 3 S_2(R_2)\,\alpha_2\right] \widetilde{\alpha}_y \alpha_v\\
    &- 24 d(R_2) \left[\alpha_u^2 + 4 \alpha_v \alpha_u+ \frac{13}{3} \alpha_v^2 \right]\, \widetilde{\alpha}_y + 2 d(R_2) \left[4\,\alpha_u  - 9\,\alpha_v + 2\,\widetilde{\alpha}_y\right] \widetilde{\alpha}_y^2 - \frac12 I_\kappa\, \alpha_\kappa \alpha_y \alpha_v \\
    &+ 4\left[\alpha_1 +  3\,\alpha_2 - 3\,\alpha_t - 3\,\alpha_b - 3\, \alpha_\kappa - 5\, \alpha_v - 2\,\alpha_\delta\right] \alpha_\delta^2\,,\\
    \beta_\delta^{(2)} = &\left[\frac{557}{48} + 5\, Y^2 d(R_2)\right] \alpha_1^2 \alpha_\delta + \frac{15}{8} \alpha_1 \alpha_2 \alpha_\delta + \left[- \frac{145}{16} + 15\,S_2(R_2) \right] \alpha_2^2 \alpha_\delta + \left[\frac{85}{12} \alpha_t + \frac{25}{12} \alpha_b \right] \alpha_1  \alpha_\delta \\
    &+ \frac{45}{4} \left[\alpha_t + \alpha_b\right]\alpha_2  \alpha_\delta + 40\,\left[\alpha_t + \alpha_b\right] \alpha_3  \alpha_\delta + 10 \left[Y^2 d(R_2)\, \alpha_1 + 3\,S_2(R_2)\,\alpha_2\right] \widetilde{\alpha}_y \alpha_\delta\\
    &- \frac{27}{2} \left[\alpha_t^2 + \alpha_b^2\right] \alpha_\delta  - 21\, \alpha_t \alpha_b \alpha_\delta -  d(R_2) \left[48\,\alpha_u + 80\,\alpha_v + 9 \widetilde{\alpha}_y \right]\widetilde{\alpha}_y \alpha_\delta - \frac{17}{24} I_\kappa\, \alpha_\kappa \alpha_y \alpha_\delta  - \frac{27}{12} J^\lambda_{\kappa\kappa} \alpha_\kappa^2 \alpha_\delta\\
    &- 200\,\left[\alpha_u^2 + \frac{6}{5} \alpha_u \alpha_v + \alpha_v^2\right]\alpha_\delta  +12 \left[2\,\alpha_1 + 6\, \alpha_2 - 6\, \alpha_t - 6\, \alpha_b - 6\, \alpha_\kappa  - 5 \alpha_\lambda\right] \alpha_\lambda \alpha_\delta\\
    &+ \left[\alpha_1 + 3 \alpha_2 -12 \alpha_t - 12 \alpha_b - 144 \alpha_u - 240 \alpha_v -I_\kappa\, \alpha_\kappa - 4\,d(R_2) \,\widetilde{\alpha}_y - 19 \alpha_\delta - 72\, \alpha_\lambda\right] \alpha_\delta^2\\
    &+\frac{15}{4} \left[\frac{1}{3}\,D_{1\kappa} \alpha_1 + D_{2 \kappa} \alpha_2\right]  \alpha_\kappa \alpha_\delta +\frac{5}{2} I_\kappa\, \alpha_\kappa \alpha_y \widetilde{\alpha}_y + H^\delta_{\kappa \kappa y}\,\alpha_\kappa^2 \alpha_y \\
    &-12\, Y^2 d(R_2)\, \alpha_1^2 \widetilde{\alpha}_y -36\,\left[\frac{5}{3} d(R_2) - 4\right] S_2(R_2)\, \alpha_2^2 \widetilde{\alpha}_y - L^\delta_{1y}\,\alpha_1 \alpha_y \alpha_\kappa - L^\delta_{2y}\,\alpha_2 \alpha_y \alpha_\kappa\,,
  \end{aligned}
\end{equation}
with model-specific loop coefficients tabulated in  Tab.~\ref{tab:gauge-yuk_2}.
\begin{table*}
	\renewcommand{\arraystretch}{1.2}
	\setlength{\tabcolsep}{4pt}
	\rowcolors{2}{lightgray}{}
	\begin{tabular}{`cccccccc`}
		\toprule
		\rowcolor{LightBlue}
		\ \ \bf Model\ \  & $P^y_{\kappa \kappa}$ & $P^\kappa_{\kappa \kappa}$ & $Q^y_{1\kappa}$ & $Q^y_{2\kappa}$ & $Q^\kappa_{1y}$ & $Q^\kappa_{2\kappa}$ & $R^\kappa_{11}$
		\\ \midrule	%
		\bf A & 19/2 & 24 &37/4 & 51/4 & 6 & 225/8 & 721/12 \\
		\bf B & 57/32 & 59/8  & 37/16 & $-$101/16 & 6 & 1343/32 & 1249/12\\
		\bf C & 5 & 24 & 55/8 & 33/8 & 15 & 225/8 & 589/12\\
		\bf D & 5 & 24  & 95/8 & 33/8 & 27 & 225/8 & 4541/12\\
		\bf E & 19/2 & 24  & 17/4 & 51/4 & 0 & 225/8 & 35/12\\
		\bf F & 57/32 & 59/8  & 17/4 & -101/16 & 0 & 1343/32 & 35/12\\ \toprule
	\end{tabular}
	\caption{Model-specific  two loop coefficients for the BSM Yukawa beta functions  \eqref{eq:BSM-yuk-2L}.}
	\label{tab:gauge-yuk_3}
\end{table*}

\subsection{\label{sec:appg-2} BSM Contributions to   $g-2$}

Results for weak corrections to $g-2$ in general gauge models can be found in \cite{Leveille:1977rc}. In this work the relevant BSM contribution comes from a neutral scalar-$\psi$ loop. 
Using the general Yukawa Lagrangean with chiral projectors $P_{L/R}=(1 \mp \gamma_5)/2$
\begin{equation}\label{eq:lag_g-2}
\mathcal{L}_Y = \overline{\psi}\, (c_{Li} P_L + c_{Ri} P_R)\,\ell_i H +h.c.\,,
\end{equation}
where $\psi$ is a fermion with charge $Q_F=-1$, $H$ is a neutral scalar and $\ell_i$ is a charged lepton of flavuor $i$
\begin{equation}\label{eq:g-2full}
\begin{aligned}
a_{i}^{\text{NP}} &= \frac{m^2_{i}}{8\pi^2} \int_0^1 \text{d}x 
\frac{ \frac{1}{2}(c_{Li}^2+c_{Ri}^2)\big( x^2-x^3\big) + \frac{M_F}{m_{i}} \,c_{Li}c_{Ri} \, x^2
}{m_{i}^2x^2+(M_F^2-m_{i}^2)x+m_H^2(1-x)} \\ &= 
\frac{m_{i}^2}{16\pi^2 m_H^2}\Bigg[\frac{1}{2}(c_{Li}^2+c_{Ri}^2)\,I_1(M_F^2/m_H^2) 
+\frac{M_F}{m_{i}}c_{Li}c_{Ri}\,I_2(M_F^2/m_H^2)\Bigg]  \,,
\end{aligned}
\end{equation}
where we assumed real couplings $c_{Li}, c_{Ri} $. For $m_{i}\rightarrow 0$ in the integrals with  $t = M_F^2/m_H^2$ one obtains
\begin{equation}
\begin{aligned}
I_1(t) & = \frac{t^3-6 t^2+3 t+6 t \ln(t)+2}{3 (t-1)^4}\,,\\
I_2(t) & = \frac{t^2-4 t+2 \ln(t)+3}{(t-1)^3}\,.
\end{aligned}
\end{equation}
The limits $t\rightarrow\infty$ (heavier fermion) and $t\rightarrow 0$ (heavier scalar) yield
\begin{equation}\label{eq:g-2app}
\begin{aligned}
a_{i}^{\text{NP}, F} &=  \frac{1}{16\pi^2} \frac{m^2_{i}}{M_F^2}\, \Bigg[ 
 \frac{1}{6}(c_{Li}^2+c_{Ri}^2) + \frac{M_F}{m_{i}} c_{Li}c_{Ri}
\Bigg]\,, \\
a_{i}^{\text{NP}, H} &=  \frac{1}{16\pi^2} \frac{m^2_{i}}{m_H^2}\, \Bigg[ 
 \frac{1}{3}(c_{Li}^2+c_{Ri}^2) 
- \frac{M_F}{m_{i}}\, c_{Li}c_{Ri}\,\left(2\ln \frac{M_F^2}{m_H^2}+3\right)
\Bigg]
\end{aligned}
\end{equation}
respectively. For $t=1$, the integrals are well defined, $I_1(1) = 1/6$, $I_2(1) =2/3$. 
The presence of both Yukawas $c_{Li}, c_{Ri} \neq0$ switches on the rightmost terms in  \eqref{eq:g-2app} with enhancement factors $M_F/m_{i}$.

\subsection{\label{sec:appLFV} LFV Branching Ratios}

Here we provide  the $\ell_i\rightarrow \ell_j\gamma$ decay rate mediated by Yukawa interactions with a neutral scalar for a general Lagrangean \eqref{eq:lag_g-2}. We consider only the cases where either the fermion $F$ or the boson $H$ propagating in the loop are much heavier than the leptons. If the interaction is purely left- or right-handed (either $c_{Li} = 0$ or $c_{Ri} =0 $ for all $i$), the decay rate is \cite{Lavoura:2003xp} 
\begin{equation}\label{eq:LFVrates2}
\begin{aligned}
\Gamma(\ell_i\rightarrow \ell_j\gamma) = 
&\frac{\alpha_e}{4 m_i^3}\left(m_i^2-m_j^2\right)^3 \left(m_i^2+m_j^2\right) 
\left(c_{Xj}^*c_{Xi}^{\phantom *}\right)^2  |F_1(M_F^2/m_H^2)|^2\,,
\end{aligned}
\end{equation}
where $X=L,R$, and $F(t)$ in the limit $m_i^2, m_j^2 \rightarrow 0$ reads
\begin{equation}
\begin{aligned}
F_1(t) &=\frac{i}{16\pi^2 m_H^2}\bigg[ \frac{t^2-5t-2}{12(t-1)^3} + \frac{t \ln(t)}{2(t-1)^4}\bigg]\,.\\
\end{aligned}
\end{equation}
Taking $t\rightarrow\infty$ and $m_i \gg m_j$ one recovers equation \eqref{eq:LFVrateF}. For a scalar more massive than the fermion $F$, taking $t\rightarrow 0$ and $m_i \gg m_j$ we obtain
\begin{equation}\label{eq:LFVrateS}
\Gamma^H(\ell_i\rightarrow \ell_j\gamma) = 
\frac{\alpha_e}{144}\left(\frac{c_{Xj}^*c_{Xi}^{\phantom *}}{16 \pi^2}\right)^2 \frac{m_i^5}{m_H^4}\,.
\end{equation}

If  both left- and right-handed interactions are present, the leading contribution reads 
\begin{equation}\label{eq:LFVrates3}
\begin{aligned}
\Gamma(\ell_i\rightarrow \ell_j\gamma) = 
&\frac{\alpha_e}{4 m_i^3} \left(m_i^2-m_j^2\right)^3 M_F^2  |F_2(M_F^2/m_H^2)|^2   
\left[\left(c_{Rj}^* c_{Li}^{\phantom *}\right)^2+\left(c_{Lj}^* c_{Ri}^{\phantom *}\right)^2\right]\,\,,
\end{aligned}
\end{equation}
with 
\begin{equation}
\begin{aligned}
F_2(t) &=\frac{i}{16\pi^2 m_H^2}\bigg[ \frac{t-3}{2(t-1)^2} + \frac{ \ln(t)}{(t-1)^3}\bigg]\,.\\
\end{aligned}
\end{equation}
For  $t\rightarrow 0$, $t \rightarrow \infty$ and  $m_i \gg m_j$  yields, respectively,
\begin{equation}\label{eq:LFVrates4}
\begin{aligned}
\Gamma^F(\ell_i\rightarrow \ell_j\gamma) &= \frac{\alpha_e}{16}
\left[
\left(\frac{  c_{Rj}^* c_{Li}^{\phantom *}}{16 \pi^2}\right)^2+
\left(\frac{c_{Lj}^* c_{Ri}^{\phantom *}}{16 \pi^2}\right)^2
\right]\, \frac{m_i^3}{M_F^2}\,,\\
\Gamma^H(\ell_i\rightarrow \ell_j\gamma) &= \frac{\alpha_e}{4}\left[
\left(\frac{  c_{Rj}^* c_{Li}^{\phantom *}}{16 \pi^2}\right)^2+
\left(\frac{c_{Lj}^* c_{Ri}^{\phantom *}}{16 \pi^2}\right)^2
\right]
\frac{m_i^3M_F^2}{m_H^4} \bigg(\frac{3}{2} +\ln \frac{M_F^2}{m_H^2}\bigg)^2\,.
\end{aligned}
\end{equation}
\end{widetext}
Here we neglected terms proportional to $\left(c_{Xj}^*c_{Xi}^{\phantom *}\right)^2$; assuming $\left(c_{Xj}^*c_{Xi}^{\phantom *}\right)^2= \mathcal{O}\left( c_{Rj}^* c_{Li}^{\phantom *},\ c_{Lj}^* c_{Ri}^{\phantom *} \right)^2$, the results \eqref{eq:LFVrates3}-\eqref{eq:LFVrates4} are valid up to corrections of order $m_i/M_F$.
The results apply for the BSM Yukawa couplings with the physical Higgs in models A,C,D and in model B for $\alpha_{\kappa}^{ij} \rightarrow \alpha_{\kappa}^{ij}/2$.

\subsection{\label{sec:appscalar} Mass Matrices and Scalar Potential}
The VEVs in terms of the parameters of the potential \eqref{eq:scalarV} are obtained as
\begin{equation}\label{eq:vevsdef}
\begin{aligned}
&v_s^2 = \frac{\mu_s^2 - \frac{\delta}{2\lambda}\ \mu^2 }{u+n v - n \frac{\delta^2}{4\lambda}}\,,\\
&v_h^2 =   \frac{\mu^2 - \frac{\delta n }{2 (u+n v)}\mu_s^2 }{\lambda - n \frac{\delta^2}{4 (u+ n v) }}        = \frac{1}{\lambda}\bigg(\mu^2-n\delta \frac{v_s^2}{2}\ \bigg)\,,
\end{aligned}
\end{equation}
with $n=1,3$ for the vacuum solutions $V^-$ and $V^+$, respectively. If the trilinear term $\mu_{\rm det}$ is switched on, for $V^+$ one should replace $\mu_s^2\rightarrow \mu_s^2 + \mu_{\rm det}v_s/\sqrt{2}$ and solve accordingly for $v_s$. A detailed analysis of the vacuum structure can be found in \cite{Bai:2017zhj} for a similar case. Before the scalars acquire these VEVs the potential is symmetric under the transformation $S \rightarrow U_{\psi_L} S\ U^{\dagger}_{\psi_R}$, where $U_i$ are $3\times3$ unitary matrices, each with $9$ degrees of freedom. In the case of a muon-aligned $V^-$, the VEV in $s_{22}$ breaks this symmetry into $U(2)_{\psi_L}\times U(2)_{\psi_R}\times U(1)$. The number of massless modes in $S$ is then  $2\cdot 9 - 2\cdot 4 - 1 = 9$. 
In $V^+$, the universal VEVs break $U(3)_{\psi_L}\times U(3)_{\psi_R}\to U(3)_{diag}$, yielding 9 Goldstone modes as well.
We  assume that additional mass terms prohibit the presence of massless Goldstones.
The symmetries involving the Higgs are the same as in the SM, rendering 3 massless states, which are eaten by $W^\pm, Z$.

\begin{table*}
	\centering
	\setlength{\extrarowheight}{4pt}
	\renewcommand{\arraystretch}{1.2}
	\rowcolors{1}{}{lightgray}
	\begin{tabular}{`ccc|ccc`}
		\toprule
		\rowcolor{LightBlue}
		{\ \ \bf $\bm{\overline{f^{\,Q}},\, f'^{\,Q+1}}$\ \ }& $\bm{c_L^W}$ & $ \qquad \bm{c_R^W}\qquad $  & {\ \ \bf $\bm{\overline{f^{\,Q}},\, f'^{\,Q+1}}$\ \ } & $\bm{c_L^W}$ & $\qquad \ \bm{c_R^W}\qquad $\\  \midrule	
		$\overline{\ell},\,\nu$  & $c_{\theta_L} c_{\theta_L^0} + C_{0} s_{\theta_L}s_{\theta_L^0}$ & 0  & $\overline{\psi^0},\,\psi^{+1}$  & $\ \ C_1 c_{\theta_L^0} $ & $C_1$\\
		$ \overline{\psi^{-1}}$, $\psi^{0} $ & $s_{\theta_L} s_{\theta_L^0} + C_{0} c_{\theta_L}c_{\theta_L^0}$ & $\ \ C_0 c_{\theta_R}$ & 
		$\overline{\nu},\,\psi^{+1}$  & $-C_1 s_{\theta_L^0} $ & 0\\
		$\overline{\ell}$, $\psi^0 $  & $c_{\theta_L} s_{\theta_L^0} - C_{0} s_{\theta_L}c_{\theta_L^0}$ & $-C_0 s_{\theta_R}$  & 
		$\overline{\psi^{-2}},\,\psi^{-1}$  & $\ \ C_{-1} c_{\theta_L} $ & $\ \ C_{-1}c_{\theta_R}$\\
		$\overline{\psi^{-1}},\,\nu$ & $ s_{\theta_L}c_{\theta_L^0} - C_{0} c_{\theta_L}s_{\theta_L^0}$ & 0&
		$\overline{\psi^{-2}},\,\ell$ & $- C_{-1} s_{\theta_L} $ & $-C_{-1}s_{\theta_R}$\\
		\bottomrule
	\end{tabular}
	\caption{Coefficients of the $W^{-}$ boson interactions with fermions in the mass basis, see Eq.~\eqref{weaklag}. The non-vanishing Clebsch-Gordan coefficients are $C_0^B = -C_{-1}^B = \sqrt{2}$,  $C_0^C = C_{-1}^D = 1$ and  $C_1^F = -C_{0}^F = \sqrt{2}$. Angles should be taken from Tab.~\ref{tab:fangles} according to the vacuum structure and the lepton flavor $\ell$.}
	\label{tab:Wint}
\end{table*}

 The $S-H$ mixing in  the mass Lagrangean $V^{\text{mass}}$  in vacuum $V^-$ is obtained from
\begin{equation}\label{eq:scalarmasses}
\begin{aligned}
\frac{\partial^2V}{\partial h\partial h }\bigg\vert_{S,H=0} &={m_h^2} = -\mu^2 + 3v_h^2\lambda+ \frac{1}{2}\delta v_s^2  \\
&=   \frac{2(u+v)\mu^2-\delta\mu_s^2 }{ (u+v)-\delta^2/4\lambda}\,, \\
\frac{\partial^2V}{\partial s_{22} \partial s_{22}  }\bigg\vert_{S,H=0} &= {m_s^2} =  -\mu_s^2 + 3v_s^2(u+v) + \frac{1}{2}\delta v_h^2\\
&= \ \frac{ 2\lambda \,\mu_s^2 -  \delta\mu^2}{\lambda - \delta^2/ 4(u+v) } \,, \\
\frac{\partial^2V}{\partial h\partial s_{22} }\bigg\vert_{S,H=0} &= m_{sh} = \delta \,v_s v_h \\
&=  \frac{\delta}{2\sqrt{\lambda(u+v)}} \ m_sm_h \,.
\end{aligned}
\end{equation}
Thus, $h$ and $s_{22}$ mix according to  
\begin{equation}\label{eq:mixingMscalars}
\begin{aligned}
V^{\text{mass}}(s_{22},h) =  \frac{1}{2} \begin{matrix}\begin{pmatrix} s_{22} \, , & h \end{pmatrix}\end{matrix}
 \left( \begin{array}{cc} m_s^2 & m_{sh}  \\ m_{sh} & m_h ^2\end{array} \right)
\begin{pmatrix} s_{22} \\ h \end{pmatrix}\,,\quad
\end{aligned}
\end{equation} 
with  eigenvalues 
\begin{equation}
m_{_{2}^{1}}  = \frac{1}{2}\, \Big[ m_s^2 +m_h^2\pm \sqrt {\big(m_s^2-m_h^2\big)^2 + 4m_{sh}^2} \ \Big] \,.
\end{equation}
The  masses of the  fields which do not get a finite VEV are obtained as
\begin{equation}
\begin{aligned}
\frac{\partial^2V}{\partial (S_{ii}^2) }\bigg\vert_{S,H=0} &=  \bar m_s^2 = -\mu_s^2+v_s^2\,v +\frac{1}{2}\delta v_h^2 \\
&= -\frac{u}{2(u+ v)}{m_s^2} \quad \text{ for } i = 1,3  \, .
\end{aligned}
\end{equation}
Note, $\bar m_s^2$ is positive since $u<0$ for $V^-$ (\ref{eq:vstab}).

The mass eigenstates $h_1,h_2$ can be expressed as in  \eqref{eq:rotation} in terms of the angle $\beta$, where
\begin{equation}\label{eq:tanbeta}
\tan 2\beta = \frac{2m_{sh}}{ m_{s}^2 -{m_h^2} }
 =\frac{2\delta v_h v_s}{m_s^2-m_h^2}\,.
\end{equation}
Expanding for $m_h\ll m_s$ yields (\ref{eq:s2b}).

The VEVs of the $S$ and $H$  scalars induce mixing between the BSM fermions and the leptons. Defining   $f_X = (e_X, \mu_X, \tau_X, \psi_{X1}, \psi_{X2}, \psi_{X3})^T$, 
$X=L,R$, the corresponding mass mixing term  for model A can be written as
\begin{equation}
\begin{aligned}
  \overline{f}_L \mathcal{M}_f f_R = &{\phantom +}\frac{v_h}{\sqrt{2}} \ \overline{e}_L\,Y_{e}  e_{R}  + \frac{v_h}{\sqrt{2}} \, \kappa \ \overline{e}_L\,\psi_R \\ 
  &+ \frac{v_s}{\sqrt{2}} \, \kappa' \ \overline{\psi}_{L 2}\,\mu_R + \frac{v_s}{\sqrt{2}} \, y\ \overline{\psi}_{L2}\,\psi_{R2}\\ 
  &+ M_F\ \overline{\psi}_L\, \psi_R \,,
\end{aligned}
\end{equation}
where $\frac{v_h}{\sqrt{2}} Y_e = \frac{v_h}{\sqrt{2}} \text{diag}(y_e,y_{\mu},y_{\tau})$.
Diagonalizing 
$\mathcal{M}_f  {\cal{M}}^\dagger_f$ and ${\cal{ M}}^\dagger_f \mathcal{M}_f$ to get rotations  for $f_L$ and $f_R$, respectively,
with $m_2 = M_F + \frac{v_s}{\sqrt{2}} y$, reveals mixing angles  at the order
\begin{equation}
\label{eq:nofermionmixing}
\begin{aligned}
\theta_L^A \simeq  \frac{\kappa v_h}{ \sqrt{2}M_F},  \theta_R^A \simeq  \frac{\kappa'v_s}{\sqrt{2} m_2}  
\end{aligned}
\end{equation}
for $\ell_{iL}-\psi_{Li}$, and $\mu_R-\psi_{R2}$, respectively. The mixing angles (up to order of magnitude) for the different models are given in table \ref{tab:fangles}. In models B, C, E and F, where the $\psi$ multiplets contain $Q_F = 0$ states, left-handed rotations are introduced between the $\nu_L-\psi_L^0$.

\begin{table*}[t]
	\centering
	\setlength{\extrarowheight}{2pt}
	\renewcommand{\arraystretch}{1.2}
	\rowcolors{1}{}{lightgray}
	\begin{tabular}{`ccc`}
	\toprule
		\hline
		\rowcolor{LightBlue}
		{\ \ \bf $\bm{\overline{f^{\,-1}}, f'^{\,-1}}$\ \ } & $\bm{g_V}$ & $\bm{g_V}$  \\  \midrule
		$\overline{\ell},\,\ell$  & $-\frac12 + 2s^2_{w} + \Delta g_V^\ell$  & $-\frac12 + \Delta g_A^\ell$ \\
		$\overline{\psi^{-1}}$, $\psi^{-1} $ &  $ 2\left(T^{3 }_{\psi^{-1}} + s^2_w\right) - \Delta g_V^\ell$  & $ - \Delta g_A^\ell$ \\
		$\overline{\psi^{-1}}$, $\ell $ & $\ \  -\frac{1}{2}\left[ \sin{2\theta_L}(T^{3 }_{\psi^{-1}} +\frac12) + \sin{2\theta_R} T^{3 }_{\psi^{-1}}\right] \ \ $ 
		& $\ \ -\frac{1}{2}\left[ \sin{2\theta_L}(T^{3 }_{\psi^{-1}} +\frac12) - \sin{2\theta_R} T^{3 }_{\psi^{-1}}\right] \ \ $  \\
				\midrule
		\rowcolor{LightBlue}
		{\ \ \bf $\bm{\overline{f^{\,0}}, f'^{\,0}}$\ \ } & $\bm{g_V}$  & $\bm{g_A}$ \\  \midrule	
		$\overline{\nu},\,\nu$  & $\frac12 + \Delta g^\nu$  & $\frac12 + \Delta g^\nu$ \\
		$\overline{\psi^0}$, $\psi^{0} $ &  $ 2T^{3 }_{\psi^{0}} - \Delta g^\nu$  & $ - \Delta g^\nu$ \\
		$\overline{\psi^0}$, $\nu $ & 	$\qquad \quad-\frac{1}{2}\sin{ 2\theta_L^0}( T^{3 }_{\psi^{0}} -\frac12)\qquad \quad $& 	$\qquad  -\frac{1}{2}\sin{2\theta_L^0}( T^{3 }_{\psi^{0}} -\frac12) \qquad \quad$\\ 
		\bottomrule
	\end{tabular}
	\caption{Coefficients of the $Z$ boson interactions with $Q=-1$ and $Q=0$ fermions in the mass basis, see~\eqref{weaklag}, with $\Delta g_{_{A}^{V}}^\ell =  s^2_{\theta_{L}}(T^{3 }_{\psi^{-1}} +\frac12 ) \pm  s^2_{\theta_{R}}T^{3 }_{\psi^{-1}}$ and $\Delta g^\nu =  s^2_{\theta_L^{ 0}}\left[ T^{3 }_{\psi^{0}} -\frac12\right]$. Angles should be taken from Tab.~\ref{tab:fangles} according to the vacuum structure and the lepton flavor $\ell$.}
\label{tab:Zint1}
\end{table*}

In $V^+$, where all diagonal components of $S$ acquire a VEV, one obtains ($i,j$: no sum)
\begin{equation}\label{eq:scalarmasses2}
\begin{aligned}
\frac{\partial^2V}{\partial h\partial h }\bigg\vert_{S,H=0} &= {m'_h}^2 = -\mu^2 + 3v_h^2\lambda+ \frac{3}{2}\delta v_s^2 \,, \\
\frac{\partial^2V}{\partial s_{ii} \partial s_{ii}  }\bigg\vert_{S,H=0} &= {m'_s}^2 =  -\mu_s^2 + v_s^2(3u +5v) + \frac{1}{2}\delta v_h^2\,, \\
\frac{\partial^2V}{\partial s_{ii} \partial s_{jj}  }\bigg\vert_{S,H=0} &= {m_{ss}} =  2v v_s^2-\mu_{\rm det}\frac{v_s}{\sqrt{2}}\quad (i \neq j),\\
\frac{\partial^2V}{\partial h\partial s_{ii} }\bigg\vert_{S,H=0} &= m_{sh} = \delta \,v_s v_h \,.
\end{aligned}
\end{equation}
The  normalized mass eigenstates in the basis $(s_{11},s_{22},s_{33},h)$ read
\begin{equation} \label{eq:evplus}
\begin{aligned}
h_1' &= \frac{1}{\sqrt{3}}\left(\cos\beta',\ \cos\beta',\ \cos\beta',\,\sqrt{3}\sin\beta' \right)\,,\\ 
h_2' &=- \frac{1}{\sqrt{3}}\left(\sin\beta',\ \sin\beta',\ \sin\beta',\ -\sqrt{3} \cos\beta' \right)\,,\\
h_3' &= \frac{1}{\sqrt{2}}(-1,0,1,0)\,,\\
h_4' &= \frac{1}{\sqrt{2}}(-1,1,0,0)\,,\\
\end{aligned}
\end{equation} 
with corresponding eigenvalues 
\begin{equation}
\begin{aligned}
m'_{_{2}^{1}}  &= \frac{1}{2}\, \bigg( {m'_s}^2 +{m'_h}^2+2m_{ss} \\
& {\phantom {+ }} \pm \sqrt {\big({m'_s}^2-{m'_h}^2+2m_{ss}\big)^2 + 12\,m_{sh}} \ \bigg)\,, \\\
m'_{_{3}^{4}}  &={m'_s}^2-\,m_{ss}\,.
\end{aligned}
\end{equation}

Thus, the mixing of the Higgs with the BSM scalars occurs only for the states $h'_{1,2}$, and is universal. Due to the degeneracy of $m'_{_{3}^{4}} $, any linear combination of the states $h'_{3,4}$ is an eigenvector, too. In the limit $\mu_{\rm det}, v\rightarrow 0$, the angle $\beta'$ can be easily expressed as
\begin{equation}\label{eq:tanbeta'}
\tan 2\beta' = \frac{2\sqrt{3}\,m_{sh}}{ m_{s}'^2 -{m_h'^2} }
=\frac{2\sqrt{3}\,\delta v_h v_s}{m_s'^2-m_h'^2}\,.
\end{equation}
For $m_h'\ll m_s'$ one obtains (\ref{eq:betprime}). For fermion mixing, we find, similar to \eqref{eq:nofermionmixing},
\begin{equation}
\label{eq:nofermionmixingVplus}
\begin{aligned}
\theta_L ^A\simeq  \frac{\kappa v_h}{ \sqrt{2} m_2},  \theta_R^A \simeq  \frac{\kappa'v_s}{\sqrt{2} m_2}  
\end{aligned}
\end{equation}
for $\ell_{iL}$- $\psi_{Li}$, and $\ell_{iR}$- $\psi_{Ri}$,  respectively.

\subsection{\label{sec:app-mixing} Weak Interactions after EWSB}

Chiral mixing between vector-like fermions and leptons  modifies their couplings with the weak bosons. Explicit rotations to the mass basis yield
\begin{equation}
\begin{aligned}
\psi^{-1, \, \rm gauge}_{X} &= c_{\theta_{X}} \psi^{-1}_{X} - s_{\theta_{X}} \ell_{X}\,,\\
 \ \ell^{ \,\rm gauge}_{X} &= c_{\theta_{X}} \ell_{X} + s_{\theta_{X}}  \psi^{-1}_{X}\,,\\
\psi^{0, \, \rm gauge}_{L} &= c_{\theta_{L}^0} \psi^{0}_{L} - s_{\theta_{L}^0} \nu_L\,,\\
 \ \ \ \nu^{ \,\rm gauge}_{L} &= c_{\theta_{L}^0} \nu_{L} + s_{\theta_{L}^0}  \psi^{0}_{L}\,,
\end{aligned}
\end{equation}
where $X= L,R$ and the angles $\theta$ are positive and can be found for all models in Tab.~\ref{tab:fangles}. 
After rotating to the mass basis, weak interactions are described by the Lagrangean
\begin{equation}\label{weaklag}
\begin{aligned}
\mathcal{L_{\rm W}}& = \frac{g_2}{2 \cos\theta_w} \overline{f^{\,Q}}\gamma^\mu(g_V - g_A\gamma^5)f'^{\,Q} Z_\mu \\
&+ \frac{g_2}{\sqrt{2}} \overline{f^{\,Q}}\gamma^\mu(c_L^W P_L + c_R^WP_R )f'^{\,Q+1} W^-_\mu + {\rm h.c.}\,,
\end{aligned}
\end{equation}
and for all possible combinations of fermions $f, f'$  in our models. The coefficients $c_{L,R}^W$ can be found in Tab.~\ref{tab:Wint}. Expressions for the couplings   $g_{V,A}$  are collected in Tab.~\ref{tab:Zint1}.

\bibliography{draft}
\bibliographystyle{mystyle.bst}

\end{document}